\documentclass[superscriptaddress,prd,showkeys,twocolumn,nofootinbib,longbibliography]{revtex4-1}
\usepackage[utf8]{inputenc} 
\usepackage{graphicx,xcolor,overpic,mathtools}
\usepackage{amsthm,amsmath,amssymb}
\definecolor{coolblack}{rgb}{0.0, 0.18, 0.39}
\usepackage[colorlinks]{hyperref}
\hypersetup{
    colorlinks = true,
	linkcolor= coolblack, 
	filecolor=black, 
	citecolor = coolblack,       
	urlcolor=blue, 
} 
\usepackage{textcomp}
\PassOptionsToPackage{normalem}{ulem}
\usepackage{ulem} 
\usepackage{sidenotes}
\usepackage{bm}
\usepackage{url}
\usepackage{todonotes}
\usepackage{physics}
\usepackage[makeroom]{cancel}
\usepackage{cleveref}
\usepackage{tensor}
\usepackage{mathrsfs}
\usepackage{slashed}
\usepackage{caption}
\usepackage{subcaption}
\usepackage[mode=buildnew]{standalone}

\newcommand{\comment}[1]{}

\usepackage{pgfplots}
\usepackage{diagbox} 
\usepackage{diagbox}
\usepackage{stackengine,amssymb,graphicx}

\NewDocumentCommand{\evat}{sO{\bigg}mm}{%
  \IfBooleanTF{#1}
   {\mleft. #3 \mright|_{#4}}
   {#3#2|_{#4}}%
}
\usepackage{enumerate}
\definecolor{azure}{rgb}{0.0, 0.5, 1.0}

\newcommand{\nlsm}{NL$\sigma$M }
\begin{document}

\title[]{Spinning boson stars in nonlinear sigma models and Universal Relations}

\author{Christoph Adam}
\author{Jorge Castelo Mourelle}

\affiliation{%
Departamento de F\'isica de Part\'iculas, Universidad de Santiago de Compostela and Instituto
Galego de F\'isica de Altas Enerxias (IGFAE) E-15782 Santiago de Compostela, Spain
}%

\author{Alberto Garc\'ia Mart\'in-Caro }
\affiliation{
Department of Physics, University of the Basque Country UPV/EHU, Bilbao, Spain
}%
\affiliation{EHU Quantum Center, University of the Basque Country, UPV/EHU,}

\author{Andrzej Wereszczynski}
\affiliation{
Institute of Physics, Jagiellonian University, Lojasiewicza 11, Krak\'ow, Poland
}%
\affiliation{International Institute for Sustainability with Knotted Chiral Meta Matter (WPI-SKCM2), Hiroshima University, Higashi-Hiroshima, 739-8526, Japan}

\date[ Date: ]{\today}
\begin{abstract}
Boson stars are hypothetical compact objects derived from solutions of a self-gravitating complex scalar field. In this study, we extend the traditional models by generalizing the kinetic term of the scalar field to that of a nonlinear sigma model. 
Concretely, we obtain spinning boson star solutions for a family of models parametrized by the curvature of their two-dimensional target manifold, as well as for various self-interaction potentials. We derive the global properties and multipolar structure of these solutions as a function of both the curvature of the target space and the strength of self-interactions. Our results suggest that a nonzero curvature in the target manifold can have a significant impact on the structure of the solutions, allowing for a range of notably different masses and degrees of compactness. However, we find that the relations between different multipoles are consistent with those for the standard complex scalar stars, and hence the universality of such relations is extended to curved target spaces.



\end{abstract}

\maketitle

\begin{quote}
 
\end{quote}

\tableofcontents

\section{Introduction}
In recent years, the study of boson stars --- gravitating, non topological solitons sourced by a scalar field condensate--- has enjoyed a great expansion due to their phenomenological potential in relativistic astrophysics as black hole mimickers \cite{PhysRevD.80.084023,Herdeiro:2021lwl,Rosa:2022tfv,Rosa:2023qcv,Sengo:2024pwk,Adam:2010rrj,Adam:2024zqr} or as sources of gravitational waves in merger events \cite{Bustillo:2020syj,Luna:2024kof,Siemonsen:2023hko}, as well as their cosmological implications as dark matter candidates \cite{Schunck:1998nq,Urena-Lopez:2010zva,Broadhurst:2019fsl,Chen:2020cef,Annulli:2020ilw,Amruth:2023xqj,Pozo:2023zmx,Mourelle:2024dlt} and for the evolution of the Early Universe \cite{Cotner:2017tir,Kusenko:2019kcu}.

Whilst no regular, asymptotically flat, static (or stationary) solutions of the Einstein equations sourced by a real scalar field are known \footnote{Although there are almost stationary solutions, known as oscillatons, which are also phenomenologically interesting \cite{Fodor:2019ftc}.}, the simplest models of boson stars require a single complex scalar field with a generic self-interacting scalar potential that preserves the global $U(1)$ symmetry. Stability under their own gravitational collapse is then achieved by endowing the scalar condensate with a non-vanising, conserved Noether charge associated with the corresponding global symmetry. Various scalar potentials have been explored in the literature, leading to different types of boson stars \cite{Schunck:2003kk,Grandclement:2014msa,Choi:2019mva,Guerra:2019srj,Delgado:2020udb,Adam:2022nlq}, which however satisfy a set of potential-independent universal relations involving their moments of inertia and other metric multipoles \cite{Adam:2023qxj}.

However, the requirement of a $U(1)$ symmetry for the stabilization of the scalar condensate can be extended to more general models. Indeed, this observation has allowed us to find boson star solutions in simple nonlinear sigma models \cite{Cano:2023bpe,Herdeiro:2018djx,Verbin:2007fa}. 

A non-linear sigma model (\nlsm) is a scalar field theory describing mappings between spacetime and a target manifold,
 \begin{equation}
     \phi:\mathbb{R}^{(3,1)}\longmapsto \mathcal{T}.
 \end{equation}
 
The requirement for the existence of boson star solutions within a given \nlsm can be stated in geometric terms as the following conjecture for the general condition for the existence of BSs:
{\it For a \nlsm to present BS solutions when minimally coupled to gravity, the target space must admit a toric action with at least one fixed point.}

Indeed, the necessity of a Noether charge, or equivalently, the associated global $U(1)$ symmetry for field configurations implies that the target manifold must admit a $U(1)$ group action. The requirement of asymptotic flatness (i.e., finite ADM mass) implies that the field must tend to vacuum at large distances, which in particular must have zero Noether charge. In other words, the vacuum manifold of the model must be contained in the set of fixed points under the torus action of $\mathcal{T}$.

Toric geometry deals with the study of manifolds presenting multiple toric actions and fixed points \cite{Leung:1997tw}. The simplest example is the complex plane $\mathbb{C}$, which presents a toric action:
\begin{equation}
    \mathbb{C}\ni z\to e^{i\theta}z,\qquad \theta \in \mathbb{R}.
\end{equation}
with the origin being a fixed point. The complex plane endowed with such action can be seen as a half-line with a circle fiber, which shrinks at the origin.

The next nontrivial example of toric geometry is the complex projective line, $\mathbb{C}P^1$, or Riemman sphere, which can be seen as a finte interval with a $T^1$ fiber that shrinks at the two endpoints. 

Although we will restrict our study to two-dimensional target spaces, toric geometry allows one to generalize the above construction to arbitrary dimensions. In fact, the manifold $\mathbb{C}P^{N-1}$ can be represented as an $(N-1)$-simplex where each points presents a ($N-1$)-torus $T^{N-1}$ fiber. One of the $T^1$ fibers shrinks on each of the $(N-2)$-simplexes at the boundaries of the original $(N-1)$-simplex, while at its vertices the whole $T^{N-1}$ fiber shrinks. A simple example would be the complex projective plane $\mathbb{C}P^2$, whose toric diagram is a triangle with a $T^2$ fiber, whose sides present a $T^1$ fiber that shrinks at the vertices. Therefore, a \nlsm with such a space as the target manifold would allow for Q-ball or boson star solutions with different kinds of Noether charges \cite{Sawado:2020ncc}. Another interesting type of \nlsm that has been considered in the astrophysical context are the Skyrme-type models, which, on top of a Noether charge, can present a topological charge. Hence, such models have been proposed both as models for boson stars \cite{Kirichenkov:2023omy} and also for neutron stars \cite{Adam:2015lpa,Adam:2020yfv}. In the latter case, solutions were found to satisfy universal relations similar to those of other perfect fluid stars \cite{Adam:2020aza}.

Nonlinear sigma models are also interesting from the fundamental physics point of view, as they generically emerge as low energy effective actions from dimensional compactification in supergravity and string theory. In particular, scalar field multiplets emerge as moduli of Calabi-Yau manifolds \cite{Ortin:2015hya} or components of chiral and higher-dimensional gauge fields in flux compatification scenarios \cite{Roest:2004aqa}. In this context, star-like solutions are known as \emph{moduli stars} \cite{Krippendorf:2018tei}. Furthermore, in the cosmological context, scalar fields with nonstandard kinetic terms and nonlinear self-couplings are interesting for modified theories of gravity. In fact, theories such as massive gravity can be reformulated as a \nlsm coupled to Einstein gravity \cite{deRham:2015ijs}. 

 Static boson star solutions in \nlsm have already been studied in \cite{Cano:2023bpe}, and the spinning case was also constructed for some target manifold curvatures, where also some first multipoles were studied \cite{Collodel:2019uns}. In this paper we will focus on the physically more relevant case of spinning boson stars \cite{schunck1998rotating,Yoshida:1997qf}.  In particular, we will show that a nontrivial topology of the target space manifold has a noticeable effect on the properties of the corresponding stars, namely, on their compactness and energy distribution, as well as on the relations satisfied by their multiples, when the values for the target manifold curvatures are sufficiently distinct. We therefore will be able to answer the question how the topology of the target space affects the universal relations of the corresponding Boson Stars. We also focus our research on the multipolar decomposition, intending to study how the universality is affected by the distinct types of target field manifolds.


\section{Theoretical set-up}\label{setup}

\subsection{Nonlinear sigma models }
\label{secTSet}

One of the most studied examples of a non-linear sigma model is the $S^2$-sigma model, in which the target space is the 2-sphere. It can be constructed as a theory of three real scalar fields $\phi_a$, $a=1,2,3$ subject to the constraint $\phi_ a\phi_ a=\rho^2$, where $\rho $ is the radius of the $2$-sphere. This parameterization corresponds to the Cartesian coordinates of the sphere as an embedded surface on a 3-dimensional Euclidean space. In such representation, the Lagrangian for the massless $S^2$-sigma model (in flat spacetime) is
\begin{equation}
    \mathcal{L}=\frac{1}{2}\partial_\mu \phi_a\partial^\mu\phi_a\, .
    \label{S2nlsmCart}
\end{equation}

Alternatively, we can use stereographic projections from either the north or the south pole of $S^2$ to the (complex) plane, which provide a minimal atlas for the sphere, to describe the dynamics. Indeed, using the south chart 
\begin{equation}
    \Phi=\rho\frac{\phi_1+i\phi_2}{\rho-\phi_3}\in\mathbb{C},
\end{equation}
with inverse mappings given by 
\begin{equation}
    \phi_1+i\phi_2=\frac{2\rho^2\Phi}{\rho^2+|\Phi|^2},\quad \phi_3=\rho\frac{|\Phi|^2-\rho^2}{\rho^2+|\phi|^2},
\end{equation}
we can rewrite \eqref{S2nlsmCart} as
\begin{equation}
    \mathcal{L}=\frac{1}{2}\frac{4\rho^4}{(\rho^2+|\Phi|^2)^2}\partial_\mu\Phi^*\partial^\mu\Phi
\end{equation}
which is the usual $\mathbb{CP}^1$ sigma model. 

Nonlinear sigma models are invariant under the isometry group of the corresponding target manifold. In particular, the $S^2$ sigma model is invariant under $O(3)$ transformations of the fields $\phi_a$, and, after stereographic projection, this symmetry is explicitly broken by a suitable potential to a global $U(1)$ of the complex field $\Phi$. 

In the present work we will analyze the following fa\-mily of sigma models with two dimensional target space manifold:
\begin{equation}
 \mathcal{L}_{\Phi}=-\frac{K(|\Phi|^2)}{2}\partial_{\alpha}\Phi^*\partial^{\alpha}\Phi-V\left(|\Phi|^2\right) .
    \label{lagrangian}
\end{equation}
where
\begin{equation}
    K(|\Phi|^2)=\frac{1}{\left(1-\frac{\kappa^2}{4}|\Phi|^2\right)^2}\,.
    \label{K}
\end{equation}
is the most general metric for a two dimensional \nlsm with constant curvature target space and a $U(1)$ isometry \cite{Cano:2023bpe}. Indeed, it coincides with the above-mentioned $\mathbb{CP}^1$ sigma model (with $\mathbb{S}^2$ as the target manifold) with the identification $\kappa^2=-4/\rho^2<0$. On the other hand, for $\kappa^2>0$, it describes the metric of a negative constant curvature target space, i.e. the hyperbolic plane $\mathbb{H}^2$\footnote{We remark that the $\mathbb{H}^2$ model is equivalent to the axion-dilaton model which appears generically in superstring theory \cite{Lidsey:1999mc}. (See \cite{Cano:2023bpe} for an explicit derivation of this equivalence)}. Finally, for $\kappa=0$, we arrive at the flat target space case.  

On the other hand, we will consider two qualitatively different families of potentials. Obviously, since they respect the global $U(1)$ invariance of the model, they can depend only on the absolute value of the scalar field.  The first one is the widely used $\phi^4$ potential 
\begin{equation}
    V(|\Phi|^2)=\mu^2 \Phi^*\Phi+\Lambda (\Phi^*\Phi)^2\,,
    \label{potential_quartic}
\end{equation}
where $\mu$ is the mass parameter of the perturbative excitations and $\Lambda$ is a dimensionless self coupling parameter. The second one is a two vacuum potential
\begin{equation}
    V(|\Phi|^2)=\mu^2\frac{|\Phi|^2}{(1+\beta|\Phi|^2)^2}
\end{equation}
with $\beta$ being a positive parameter. 
In terms of the $O(3)$ iso-scalar field $\phi_a$ it can be written as
\begin{equation}
    V(\phi_a)=\frac{\mu^2}{(1+\beta \rho^2)^2} \frac{(\rho-\phi_3^2)}{(1- \frac{\phi_3}{\rho} \frac{1-\beta \rho^2}{1+\beta \rho^2} )^2}\,,
\end{equation}
which has two vacua at the north and south poles, ${\phi_3=\pm \rho}$.

\subsection{\nlsm stars}
\label{sigmaBS}

The boson stars considered in this paper are described by the Einstein-$\sigma$ model action, 
\begin{equation}
    \mathcal{
    S}_{\Phi}=\frac{1}{16\pi G}\int \left(R+\mathcal{L}_{\Phi}\right)\sqrt{-g}d^4x.
    \label{action}
\end{equation}
Here, $g$ is the metric determinant, $R$ the Ricci scalar, and $\mathcal{L}_{\Phi}$ is the minimally coupled version of \eqref{lagrangian}. Furthermore, we have rescaled the scalar field as
\begin{equation}
    \Phi\to \sqrt{8\pi G}\Phi\equiv M_P^{-1}\Phi
\end{equation}
so that the new scalar field variable is dimensionless, and so is the rescaled curvature parameter, $\gamma^2\equiv M_P^{2}\kappa^2$. Then its value can be interpreted as a ratio $\abs{\gamma}=M_P/E$ between the characteristic energy scale $E$ of the target manifold (dimensionful) curvature and the (reduced) Planck mass $M_P=(8\pi G)^{-1/2}$.

By varying the action \eqref{action} the following equations arise,
\begin{align}
    &R_{\alpha\beta}-\frac{1}{2}Rg_{\alpha\beta}=8\pi T_{\alpha\beta}, \notag\\[2mm]
g^{\alpha\beta}\nabla_\alpha \left(K \nabla_{\beta}\Phi \right) &=\frac{dV}{d|\Phi|^2}\Phi -\frac{1}{2}g^{\alpha\beta}\nabla_{\alpha}\Phi^*\nabla_{\beta}\Phi \frac{dK}{d|\Phi|^2}\Phi,
\label{kg}
\end{align}
where $R_{\alpha\beta}$ is the Ricci tensor and $T_{\alpha\beta}$ is the canonical stress-energy tensor of the scalar field,
\begin{equation}
\begin{split}
    T_{\alpha\beta}=&K(|\Phi|^2)\nabla_{(\alpha}\Phi^*\nabla_{\beta)}\Phi\\
    &-g_{\alpha\beta}\left[\frac{K(|\Phi|^2)}{2}g^{\mu\nu}\nabla_{(\mu}\Phi^*\nabla_{\nu)}\Phi+V\left(|\Phi|^2\right)\right].
    \end{split}
    \label{stress}
\end{equation}
For the above stress-energy tensor to satisfy stationarity and axial symmetry, the scalar field ansatz takes the form
\begin{equation}
     \Phi(t,r,\theta,\psi)=\phi(r,\theta)e^{-i(w t+n\psi)}.
     \label{scalar}
 \end{equation}
Here $\phi(r,\theta)$ is the profile of the star, $w \in \mathbb{R}$ is the angular frequency of the field (associated to the rotation of the field phase in internal space), and $n \in \mathbb{Z}$  is the \textit{azimutal harmonic index}, or \textit{azimutal winding number} \cite{Vaglio:2022flq,Ryan:1996nk}. This parameter enters the problem as an integer related to the star's angular momentum. 

 Finally, we assume the following ansatz for the metric, describing the stationary 
and axisymmetric space-time \cite{Herdeiro:2015gia,Ryan:1996nk},
\begin{equation}
\begin{split}
    ds^2=&-e^{2\nu}dt^2+e^{2\beta}r^2\sin^2\theta\left(d\psi-\frac{W}{r}dt\right)^2\\
    &+e^{2\alpha}(dr^2+r^2d\theta^2),
    \end{split}
    \label{metric}
\end{equation}
where $\nu, \alpha, \beta$ and $W$ are functions dependent only on $r,\theta$.

For $K=1$, that is, $\gamma=0$, we recover the usual $U(1)$ Boson Stars (BSs). We note that changing the absolute value of the parameter $\gamma\neq 0$ is effectively equivalent to changing the strength of gravity, that is, the value of Newton's constant $G$. Indeed, a rescaling 
$
    \phi\to {\phi}/{\abs{\gamma}}
$
brings us to a model where the magnitude of the curvature of the target space only appears as a modified Newton's constant
$
    G\to G|\gamma|^2
$ (accompanied with the corresponding rescaling  of the $\Lambda$ parameter in the potential). This explicitly underlines that we have three qualitatively very distinct cases corresponding to three types of constant curvature geometries of the target space: positive $\mathbb{S}^2$, negative $\mathbb{H}^2$, and flat $\mathbb{C}$. 
However, in our numerical analysis, we use the formulation with the curvature parameter $\gamma$ of the target space explicitly taken into account. 

\section{Multipolar structure and global properties}\label{multi}

In this section, we study the multipolar expansion of the spacetime metric and the related physical properties relevant for the data analysis. Within the framework of General Relativity, there are two classes of multipoles, originating from the energy density and the current density, respectively, as discussed in \cite{RevModPhys.52.299,PhysRevD.52.821}. These multipoles are indispensable from both theoretical and astrophysical perspectives. The foundational principles of metric multipole expansions were established in \cite{Geroch:1970cd,Hansen:1974zz,fodor1989multipole}. Subsequently, this concept was applied in the context of neutron stars, as referenced in \cite{Pappas:2018csu,1976ApJ...204..200B}. Our approach is consistent with these precedents.

\subsection{Multipole moments}
In the same fashion as in \cite{Adam:2023qxj}, we obtain the multipoles following the method described in \cite{1976ApJ...204..200B,morse1954methods}. 
First, we reparameterize our metric functions as $\omega=\frac{W}{r},\hspace{0.4cm}B=e^{\nu+\beta}$ and expand our new metric functions using polynomial bases for spherical coordinates, that is, the Legendre polynomials $P_l(\cos\theta)$ and Gegenbauer polynomials $T_l^{\frac{1}{2}}(\cos\theta)$. The radial coefficients are also expanded in inverse powers of the radial coordinate,
\begin{eqnarray}
        \nu&=&\sum_{l=0}^{\infty}\bar{\nu}_{2l}(r)P_{2l}(\cos\theta),\hspace{0.4cm}   \hspace{0.9cm} \bar{\nu}_{2l}(r)= \sum_{k=0}^{\infty}\frac{\nu_{2l,k}}{r^{2l+1+k}}, \nonumber  \\
         \omega&=&\sum_{l=0}^{\infty}\bar{\omega}_{2l-1}(r)\frac{dP_{2l-1}(\cos\theta)}{d\cos\theta},     \hspace{0.2cm} \bar{\omega}_{2l-1}(r)= \sum_{k=0}^{\infty}\frac{\omega_{2l-1,k}}{r^{2l+1+k}}, \nonumber  \\
          B&=&1+\sum_{l=0}^{\infty}\bar{B}_{2l}(r)T_{2l}^{\frac{1}{2}}(\cos\theta),     \hspace{0.2cm}     \hspace{0.2cm} \bar{B}_{2l}(r)= \frac{B_{2l}}{r^{2l+2}},       
    \label{multipoles}
\end{eqnarray}

For more details, we refer to \cite{Adam:2023qxj,Ryan:1996nk,Doneva:2017jop}. Once we know the functions $\nu$, $\omega$, and $B$, the coefficients are easily extracted. The combinations that provide the relevant multipoles \cite{Pappas:2014gca} are

\begin{equation}
\begin{split}
 &M_0=\bar M = -\nu_{0,0},\\
 &S_1=J =\frac{\omega_{1,0}}{2},\\
 &M_2=Q=\frac{4}{3}B_{0}\nu_{0,0}+\frac{\nu_{0,0}^3}{3}-\nu_{2,0},\\
 &S_3=-\frac{6}{5}B_0\omega_{1,0}-\frac{3}{10}\nu^2_{0,0}\omega_{1,0}+\frac{3}{2}\omega_{3,0},\\
 &M_4 =-\frac{32}{21}b_0\nu^3_{0,0}-\frac{16}{5}b^2_0\nu_{0,0}+\frac{64}{35}b_2\nu_{0,0}+\frac{24}{7}b_0\nu_{2,0}\\
   &+\frac{3}{70}\nu_{0,0}\omega^2_{1,0}-\frac{19}{105}\nu^5_{0,0}+\frac{8}{7}\nu_{2,0}\nu^2_{0,0}-\nu_{4,0}.
    \end{split}
    \label{multipoles_orders}
\end{equation}

A brief discussion of the numerical errors associated with the multipole extraction and the simulations is provided in \Cref{errors_numerical}.

Here we have defined the dimensionless mass:
\begin{equation}
       \bar{M}=\frac{M\mu }{8\pi M_P^2}\equiv G M \mu 
\end{equation}
Let us now, and for the rest of the text, take a geometrized system of units in which $G=1$. 

Furthermore, it will become convenient to work with dimensionless multipolar quantities, or \emph{reduced multipoles}, as is standard in the literature \cite{2013Sci...341..365Y,Yagi:2014bxa}. Following the conventions established in \cite{Adam:2022nlq,Adam:2023qxj}, we employ the following definition of dimensionless quantities,
\begin{equation}
\begin{split}
    &m_{2m}\equiv (-1)^m\frac{M_{2m}}{\chi^{2m}M_0^{2m+1}}\,,\\
    &s_{2m-1}\equiv (-1)^{m+1}\frac{S_{2m-1}}{\chi^{2m-1}M_0^{2m}}\,.
\end{split}
\end{equation}

In the above expression, $m$ can take any integer value, establishing a consistent, dimensionless formula for each multipolar order. These quantities are connected to those obtained by the multipolar expansion of \cref{multipoles_orders}. Although detailed calculations from the metric in \cref{metric} to the expansion of metric functions in \cref{multipoles_orders} are technically complex, the entire process is comprehensively described in \cite{Pappas:2012ns,Pappas:2012qg} and thoroughly reviewed in Section 2.2 of \cite{Sukhov:2023rln}. 

Owing to the infinite extension of the fields that form the BSs, the notion of the surface of the star is not well defined. Usually, the surface is assumed to delimit a volume that encloses $99\%$ of the mass, $M_{99}$. We use this definition. On the other hand, for the mass of the BSs we use the total mass of the gravitating field, $M\equiv M_0$. The relations for the usual BSs were shown to be fulfilled when this choice is made \cite{Adam:2022nlq}.

The quantities under our scope for the next subsections are thus

\begin{equation}
\begin{split}
    &\chi=\frac{S_1}{M_{0}^2},
    \hspace{0.3cm}\bar{Q}=m_2=\frac{M_2}{M_{0}^3\chi^{2}}\,,\hspace{0.1cm}\\
    &\bar{s}_3=-\frac{S_3}{M_{0}^4\chi^3}, \hspace{0.3cm}\bar{m}_4=\frac{M_4}{M_0^5\chi^4}\,.
    \label{reduced}
\end{split}
\end{equation}

\subsection{Moment of inertia and differential rotation}
As shown in \cite{Adam:2022nlq,Adam:2023qxj}, and also discussed in \cite{DiGiovanni:2020ror}, unlike regular perfect fluid stars, rotating BSs cant be treated as rigidly rotating objects in a perturbative expansion, but require a non-perturbative treatment \cite{Silveira:1995dh,Ferrell:1989kz}. In order to define the corresponding angular velocity, we take advantage of the fact that there is a natural four-vector associated with the global $U(1)$ symmetry of the Lagrangian, i.e., the associated Noether current:
\begin{equation}
    j^{\mu}=\frac{i}{2}K(|\Phi|^2)\left[\Phi^*\partial^{\mu}\Phi-\Phi\partial^{\mu}\Phi^*\right],
\end{equation}
which gives rise to a conserved particle number, $N=\int  j^0 \sqrt{-g}d^3x$. Then, the differential angular velocity of BSs can be defined as $ \Omega=j^{\psi}/j^{t}$, which agrees with that obtained by Ryan \cite{Ryan:1996nk} in the strong-coupling approximation.
Having $\Omega$ as a function of $r$ and $\theta$, we compute the moment of inertia for our differentially rotating system as the ratio between angular momentum density and angular velocity, integrated over the whole space
\begin{equation}
    I=\int_0^{\pi}\int_0^{\infty}\frac{T^t_{\psi}(r,\theta)}{\Omega(r,\theta)}r^2\sin\theta e^{\nu+2\alpha+\beta}drd\theta \ .
\end{equation}
As with the multipole moments, we may define a reduced moment of inertia that is normalized by an appropriate power of the mass such that the result is dimensionless:
\begin{equation}
    \bar{I}=\frac{I}{M_{0}^3}\,.
    \label{dimensionlesI}
\end{equation}

\section{Solutions for quartic self-interactions}

Solutions involving rotating boson stars with a quartic self-interaction potential have been extensively explored \cite{Vaglio:2022flq,Vaglio:2023zpm,Adam:2022nlq,Adam:2023qxj,Khlopov:1985fch}. Given its simplicity and the range of solutions it offers for boson stars with flat target spaces, we have chosen the quartic potential as the prototypical potential for our analysis. This choice is particularly re\-levant because varying the parameter $\Lambda$ in \cref{potential_quartic}, which can have both positive and negative values, allows us to explore a broad spectrum of outcomes. Nevertheless, we will investigate a different potential with two vacua later on. 

In this section, we will investigate how the existence and properties of BSs (masses, compactness,  occurrence of the ergoregions, etc.) depend on the value of the free parameters of our models, that are the curvature $\gamma^2$ of the target space and the quartic self-interaction $\Lambda$. Specifically, we want to understand how the topology of the target space (spherical $\gamma^2>0$, flat $\gamma=0$ and hyperbolic $\gamma^2<0$) modifies the properties of gravitating solutions.  Furthermore, the solutions depend on the angular frequency $w$, whose impact will also be studied. We remark that, in principle, the most general solutions of rotating BSs are also classified by an integer, the harmonic index $n$. In the current work, we will explore only the $n=1$ branch, leaving the analysis of higher values of $n$ for a future work.

The quantities we are going to focus on are the (dimensionless) mass $\mu M$, maximal value of the modulus of the field $\phi_{max}$, quadrupolar moment $\bar Q$ and spin octupolar moment $\bar s_3$, as defined in \cref{reduced}.

\begin{figure*}[h!]
\centering
\hspace*{-0.4cm}\includegraphics[width=0.48\textwidth]{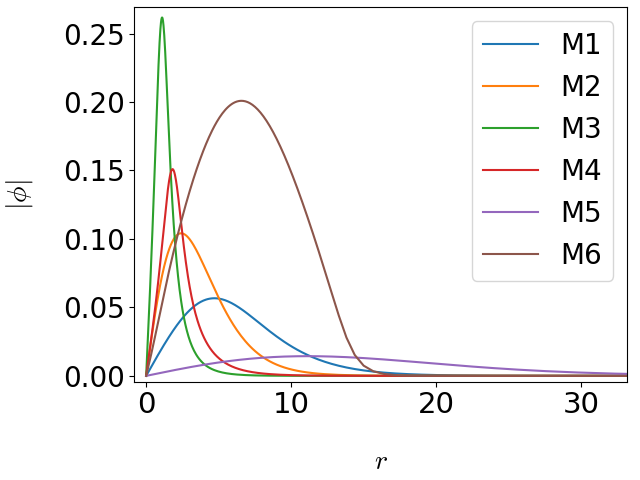}
\caption{Field modulus vs radial distance for various models: M1-$w=0.9$, $\gamma^2=0$, $\Lambda=0$; M2-$w=0.8$, $\gamma^2=10$, $\Lambda=40$; M3-$w=0.7$, $\gamma^2=-10$, $\Lambda=-100$; M4-$w=0.85$, $\gamma^2=100$, $\Lambda=100$; M5-$w=0.97$, $\gamma^2=-100$, $\Lambda=-40$;  2-vac potential M6-$w=0.2$,  $\gamma^2=0$, $\beta=400$.}
\label{maxfieldplot}
\end{figure*}

While the maximum field value $\phi_{\text{max}}$ is not commonly used as a diagnostic quantity in boson star studies—being less directly tied to conserved quantities like Noether charge or energy density—it becomes particularly meaningful in models with non-trivial target space geometry or multi-vacuum potentials \Cref{maxfieldplot}. In curved target spaces, $\phi_{\text{max}}$ indicates how deeply the field probes into regions where curvature effects become significant. This is especially relevant when the potential features multiple vacua: if $\phi_{\text{max}}$ remains small, the configuration effectively behaves as if governed by a simpler quartic interaction. However, when $\phi_{\text{max}}$ approaches the local maximum between vacua, the full structure of the potential becomes dynamically relevant. In our model, this quantity thus helps distinguish between solutions where curvature or multi-vacuum structure plays a substantial role, particularly in the case of spherical target manifolds.

Another relevant property is the compactness of the star, defined as the ratio between its mass and its radius. However, as stated above, the field configurations for\-ming the BSs extend to infinity and therefore the notion of the surface (hence the radius) is slightly diffuse.  We follow the standard prescription and define the radius $R_{99}$ as that of a sphere enclosing the $99\%$ of total ADM mass, $M_{99}$,
\begin{equation}
C=\frac{M_{99}}{R_{99}}. 
\label{compacidad}
\end{equation} 

In our numerical calculations, we vary the curvature in a rather wide range $\gamma^2 \in [-150,150]$. The value of the angular frequency $w$, or better to say, the dimensionless ratio $w/\mu$, is varied within the range of $w/\mu\sim 1$ to its initial value, which depends on the specific choice of $\gamma^2$ and $\Lambda$.

\begin{figure*}[]
\centering
\hspace*{-0.2cm}\includegraphics[width=0.54\textwidth]{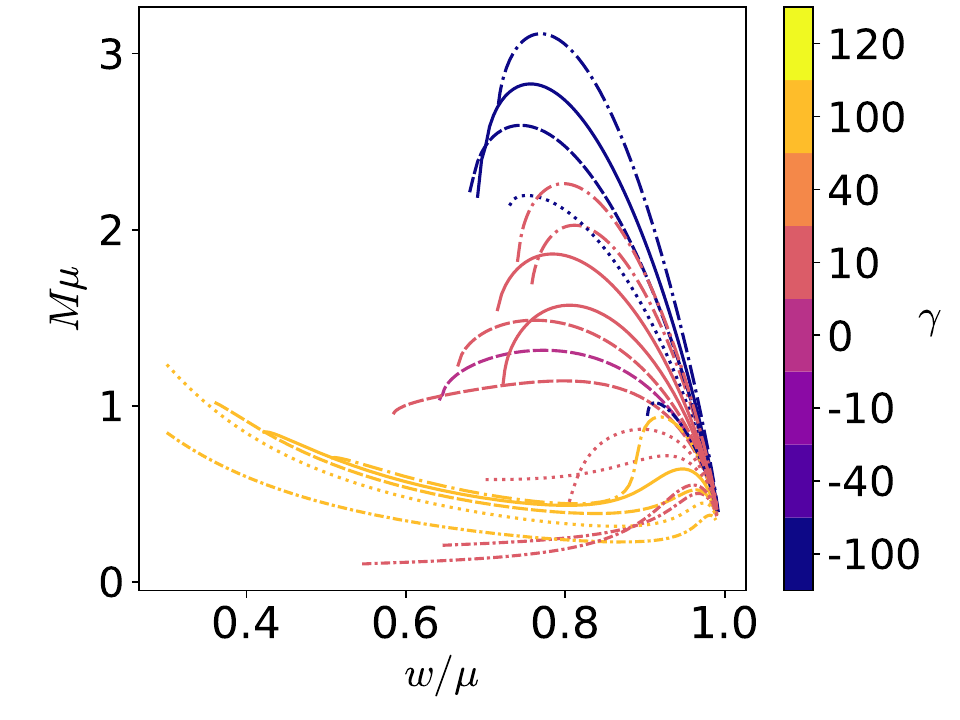}
\caption{Mass vs frequency curves for $\gamma^2=-100,-10,0,10,100$ and $\Lambda=-100,-40,0,40,100$. Both the mass and frequency are rescaled with suitable factors of $\mu$ (the boson mass) to render dimensionless quantities. For this and other plots, we have used the following linestyle codes: densely dashed for $\Lambda=0$, solid for $\Lambda=40$, dashdot for $\Lambda=100$, dotted for $\Lambda=-40$ and densely dashdoted for $\Lambda=-100$. It is important to note that, for the case $\gamma^2 = 0$, we included only the curve corresponding to $\Lambda = 0$, resulting in a total of 21 curves instead of 25.}
\label{masasc}
\end{figure*}

In \Cref{masasc} we show how the total mass of the stars depends on the field frequency as we change the curvature $\gamma$ and the parameter of the quartic self-interaction term $\Lambda$. Here we take $\gamma^2 \in [-100,100]$ and $\Lambda=-100,-40,0,40,100$. We clearly see that increasing $\Lambda$ from $-100$ to $100$ makes the maximal mass larger, while increasing $\gamma^2$ from $-100$ to $100$ acts in the opposite direction. The qualitative dependence on $w$ is the same in the cases of spherical, hyperbolic, and flat targe space.

Depending on the shape of $M-w$ curves, different criteria were used to determine the endpoint of the solution families. In cases where a decline followed a clear maximum in mass, simulations were stopped at the end of the first existence branch, beyond which solutions diverged unless a separate numerical treatment was used to access the secondary, typically unstable, branch. For curves displaying a flattened maximum, solutions were computed until a secondary turning point marked the beginning of a steeper mass decrease. In cases where the mass decayed monotonically toward zero, the endpoint was chosen based on having a sufficiently representative set of configurations. Finally, simulations were concluded when numerical convergence became inefficient in curves that reached a minimum and then increased again. These stopping conditions ensured the computed solutions covered the physically relevant regimes without unnecessary numerical overhead.

\begin{figure*}
\begin{center}
\includegraphics[clip,width=0.95\columnwidth]{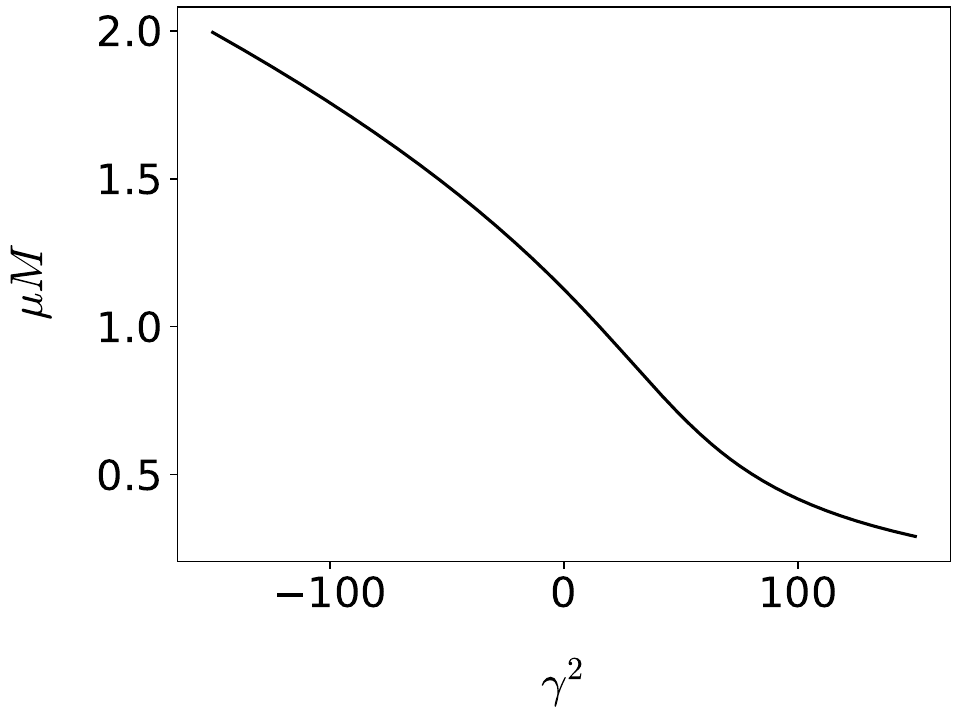}\\%
\includegraphics[clip,width=1.0\columnwidth]{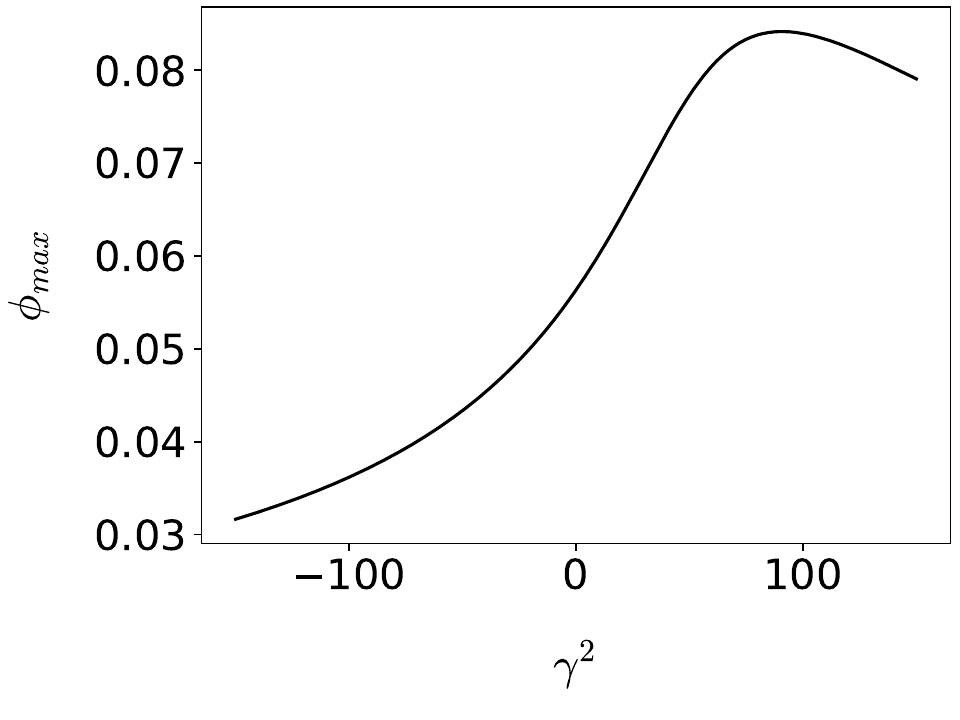}%
\caption{Mass (upper plot) and maximum field value (lower) vs $\gamma^2$ for $\omega/\mu=0.9$ and $\Lambda=1$ }
\label{changegamma}
\end{center}
\end{figure*}

To better understand the impact of the curvature of the target space, we fix the potential $\Lambda=1$, and the frequency $w/\mu = 0.9$. In \Cref{changegamma}, we show the effects of varying the dimensionless curvature $\gamma^2$ within the range of $[-150,150]$. The mass exhibits a monotonic change. It decreases as the curvature increases from negative to positive values. Specifically, at $\gamma^2=150$, the mass is $\mu M=0.29$, which is approximately $\sim 0.258$ times the mass of standard BS (where $\gamma^2=0$ and $\mu M=1.126$). For $\gamma^2=-150$, the mass increases significantly to $\mu M=1.995$. This value is about $\sim 1.772$ times the conventional BSs mass.

The behavior of the maximum of the field is more involved. Initially, as $\gamma^2$ grows from negative values, the quantity increases. However, for some value of the curvature, $\gamma^2 \approx 90$ for $\phi_{max}$, it reaches a maximum after which it decreases. We note inflection points in the mass plots and $\phi_{max}$. Their positions, approximately at $\gamma^2\approx 40$, seem to coincide. This may indicate a scaling phenomenon within this region of the parameter space.

\begin{figure*}[]
\centering
\hspace*{+0.3cm}\includegraphics[width=0.48\textwidth]{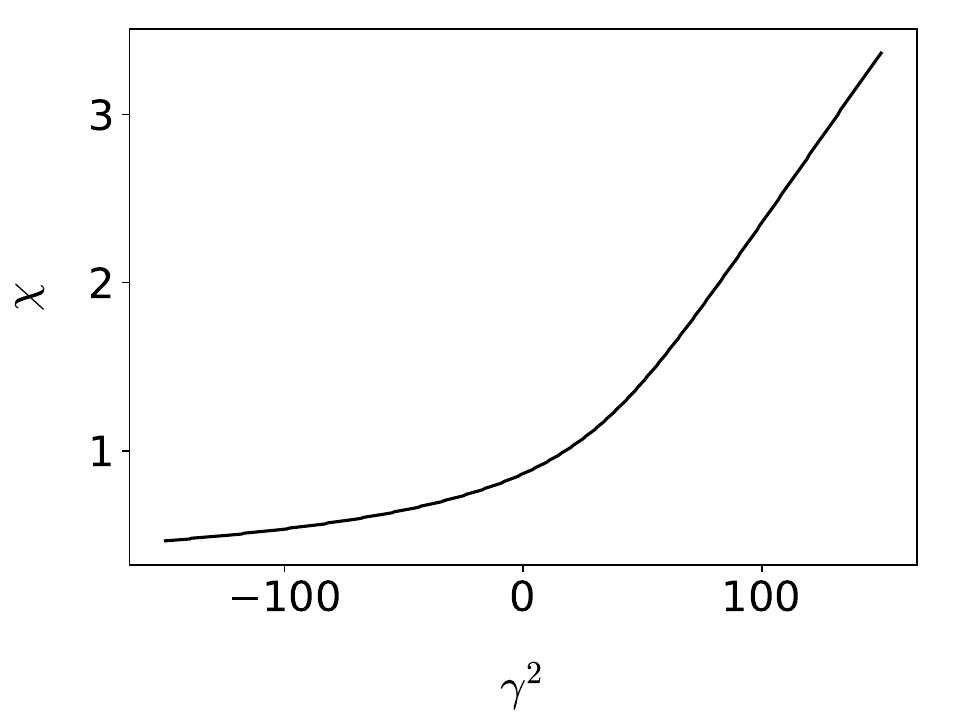}
\centering
\hspace*{-0.5cm}\includegraphics[width=0.5\textwidth]{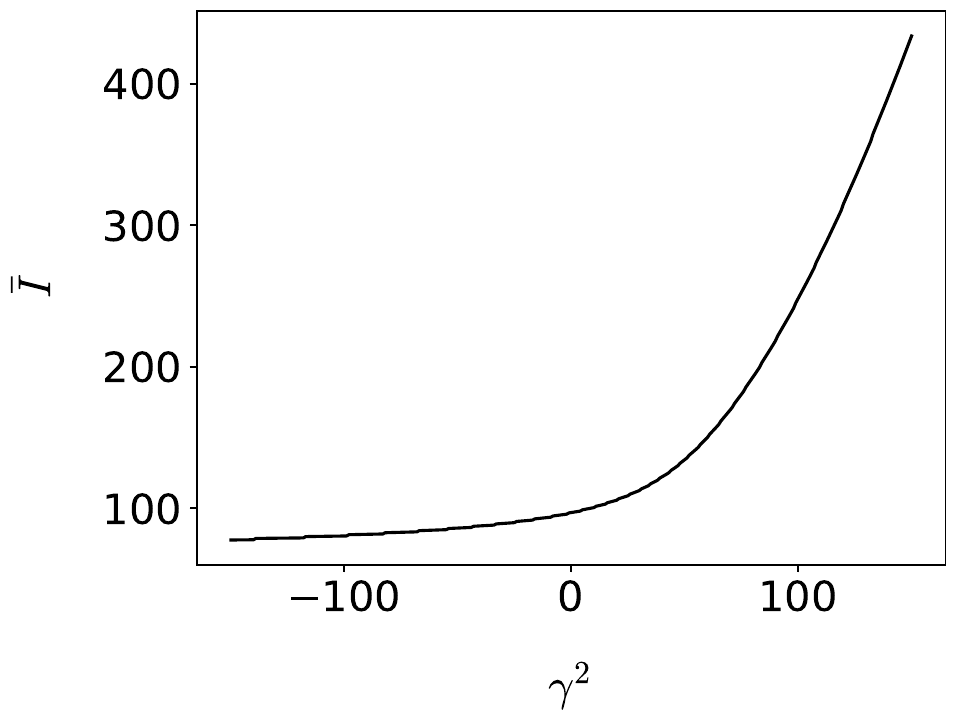}
\caption{Dimensionless spin (upper panel) and moment of inertia (lower) as functions of $\gamma^2$, for $w/\mu=0.9$ and $\Lambda=1$.}
\label{inertia_gamma}
\end{figure*}

\Cref{inertia_gamma} illustrates the behavior of the dimensionless spin and moment of inertia -defined in \Cref{reduced,dimensionlesI} - in the same range of $\gamma^2$. Both quantities exhibit a similar trend, starting at low values when $\gamma^2$ is significantly negative and increasing as $\gamma^2$ grows. In particular, the graphs can be divided into two regions: one for $\gamma^2<40$, where the increase is relatively gradual, and another for $\gamma^2>40$, where the slope becomes steep. This suggests a scaling transition around $\gamma^2\sim 40$.

\begin{figure*}[]
\centering
\hspace*{-0.7cm}\includegraphics[width=0.5\textwidth]{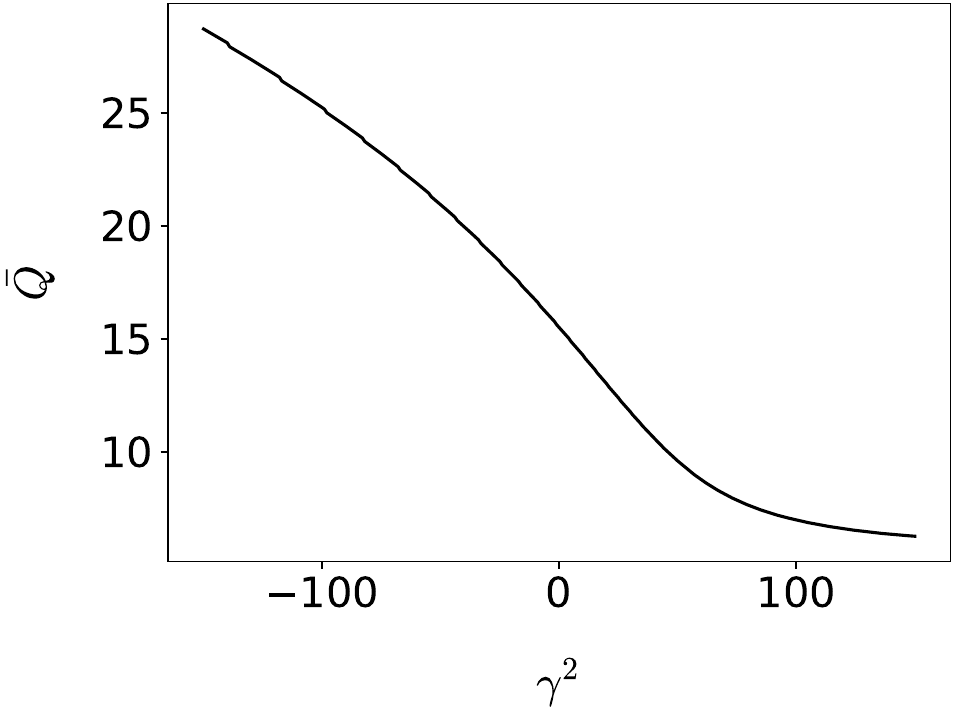}
\caption{Behavior of the dimensionless quadrupolar momentum through the variation of $\gamma^2$ for $w/\mu=0.9$ and quartic self-interaction $\Lambda=1$ \Cref{potential_quartic}.}
\label{qq_gamma}
\end{figure*}

The behavior of the dimensionless quadrupole moment, as shown in \Cref{qq_gamma}, is very similar to that of the mass. Disregarding the differences between the quadrupole moment of the mass and the mass itself, the trend is analogous—starting at its maximum values and decreasing as $\gamma^2$ increases. Interestingly, around $\gamma^2\sim 40$, an inflection point is observed, where the decrease continues but at a much lower rate.

\begin{figure*}
\includegraphics[width=0.48\textwidth]{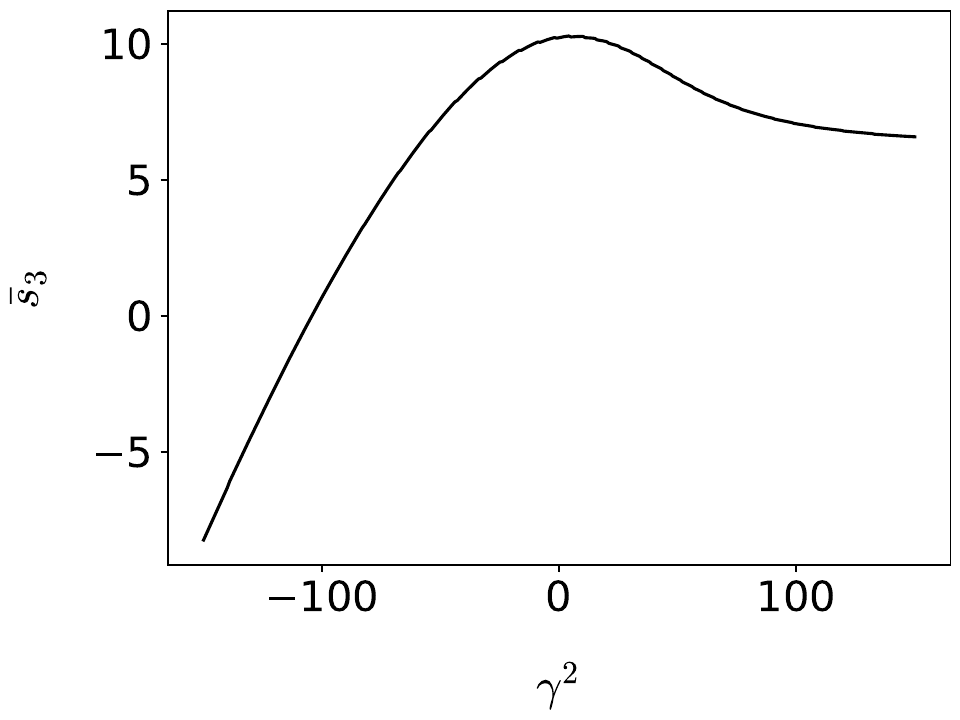}
\includegraphics[width=0.48\textwidth]{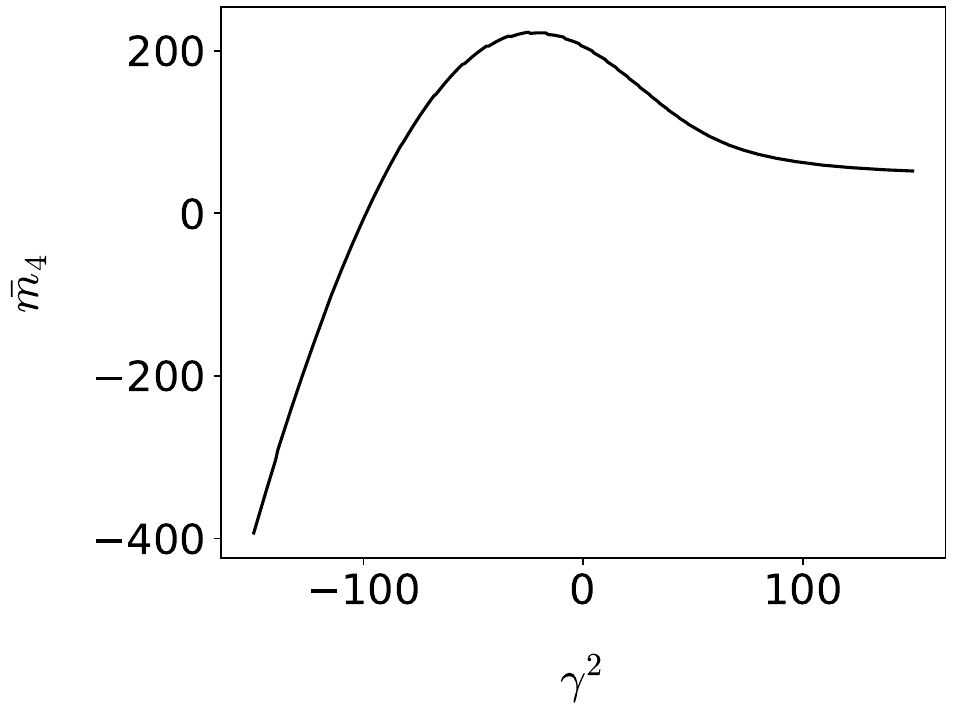}
\caption{Higher dimensionless multipoles as functions of $\gamma^2$, for $w/\mu=0.9$ and $\Lambda=1$. }
\label{Qs3_gamma}

\end{figure*}

We now perform a simultaneous analysis of the higher-order multipoles, using the dimensionless definitions for the octupolar spin and hexadecapolar mass as given in \Cref{reduced}. For the same range of $\gamma^2$ as in previous cases, the behavior of these multipoles is somewhat more complex \Cref{Qs3_gamma}. However, as we continue our analysis, we observe that their values increase steeply with $\gamma^2$, reaching an absolute maximum for a negative $\gamma^2$ value near zero (this pattern holds for both moments). Beyond this peak, the multipoles decrease in a distinct manner, showing an inflection point around $\gamma^2\sim 40$, after which a plateau is observed for higher values. The emergence of this plateau suggests that the corresponding stars approach a kind of universal limit, which may be related to the probing of the low compactness regime.

The above analysis demonstrates that variations in $\gamma^2$ significantly affect the multipolar structure of \nlsm solutions. This indicates that models with similar properties but distinct differences can be utilized in theoretical frameworks and astrophysical contexts. Such versatility opens up new avenues for modeling compact objects.

Now we will examine how solutions vary when fixing specific $\gamma^2$ and $\Lambda$ values and altering the parameter $w$. This approach is the most straightforward method to obtain solutions, and each resulting curve can be labeled as a \textit{family} of solutions, representing a distinct model. This designation is particularly apt because the frequency is an internal field parameter related to internal oscillations rather than interaction strength or kinetic terms; it essentially reflects the internal field frequency. As demonstrated later, this method is commonly used to present solutions. We have previously applied it in \Cref{masasc} to illustrate the range of masses obtained in our analysis of universality. We can enhance the analysis by deriving complete solutions families across the entire frequency range. We will maintain a constant value of $\Lambda$ in all cases, focusing on how the kinetic term influences the behavior of these families and their properties in terms of $w$. Specifically, we have chosen $\Lambda=100$ to represent the strength of self-interaction. This is a rather large value that, as we previously found, supports highly massive BSs. This is a physically interesting regime. Subsequently, we will compute the full curve of solutions for $\gamma^2=-100,10,1,0,-1,-10,-100$.

\begin{figure*}[]
\hspace{+0.6cm}
\includegraphics[clip,width=1.0\columnwidth]{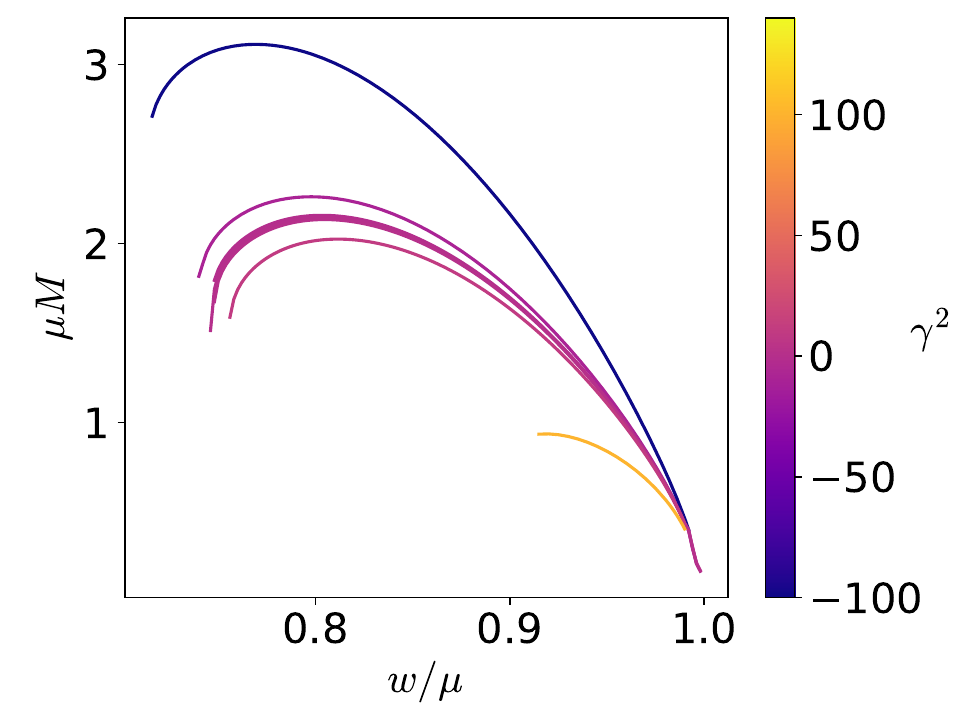}\\%
\hspace{+0.6cm}
\includegraphics[clip,width=1.0\columnwidth]{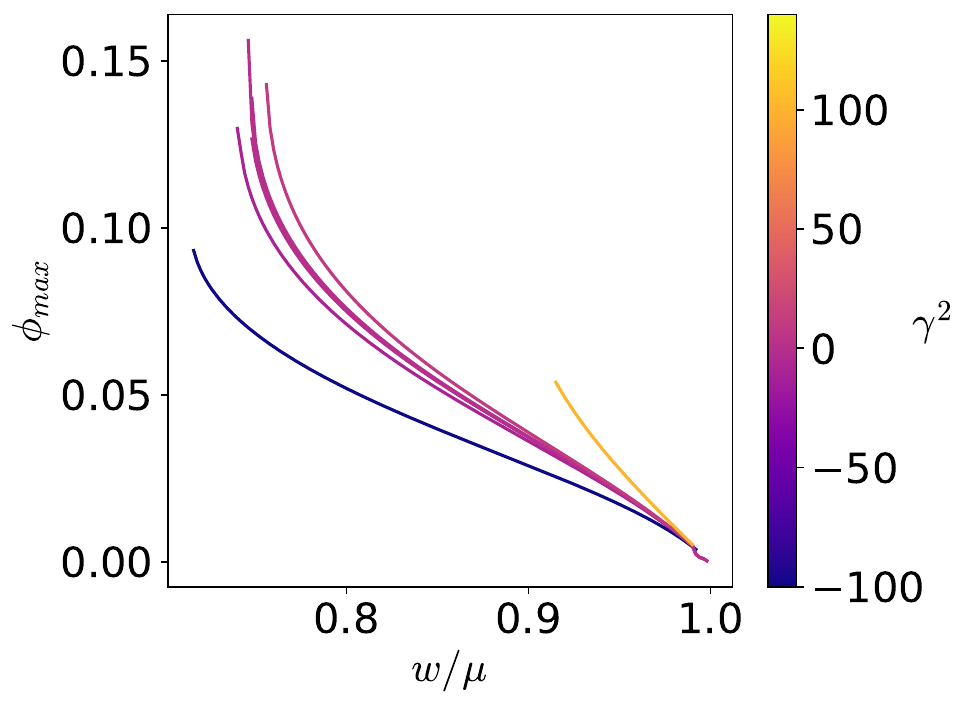}
\caption{Mass (upper plot) and  max field values (lower) versus $\omega/\mu$ for $\gamma^2=-100,10,1,0,-1,-10,-100$ and $\beta=100$.
}
\label{3gamm}
\end{figure*}

We show the masses and maximum field values in \Cref{3gamm}. We note that we do not plot the second branch of solutions,  which typically lies below the main branch and is generally linearly unstable \cite{Vaglio:2022flq,Herdeiro:2015tia}. As we already pointed out, the qualitative behavior of the mass curve does not change with curvature. However, we see a quantitative change. The masses increase as $\gamma^2$ becomes more negative. Specifically, for $\gamma^2=-100$, we observe stars with $\mu M>3$, likely located on the stable segment of the main branch. The opposite happens for growing $\gamma^2$, where we reach solutions with maximum mass values below $1 \mu M$, see, e.g., the case with $\gamma^2=100$. Thus, the spherical target space supports heavier stars than the models with hyperbolic topology.

\begin{figure*}[]
\includegraphics[clip,width=1.0\columnwidth]{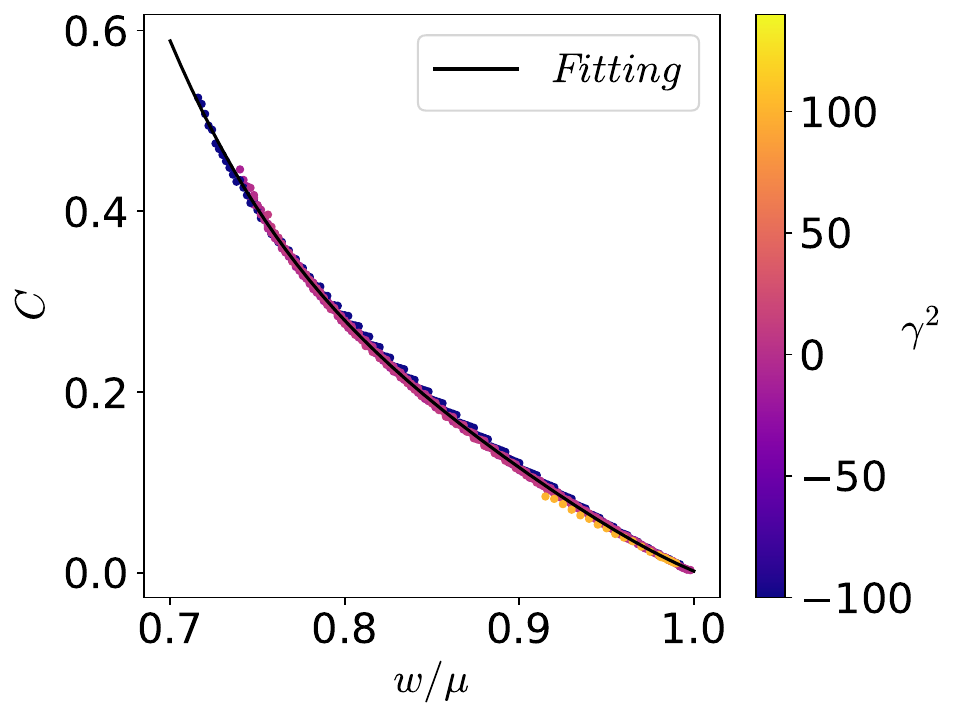}\\%
\includegraphics[clip,width=1.0\columnwidth]{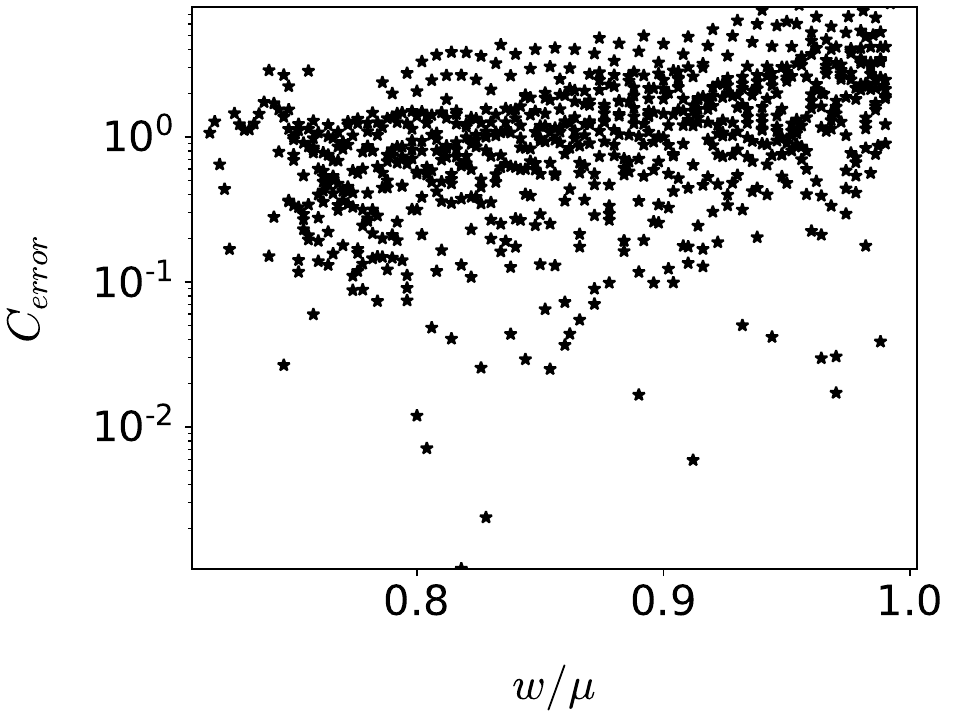}
\caption{Compactness study. The upper pannel shows the data for various models joining into a unique line, together with the solid fitting. The lower pannel shows the fitting errors.}
\label{comp_gam}
\end{figure*}

The behavior of $\phi_{max}$ follows an opposite pattern. The field modulus takes smaller values for the spherical geometry of the target space. Furthermore, independently of $\gamma^2$, $\phi_{max}$  tends to increase as $w$ decreases. Importantly, although the variation in $\phi_{max}$  due to changes in  $\gamma^2$  is rather modest, more negative $\gamma^2$ covers a broader range of solutions (at least in the stable branch), reaching smaller values of $w$ and permitting solutions with marginally smaller $\phi_{max}$.

Surprisingly, the variation in $\gamma^2$ does not directly influence the compactness, see \Cref{comp_gam}. It becomes apparent that the frequencies capable of providing a presumably stable solution within a given model typically correspond to similar compactness values. Specifically, if a particular frequency yields a solution for multiple values of $\gamma^2$, it will lead to almost identical compactness values despite differences in mass, fields, and other multipoles. Nevertheless, the value of the target space curvature constant, $\gamma^2$, does have an indirect effect. Variations in $\gamma^2$ alter the range of feasible values in $w$ for which solutions exist. Consequently, models with a wider range of $w$ achieve higher compactness values, $C$, and vice versa.

In the same plot, we observe another notable behavior in the model where $\gamma=-100$, solutions with smaller values of the angular frequency $w$ possess extremely high compactness, reaching the black hole threshold. However, these solutions are near or in the unstable region of the mass-frequency curve, indicating that although they are theoretically possible within the model, they would likely be unstable in an astrophysical context.

\begin{figure*}
\includegraphics[clip,width=0.9\columnwidth]{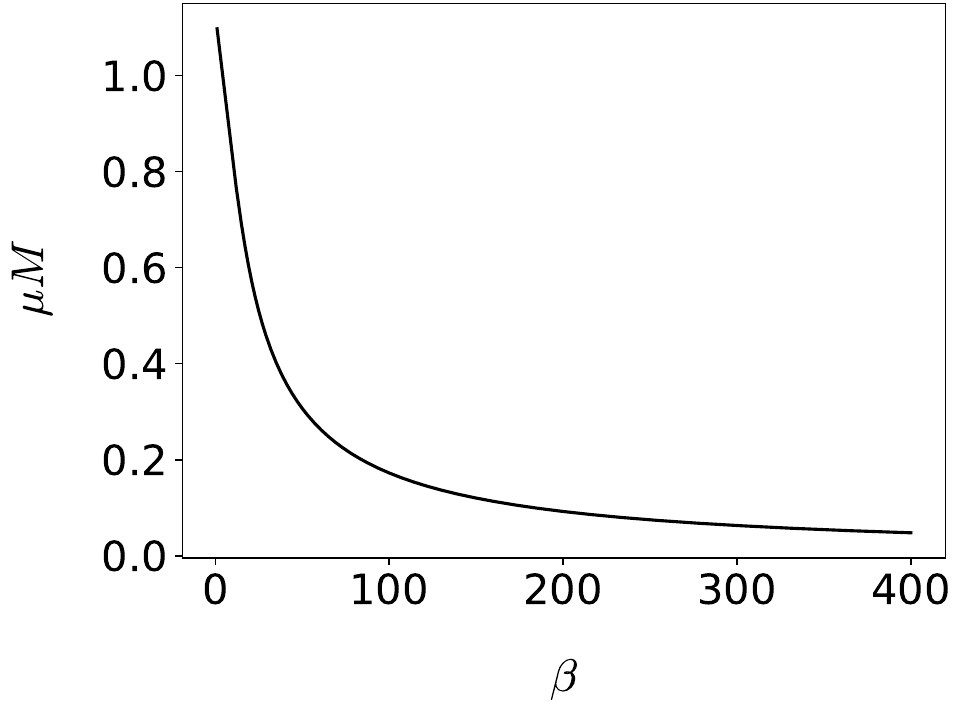}\\%
\includegraphics[clip,width=0.9\columnwidth]{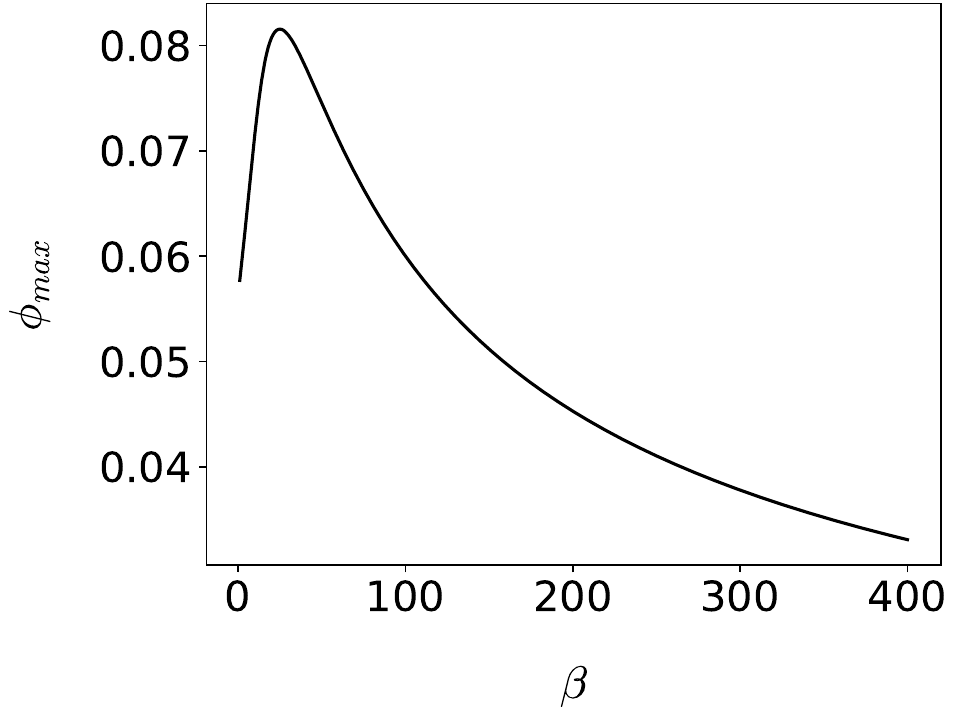}%
\caption{Mass (upper pannel) and max field (lower) vs $\beta$ of solutions for the 2-vacua potential with $w/\mu=0.9$, $\gamma^2=1$. }
\label{betass}
\end{figure*}

In a certain sense, the behavior described above can be understood as a type of universality. This is because of the fact that in the parameter space formed by $w/\mu$ and $C$, all the solutions associated with different $\gamma^2$ correspond to the same one-dimensional function (solid black line), which can be approximated by:
\begin{equation}
    C_{fitt}=\sum_{n=0}^{n=4}A_n \left( \frac{w}{\mu} \right)^n,
\end{equation}
where $A_n$ are five numerical constants shown in \Cref{coefs1d}. This $4-th$ order polynomial fitting ensures an error lower than $6\%$ for each star, as we can read from the lower plot in \Cref{comp_gam}. The error is obtained as follows,
\begin{equation}
    C_{error}=100\cdot\frac{|C-C_{fitt}|}{C_{fitt}}. 
\end{equation}

\begin{table}[]
\centering
\begin{tabular}{|c|c|c|c|c|}
\hline
 $A_0$&$A_1$  &$A_2$  &$A_3$  &$A_4$  \\ \hline
 49.6219&-210.6617  & 341.7304 &-249.1275  &68.4388  \\ \hline
\end{tabular}
\caption{Fitting coefficients.}
\label{coefs1d}
\end{table}

\section{Double vacuum potential}
\label{twovacua}
\subsection{ Role of $\beta$}
\begin{figure*}[]
\hspace{+1.3cm}
\includegraphics[clip,width=0.9\columnwidth]{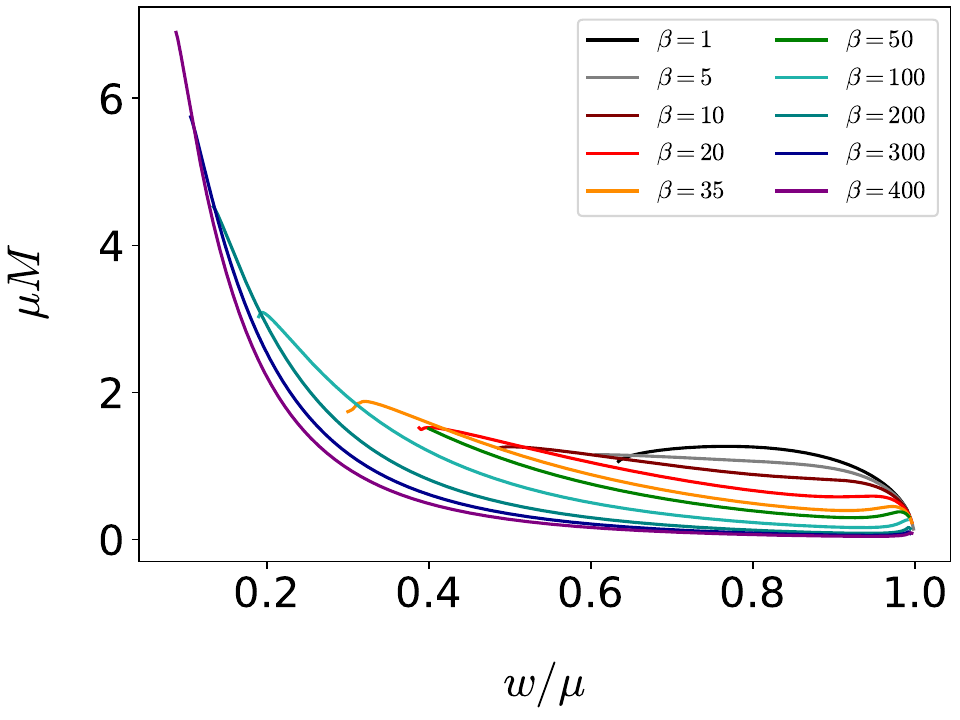}\\%
\hspace{+1.3cm}\includegraphics[clip,width=0.93\columnwidth]{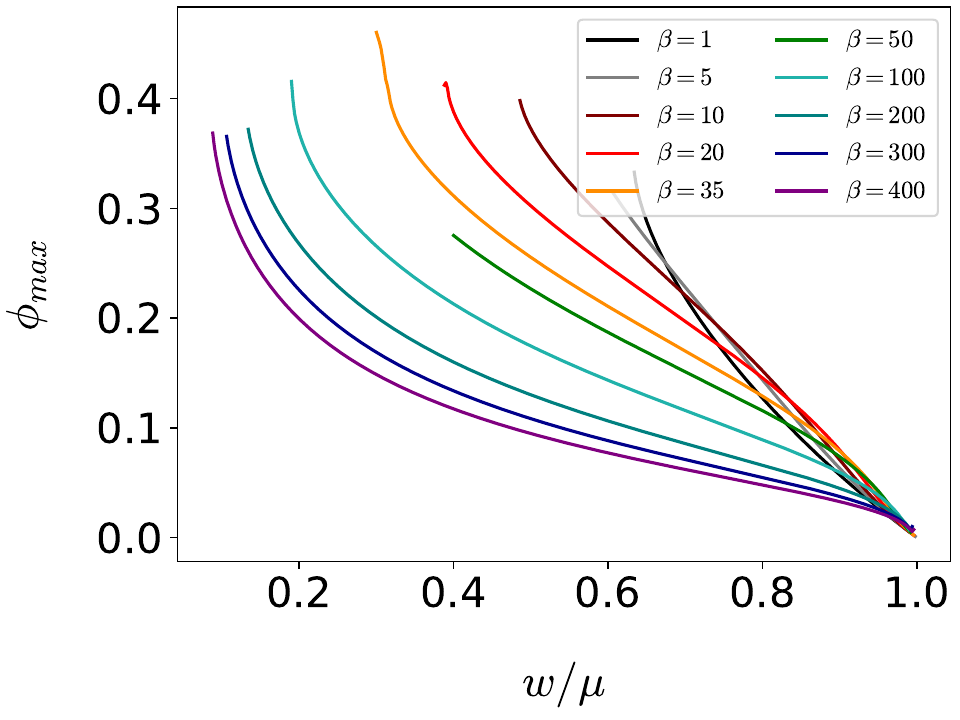}\\%
\caption{Mass (upper panel) and max field (lower panel) as a function of $w/\mu$ and for different $\beta$ in the 2-vacua potential.}
\label{masasbeta}
\end{figure*}
In this section, we will focus on the properties of the BSs for the double vacuum potential, again analyzing how they depend on the curvature of the target space. As before, we will vary the parameters of the model, which are $\gamma^2$ and $\beta$, as well as the field frequency $w$. We remark that increasing $\beta$ makes the second vacuum effectively closer, i.e., reachable
for lower field values, while having smaller $\beta$ values means that we are closer to
a single minimum potential. 

\vspace*{0.2cm}

First, we want to understand the role of the parameter $\beta$. In \Cref{betass}, upper panel, we show how the mass and the maximum of the field varies with $\beta\in[0,400]$. Here we consider the spherical target space, $\gamma^2=-1$, and assume $w/\mu=0.9$. The mass exhibits an exponential decay from an initial value of approximately $\mu M \sim 1.129$. This agrees with the mass-frequency dependence for the usual BSs with flat target space, $\gamma=0$. We present how $\phi_{max}$ is affected in the lower panel as we vary $\beta$. A non-monotonic behavior is found. 

Now, we consider solutions with different field frequencies. The curvature of the target space is still $\gamma^2=-1$, and we assume $\beta=1,5,10,20,25,30,50,100,200,300,400$. As we can see in the upper plot of \Cref{masasbeta}, the mass curves for the different potential constants $\beta$ are very different. It is clear how the existence region, that is, the values of $w$ for which we find BS solutions, varies strongly with $\beta$, having a much wider range in the cases where the constant takes high values. As an example, for $\beta=1$, the frequency range is $w/\mu\in (0.61,1)$ while high $\beta$'s, like our maximum value $400$, reach solutions in the interval $w/\mu\in (0.08,1)$. The achieved masses are also noticeably different, reaching $\mu M >6$ for $\beta > 400$. Furthermore, the shape of the mass-frequency curve changes. For low values of $\beta$, the solutions closely resemble the conventional BS solutions, with the frequency achieving maximum values ranging between $0.6$ and $0.8$. For larger $\beta$, the curves tend to flatten in the large $w$ regime. The masses in this regime are smaller in comparison with the $\beta \sim 1$ case. However, for smaller $w$, the curve grows, and the BSs attain significantly higher masses. 

The maximum value of the field modulus behaves qualitatively in the same way. See lower panel in \Cref{masasbeta}. A smaller $w$ correlates with a larger $\phi_{max}$. Interestingly, within the $1<\beta<20$ range, the maximum field value increases as $\beta$ increases for a constant frequency. Beyond these $\beta$ values, an increase in $\beta$ (while keeping $w$ constant) decreases the value of $\phi_{max}$. This agrees with the dependence of $\phi_{max}$ on $\beta$ presented in \Cref{betass}. 
It is also clear that the $\phi_{max}-\beta$ curve has a non-monotonic character only for sufficiently large $w$. For small $w$ $\phi_{max}$ always decreases with $\beta$.

\begin{figure*}[]
\hspace*{-0.5cm}
\includegraphics[clip,width=1.0\columnwidth]{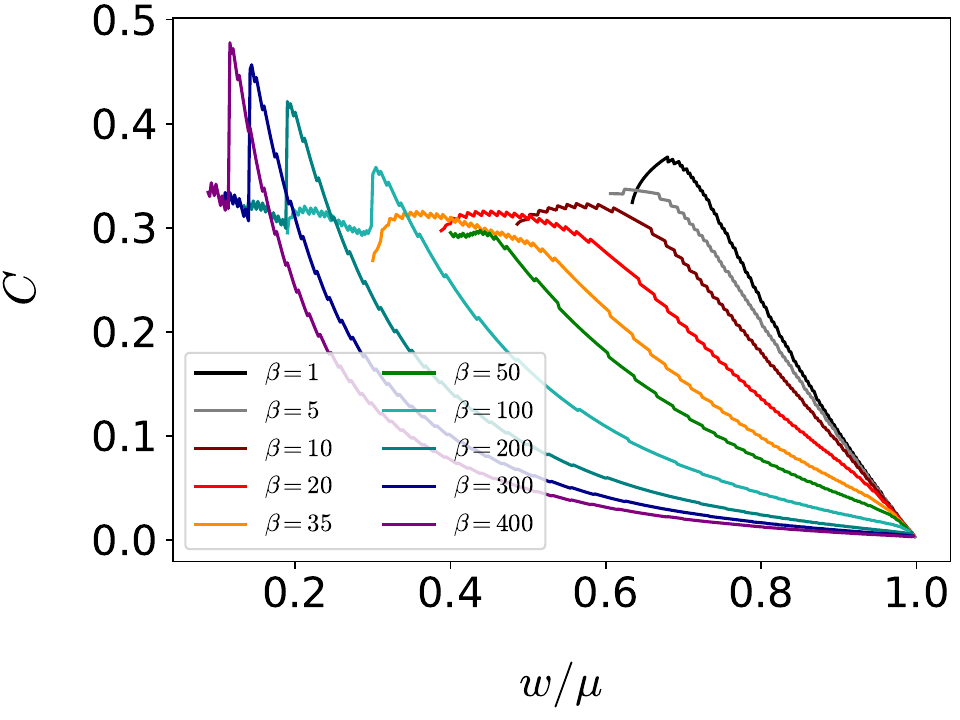}%
\caption{Compactness vs $w/\mu$ for different $\beta$ values and for the 2-vacua potential.}
\label{compactbeta}
\end{figure*}

In \Cref{compactbeta}, we present the dependence of the compactness on $w$ for families of solutions with various $\beta$. For $1<\beta<35$, we found a scaling property in which the compactness curves for different $\beta$ are slightly shifted relative to each other, yet they retain a similar shape.  However, as $\beta$ increases beyond this range, the curves are qualitatively different, exhibiting a distinct second branch at very small values $w$ and reaching notably high compactness levels. 

\begin{figure*}[]
\centering
\hspace*{-0.7cm}\includegraphics[width=0.5\textwidth]{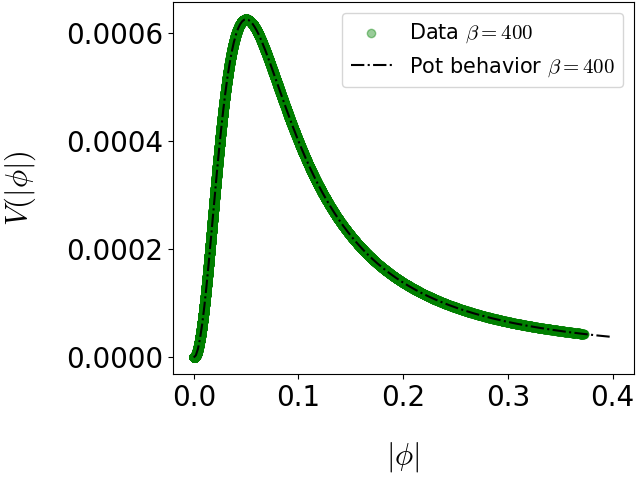}
\caption{Solution for a star with the field reaching the region near the second vacua. Set-up: $\gamma=-1$, $\beta=400$ and $w/\mu=0.088$.}
\label{second_vacua}
\end{figure*}

As we previously remarked, the second minimum (located at $\Phi=\infty$) impacts solutions only if the field takes sufficiently large values. A reasonable condition is that it passes the maximum potential at $|\Phi|=1/\sqrt{\beta}$. Hence, for small $\beta$ it requires very large fields, which is never realized in the numerical analysis. For large $\beta$, the solution indeed sees the second vacuum. As an example, we consider the model with $\beta=400$ ($\gamma^2=-1$ and $w/\mu=0.088$), for which the maximum is located at $|\Phi|=0.05$. The numerical solution reaches this value. See \Cref{second_vacua}. Here, the thick green line represents the data derived from the numerically obtained boson stars, while the dashed line indicates the potential.



\subsection{Role of $\gamma^2$}

Let us now study the impact of the curvature of the targe space $\gamma$. 

\begin{figure*}[]
\centering
\hspace*{0.4cm}\includegraphics[width=0.46\textwidth]{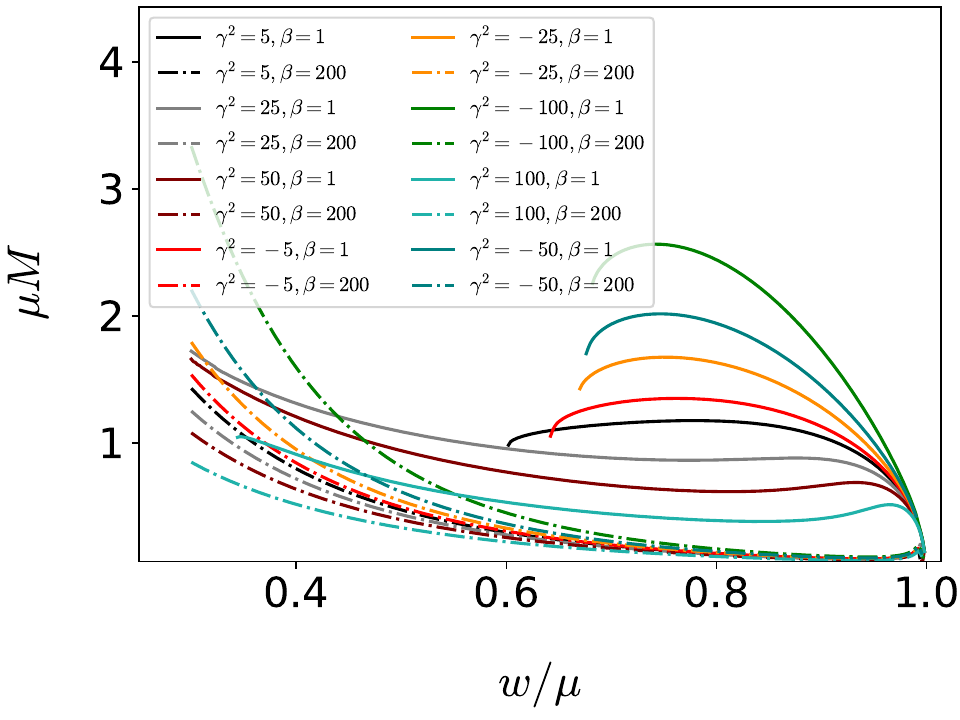}
\caption{Families of solutions for various models of \nlsm stars with a 2-vacua potential and different $\gamma^2$ and $\beta$.}
\label{two_gammas}
\end{figure*}

In \Cref{two_gammas} we present the mass-field frequency curves for $\beta=1,200$ and $\gamma^2=-100,-50,-25,-5,5,25,50,100$. First of all, this plot confirms our previous finding that the solutions are sensitive to changes of $\beta$. As we know it is due to the fact that this parameters controls the effects of the second vacua. Secondly, we see that $\gamma$ qualitatively changes the curves in the same fashion as in the case of the quartic potential. Spherical target space geometries with larger curvature (more negative $\gamma^2$) support heavier stars. It is clear that the effect of the variation of $\gamma$ is more pronounced for BS with small $\beta$. 

\subsection{Ergoregions}

\begin{figure}[]
\hspace{-0.5cm}
\includegraphics[clip,width=1.0\columnwidth]{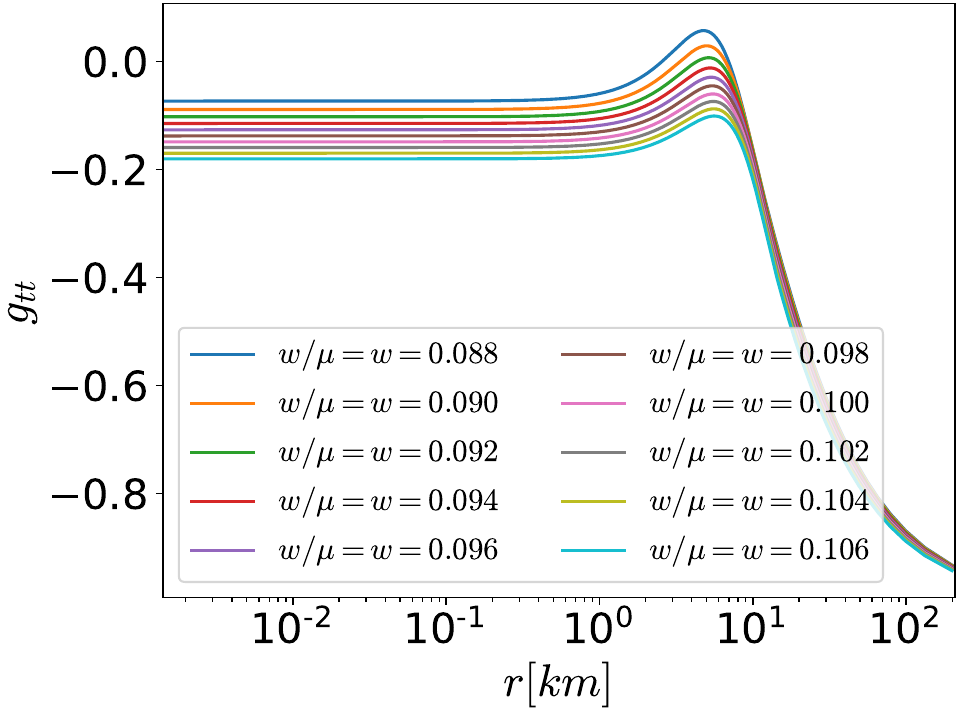}\\%
\hspace{-0.5cm}\includegraphics[clip,width=1.0\columnwidth]{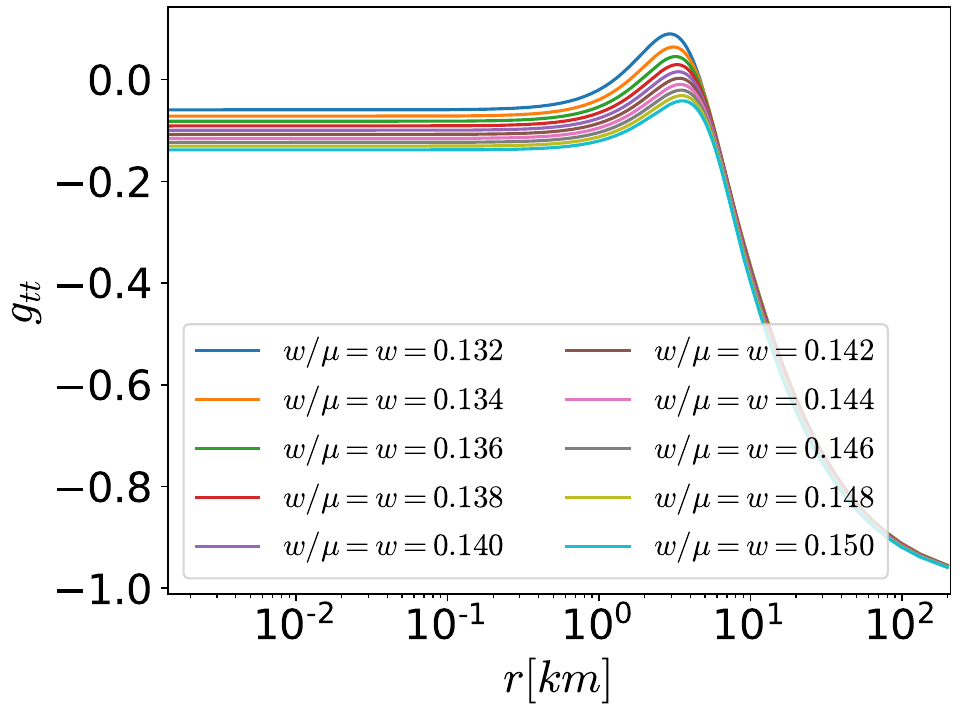}%
\caption{The upper plot shows the value of $g_{tt}$ at $\theta=\pi/2$ against the radial coordinate in $km$, for various $w$ for the model obtained through the fixing of $\gamma^2=-1$ and $\beta=400$. The lower plot shows an analogous set-up for $\beta=200$.}
\label{ergos}
\end{figure}
Ergoregion instability affects systems that contain ergoregions without horizons. This phenomenon has been documented in the standard rotating scalar boson stars \cite{Kleihaus:2007vk}, which can support ergoregions due to the self-interaction terms \cite{Vaglio:2022flq}. This can be linked to compactness levels in models with self-interacting potentials \cite{Vaglio:2022flq}. Similar observations have been made in some rapidly rotating neutron star models, where ergoregions appear for specific equations of state (EOS). In these cases, the instability timescales can exceed the Hubble time, indicating that such instabilities might not significantly alter the star's evolutionary path \cite{Tsokaros:2019mlz}. However, the scenario changes dramatically for ultracompact stars, where the instability can last from seconds to several weeks depending on their compactness \cite{Cardoso:2007az}.
Extensive studies on various exotic compact objects have further highlighted these instabilities \cite{Cardoso:2007az,Maggio:2018ivz}. 

The considerable compactness seen in certain models we studied, see \Cref{masasbeta}, suggests the formation of ergoregions in configurations with very high masses, low field frequencies, and high spins.  In \Cref{ergos}, we see how the sign of $g_{tt}$ changes twice for some solutions with small $w$, belonging to models in which $\beta$ has a high value. This change in sign indicates that there are ergoregions with toroidal topology.

In particular, these ergoregions occur in branches that are stable against radial perturbations. Yet, despite this radial stability, non-topological boson stars with ergoregions suffer long-term instability due to nonspherical modes. Future research should focus on conducting detailed stability analyses for our set-up of topological boson stars, addressing these specific instabilities.

\section{Analysis of Universality}\label{results2}
In this section, we investigate the existence of universal, that is, model-independent, relations between various observables. 

At this point, it is crucial to note that universality studies and multipolar relations for BS have been extensively investigated in recent years. For both usual rotating BSs, vector stars, and other compact stars, relations between multipoles of various kinds have been found \cite{Vaglio:2022flq,Adam:2022nlq,Adam:2023qxj,Adam:2024zqr,Aranguren:2024hds,Aranguren:2023ujo,Yagi:2013awa,Yagi:2014bxa,adam2021quasiuniversal}. The aim of these sections is to determine whether we are observing a universally general phenomenon or something more specific and limited. We will analyze our numerical data for different values of $\gamma^2$ and potential parameters. 


We remark that in the analysis of the universal relations we use both stable and also potentially unstable solutions. The later ones are assumed to emerge from not too extreme models These are defined by the following condition:
\begin{equation}
    \frac{dM}{dw}<0,
\end{equation}
however, since our main study primarily utilizes solutions that are presumably stable under this condition, some solutions may be unstable for certain models. Further referenced explanation can be found in \Cref{stability_app}. Additionally, in \Cref{IXQ1_extreme}, we briefly analyze several cases where a predominance of solutions belongs to unstable second branches.

For clarity and brevity, we will focus on the $I-\chi-Q$ case, briefly noting the existence of the other relations. We leave for \Cref{app:multipoles} a detailed study and corresponding graphs of other high-order multipolar relations.

\subsection{$I-\chi-Q$ Relations}
In the following subsections, we will check the universality for the data set shown at \Cref{masasc}, where $\gamma^2\in[-100,100]$, and using various quartic potentials with couplings $\Lambda\in[-100,100]$.

\begin{figure}[]
\includegraphics[clip,width=1.0\columnwidth]{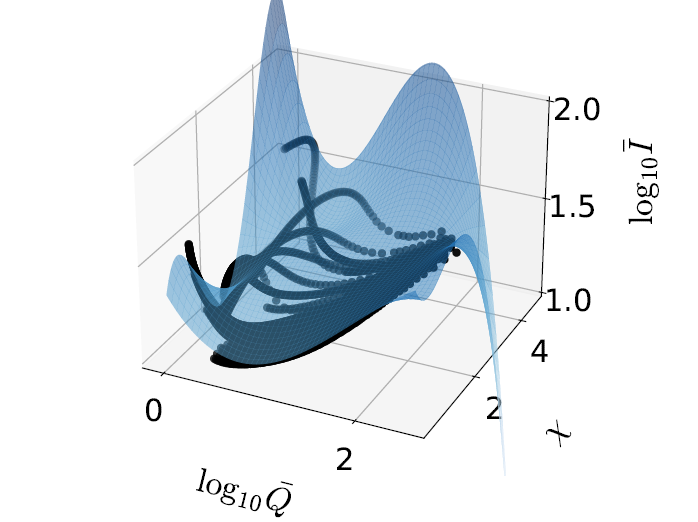}\\%
\includegraphics[clip,width=1.0\columnwidth]{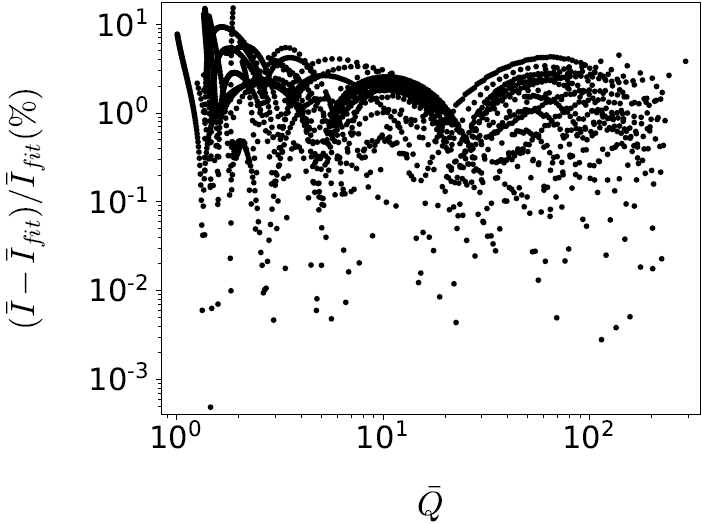}%
\caption{$\eta-\chi-\zeta$ surface for $n=1$ spinning BSs s, fitting the data points shown in \Cref{masasc}, with solutions for $\gamma^2=-100,-10,0,10,100$. The lower plot shows the relative difference between data and fitted value. The maximum error ranges the $12\%$}
\label{IXQ1}
\end{figure}

The first thing we have to do is represent our data in a space in which the dimensions are certain transformations of $\bar{Q}$ and $\bar{I}$, as well as of $\chi$ itself. Each point we see in \Cref{IXQ1} is a star in this $ 3-D$ space. Specifically, we will employ the following
\begin{equation}
  \begin{split}  
    &\eta=\sqrt[3]{\log_{10}{\bar{I}}},\\&
    \zeta=\log_{10}\bar{Q}.
    \end{split}
    \label{fit-func-par}
\end{equation}

In this new coordinate space, plotting all the star data reveals that the points lie on a clear, distinct surface in the upper plot of \Cref{IXQ1}. This is a non-trivial observation: given the numerous parameters involved—ranging from various self-interaction terms in the potential to the differing kinetic terms that generate varied curvatures in the target space manifold—one might expect the solutions to scatter randomly, forming an irregular or unbounded volume rather than a coherent surface. Instead, the data conform to a surface that the following function can accurately model,

\begin{equation}
    \begin{split}
       &\eta=A_0+A_s^j\chi^j\left(\zeta-B\right)^s,
    \end{split}
    \label{fit-func}
\end{equation}
where $s={1,2,3}$, $j={0,1,2,3,4}$. The fitting coefficients are shown in \cref{IQXBSn1fit_Kin} in \Cref{app:Coefficients}. 

It is noticeable that the difference between the fitted surface and the real data is always less than $12\%$, which means that for this data set that incorporates strong variations in $\gamma^2$ including cases with different topologies, a universal or quasi-universal $I-\chi-Q$ relation holds.

As mentioned in \cite{Adam:2022nlq,Adam:2023qxj,Adam:2024zqr}, this type of relation between the spin parameter, the moment of inertia, and the quadrupolar moment can be very useful not only from a theoretical point of view but also for observational purposes, allowing difficult quantities to be extracted from the measure at an empirical level. 

In \Cref{app:multipoles}, we also investigate universal relations linking key higher order multipoles. We have analyzed the relation between $\chi-Q$, and the spin octupolar moment $s_3$. When plotted in the corresponding parameter space, the star data lays on a smooth surface that can be accurately fitted to a function—with fitting coefficients given in \Cref{app:Coefficients} —and a maximum error below $15\%$. Notably, near the Kerr limit, both $\bar{Q}$ and $\bar{s}_3$ approach constant values, suggesting a Kerr-like spacetime in some models. A similar approach is applied to the mass hexadecapolar moment $\bar{m}_4$, where the data again define a surface fitted by an analogous function under $10\%$. The levels of accuracy in the previous cases are acceptable, given the increased numerical challenges associated with the calculations and the extreme differences between models. Again, a detailed analysis is shown in \cref{app:multipoles}. Finally, we have studied the universal relation connecting the compactness of the stars with their quadrupolar and dimensionless spin moments. By plotting the square root of the compactness against the decimal logarithms of these moments (see \Cref{app:multipoles}), the data again conform to a well-defined surface that can be accurately fitted, achieving a maximum error of approximately $11\%$. Despite some measurement uncertainties in determining the stellar radius, these relations are promising from an empirical standpoint, as they can help to infer the quadrupolar and inertia moments from observable quantities, potentially aiding in the interpretation of gravitational wave signals.
As mentioned previously, we also show the analysis for further models with potentially unstable solutions included. Although some stars in the dataset exhibit higher percentage errors between the simulated data and the fit, it remains possible to identify a quasi-universal surface, with errors generally below $21\%$ and often much lower.

\section{Comparison with other extreme models}
The existence of such universal relations is in itself a useful result from a theoretical and experimental point of view. Here, we will try to explore the limits of validity of the relations while simultaneously extracting information about their origin. Having models in which we can vary both the kinetic constant (the target space curvature) and the parameters of potentials, we will investigate how these relations behave if the parameters are taken to extreme values.

\subsection{The double vacuum potential and $I-\chi-Q$ relation}

We are going to explore whether the Boson Stars supported by the double vacuum potential leads to any universal relations and if, it is the case, whether they differ from those obtained for the quartic potential. We aim to generate a joint fit surface using all data points derived for the double vacuum potentials with a wide range of the model parameters. 


\begin{figure}[]
\includegraphics[clip,width=1.0\columnwidth]{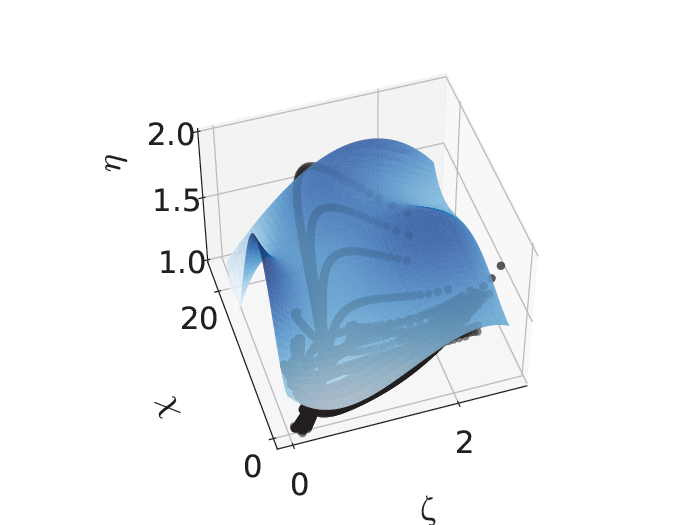}\\%
\includegraphics[clip,width=1.0\columnwidth]{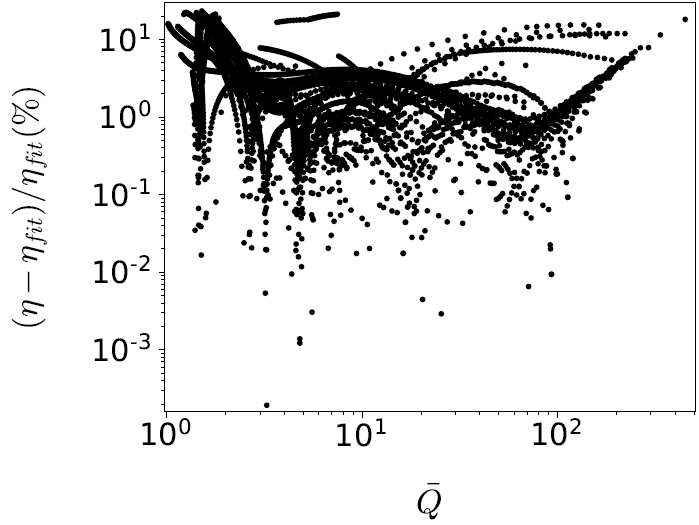}%
\caption{$\eta-\chi-\zeta$ surface for $n=1$ spinning BSs shown in \Cref{masasc} and models with $\beta\in[1,400]$ and $\gamma^2=-1$. In the lower plot, we show the relative difference between data and fitted value is always under the $22\%$ value. }
\label{IXQ_beta_all}
\end{figure}

To perform this study, we will use the families of solutions outlined in \Cref{masasbeta}. The first step is to check whether the full range of solutions obtained with the double vacuum potential for $\beta \in [1,400]$ satisfies the same relations derived from the previously studied dataset, which was based on different potentials and spanned a sufficiently general range—in other words, to determine whether the new data conform to a universality relatable or comparable with the one shown in \Cref{IXQ1}. In \Cref{IXQ_beta_all}, the data points closely follow a well-defined surface, which is the first requirement for establishing a universal relation. However, the maximum discrepancy observed between the fitted surface and the data is approximately $22\%$.

Importantly, the fitted surface resembles the one found for the quartic potential in \Cref{IXQ1}, with an additional region at higher $\chi$, where the solutions with $\beta>35$ are located. This similarity is not surprising because, in the low $\beta$ regime, the second vacuum is practically not explored by the boson star solutions, and the potential—within the range of values taken by the solutions—is not significantly different from the quartic potential. In contrast, it is in the high $\beta$ regime that new behavior is expected to emerge from the potential.

Thus, we may conclude that, for all values of $\beta$, the boson stars emerging from both potentials form a common quasi-universal surface. Although this surface cannot be fitted with high precision, it indicates that the solutions behave universally, with errors around $22\%$. Given this clear separation in regimes, we can analyze the low and high $\beta$ regimes separately.

\begin{figure}[]
\includegraphics[clip,width=1.0\columnwidth]{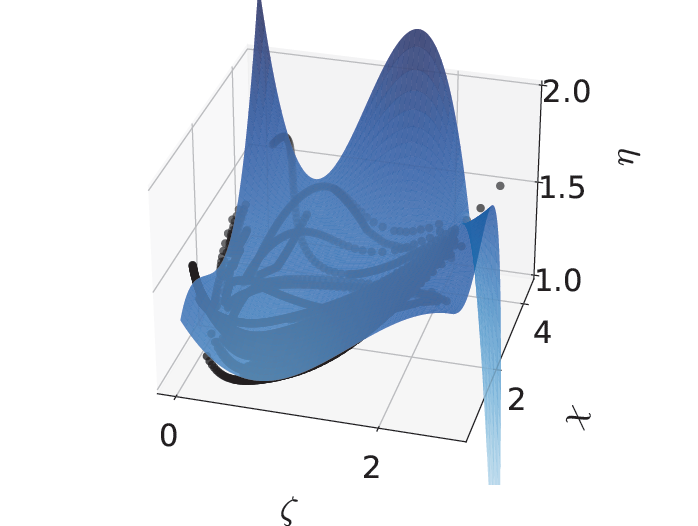}\\%
\includegraphics[clip,width=1.0\columnwidth]{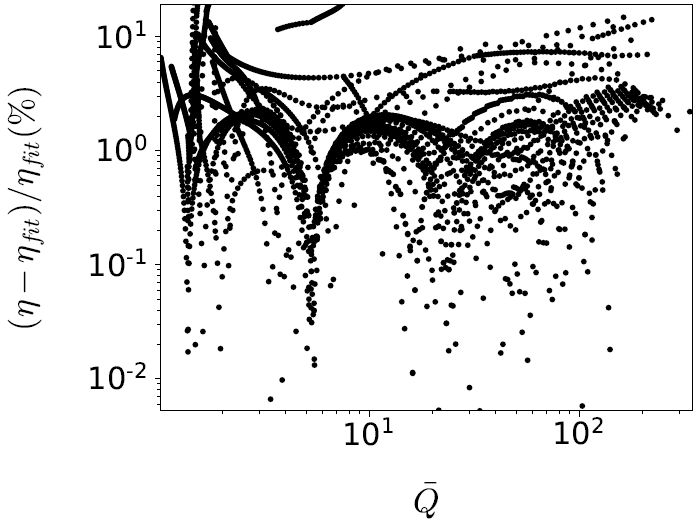}%
\caption{$\eta-\chi-\zeta$ surface for $n=1$ spinning BSs shown in \Cref{masasc} and models with $\beta<35, \gamma^2=-1$, fitting the data points. In the lower plot, we show the relative difference between data and fitted value is always under the $\sim 14\%$ value. }
\label{IXQ_low_beta}
\end{figure}

We see in \Cref{IXQ_low_beta} that a well-behaved universality is recovered by taking only the data shown in \Cref{masasc} together with solutions for $\beta<35$. Despite having a high maximal error, about $14\%$, we have reduced it significantly from the previous attempt. We can see that this surface and the one given in \Cref{IXQ1} are close at sight.

\begin{figure}[]
\includegraphics[clip,width=1.0\columnwidth]{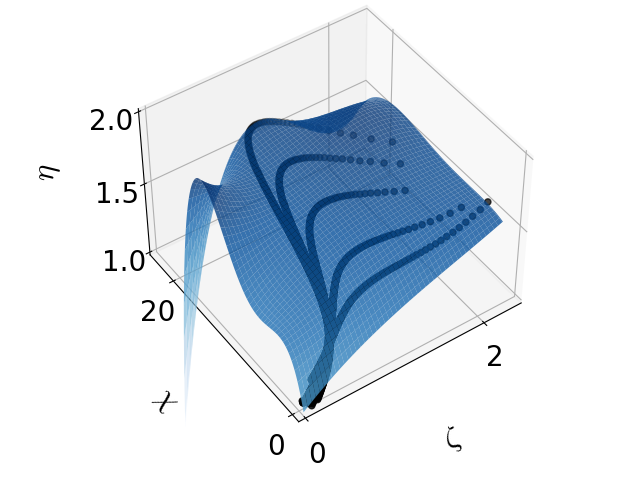}\\%
\includegraphics[clip,width=1.0\columnwidth]{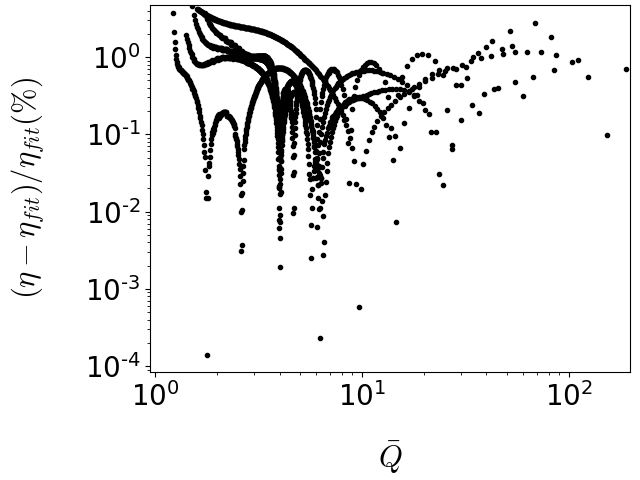}%
\caption{$\eta-\chi-\zeta$ surface for $n=1$ spinning BSs shown in \Cref{masasc} and models with $\beta>35, \gamma^2=-1$, fitting the data points. In the lower plot, we show the relative difference between data and fitted value is always under the $4.5\%$ value.}
\label{high_beta_IXQ}
\end{figure}

Let us now turn to solutions with $\beta>35$ where the second vacuum of the potential modifies the properties of the boson stars. We can see in \Cref{high_beta_IXQ}, that the points corresponding to these solutions also define a surface. However, this surface differs in shape from the previous one. We fit a function of the same type \eqref{fit-func}, but now the coefficients are completely different (see \Cref{app:Coefficients}). The lower panel of \Cref{high_beta_IXQ} shows that by this fitting, the maximal error is lower than a $\sim 4.5\%$, indicative of universality in this parameter region.

In this section, we have explored the existence of universal relations under extreme conditions. Initially, we demonstrated that, with sufficient precision, these relations persist across models, even when the kinetic constant varies significantly. Next, we introduced the topological model of stars within a curved target space and employed the two-vacua potential, confirming that the relationships still hold, albeit with slightly reduced precision. We observed that universality improves when the data is split into two regions based on $\beta$, allowing us to study restricted universalities that are amenable to a quasi-universal joint fitting, albeit with higher errors.

Alternatively, one can interpret this scenario as follows: when $\beta$ is exceedingly high, the potential effectively approximates the "massless" case. This may explain why the original universality is altered and why a new region becomes evident—one that can be described by a single surface and where the stars differ markedly from boson star models whose potential can be expanded as a mass term plus self-interactions $V(|\Phi|)\sim \mu^2\Phi^*\Phi+a_2 (\Phi^*\Phi)^2+a_3 (\Phi^*\Phi)^3+...$


\section{Conclusions}
\label{conclusions}

We have investigated the existence of approximate
universal
relations between multipole moments and principal observables for various models of scalar boson stars with different nonlinear sigma models and self-interacting potentials, finding the existence of the same kind of universal relations, presented for standard, complex scalar BSs in \cite{Adam:2023qxj,Adam:2022nlq}. 


Our calculations for spinning \nlsm stars with very different target space and various self-interactions also represent an interesting result \emph{per se}, making this paper a natural completion with respect to \cite{Collodel:2019uns}. We have studied the effective \textit{2-vacua} potential within the \nlsm framework for the first time.

For entities without horizons, relations similar to the exact no-hair theorem for black holes—known as universal or effective no-hair relations—allow the external gravitational field to be determined with high precision using only a finite number of multipole moments. This means that, even in the presence of matter, using an infinite series of multipoles to describe the gravitational system within a suitable approximation is unnecessary. We have found that these universalities also apply to our \nlsm dataset, placing boson stars with a non-trivial field target manifold inside the family of compact objects where universal behavior is observed.

This universality highlights how the compactness of these entities can diminish the influence of their matter composition in determining their physical characteristics to a certain extent. This finding is consistent with previous studies on different bosonic compact objects. We examined the multipolar properties as functions of key parameters, demonstrating how the star's properties change with variations in its kinetic or potential components.

Our research also shows that specific models exhibit extremely compact configurations that closely resemble, or even mimic, the multipolar behavior of Kerr black holes. While this supports the idea that these configurations can effectively mimic Kerr black holes, it is important to note that stability studies suggest these highly compact models may have significant stability issues, which casts doubt on their astrophysical relevance. Therefore, a balance must be found: On the one hand, stable high-compactness models align with the no-hair theorems for Kerr black holes, indicating their potential as black hole mimickers. On the other hand, many ultra-compact models suffer from stability problems, which require caution when proposing them as black hole mimickers.
We have discovered an intriguing result in this context: the compactness vs. frequency universality across different values of $\gamma^2$, revealing that the curvature of the field manifold does not influence this relationship, which remains a precise one-dimensional correlation.

Furthermore, we examine how different models of \nlsm solutions fit within a less extreme two-dimensional framework, confirming that universality exists across a significant range of solutions. However, despite a considerable range of validity, we observed that the universality becomes better after performing a split in the $3D$ space of observables for extreme models approaching the massless limit ($\beta->\infty)$ in the 2-vacua potential. This can be understood as follows: As the 2-vacua coupling $\beta$ increases, the model resembles a massless solution, causing the scalar field to spread out and the resulting star to behave more like a diffuse cloud. For these solutions, the behavior in the $I-\chi-Q$ space changes significantly, with the universality surface generally becoming flatter.  Therefore, we can distinguish between scalar BSs in a common regime and extreme bosonic objects with different properties depending on the multipolar moments and their universality. We also analyzed the intermediate region, exploring how both universal surfaces interact in neighboring space.

The existence of universal relations has practical appli\-cations. Independent measurements of different but related quantities can be used to infer the corresponding parameter. This is a powerful method for deducing physical properties and a way to cross-check them with measurements from gravitational-wave science. In our bosonic star scenario, we have identified universal or quasi-universal relations, allowing us to use them in a standard manner.

Moreover, we can take this further by using our results to predict in which region of the parameter space a third, previously unknown quantity might appear and then verify it against GW data. Given the recent advancements in GW observations, this approach could be crucial in various field areas, particularly in predicting the most likely type of observed astrophysical object.

A promising next step in our research would be to derive the tidal and rotational Love numbers in the context of rapid rotation. This is particularly relevant for gravitational wave astronomy but poses a significant challenge, as it requires perturbing the fully rotating metric as a starting point. Afterwards, extending this work to explore universal relations in more exotic compact objects, such as spinning mixed scalar boson-fermion stars or Proca-fermion stars, would be highly valuable. Additionally, investigating cases where a horizon forms within rotating boson stars—leading to the formation of hairy Kerr black holes—would deepen our understanding of universal relations in these complex systems.

Our analysis will help identify astrophysical compact objects, especially in addressing the degeneracy problem in gravitational waveforms from future binary merger events. Furthermore, this research could aid in the search for potential bosonic candidates for dark matter and contribute to a deeper understanding of the strong gravity regime in General Relativity.


\begin{acknowledgments}
AGMC wants to thank Jarah Evslin and Jose Juan Blanco-Pillado for valuable suggestions and comments.
The authors acknowledge financial support from the Spanish Research State Agency under project PID2023-152762NB-I00, the Xunta de Galicia under the project ED431F 2023/10 and the CIGUS Network of Research Centres, the María de Maeztu grant CEX2023-001318-M funded by MICIU/AEI /10.13039/501100011033, and the European Union ERDF.
AW is supported by the Polish National Science Centre,
grant NCN 2020/39/B/ST2/01553. 
 JCM thanks the Xunta de Galicia (Consellería de Cultura, Educación y Universidad) for funding their predoctoral activity through \emph{Programa de ayudas a la etapa predoctoral} 2021. AGMC acknowledges support from the PID2021-123703NB-C21
grant funded by MCIN/AEI/10.13039/501100011033/and by ERDF: “\emph{A way of making Europe}”;
and the Basque Government grant (IT-1628-22).
\end{acknowledgments}
\newpage
\bibliography{biblio}

\begin{thebibliography}{95}%
\makeatletter
\providecommand \@ifxundefined [1]{%
 \@ifx{#1\undefined}
}%
\providecommand \@ifnum [1]{%
 \ifnum #1\expandafter \@firstoftwo
 \else \expandafter \@secondoftwo
 \fi
}%
\providecommand \@ifx [1]{%
 \ifx #1\expandafter \@firstoftwo
 \else \expandafter \@secondoftwo
 \fi
}%
\providecommand \natexlab [1]{#1}%
\providecommand \enquote  [1]{``#1''}%
\providecommand \bibnamefont  [1]{#1}%
\providecommand \bibfnamefont [1]{#1}%
\providecommand \citenamefont [1]{#1}%
\providecommand \href@noop [0]{\@secondoftwo}%
\providecommand \href [0]{\begingroup \@sanitize@url \@href}%
\providecommand \@href[1]{\@@startlink{#1}\@@href}%
\providecommand \@@href[1]{\endgroup#1\@@endlink}%
\providecommand \@sanitize@url [0]{\catcode `\\12\catcode `\$12\catcode `\&12\catcode `\#12\catcode `\^12\catcode `\_12\catcode `\%12\relax}%
\providecommand \@@startlink[1]{}%
\providecommand \@@endlink[0]{}%
\providecommand \url  [0]{\begingroup\@sanitize@url \@url }%
\providecommand \@url [1]{\endgroup\@href {#1}{\urlprefix }}%
\providecommand \urlprefix  [0]{URL }%
\providecommand \Eprint [0]{\href }%
\providecommand \doibase [0]{http://dx.doi.org/}%
\providecommand \selectlanguage [0]{\@gobble}%
\providecommand \bibinfo  [0]{\@secondoftwo}%
\providecommand \bibfield  [0]{\@secondoftwo}%
\providecommand \translation [1]{[#1]}%
\providecommand \BibitemOpen [0]{}%
\providecommand \bibitemStop [0]{}%
\providecommand \bibitemNoStop [0]{.\EOS\space}%
\providecommand \EOS [0]{\spacefactor3000\relax}%
\providecommand \BibitemShut  [1]{\csname bibitem#1\endcsname}%
\let\auto@bib@innerbib\@empty
\bibitem [{\citenamefont {Guzm\'an}\ and\ \citenamefont {Rueda-Becerril}(2009)}]{PhysRevD.80.084023}%
  \BibitemOpen
  \bibfield  {author} {\bibinfo {author} {\bibfnamefont {F.~S.}\ \bibnamefont {Guzm\'an}}\ and\ \bibinfo {author} {\bibfnamefont {J.~M.}\ \bibnamefont {Rueda-Becerril}},\ }\bibfield  {title} {\enquote {\bibinfo {title} {Spherical boson stars as black hole mimickers},}\ }\href {\doibase 10.1103/PhysRevD.80.084023} {\bibfield  {journal} {\bibinfo  {journal} {Phys. Rev. D}\ }\textbf {\bibinfo {volume} {80}},\ \bibinfo {pages} {084023} (\bibinfo {year} {2009})}\BibitemShut {NoStop}%
\bibitem [{\citenamefont {Herdeiro}\ \emph {et~al.}(2021)\citenamefont {Herdeiro}, \citenamefont {Pombo}, \citenamefont {Radu}, \citenamefont {Cunha},\ and\ \citenamefont {Sanchis-Gual}}]{Herdeiro:2021lwl}%
  \BibitemOpen
  \bibfield  {author} {\bibinfo {author} {\bibfnamefont {Carlos A.~R.}\ \bibnamefont {Herdeiro}}, \bibinfo {author} {\bibfnamefont {Alexandre~M.}\ \bibnamefont {Pombo}}, \bibinfo {author} {\bibfnamefont {Eugen}\ \bibnamefont {Radu}}, \bibinfo {author} {\bibfnamefont {Pedro V.~P.}\ \bibnamefont {Cunha}}, \ and\ \bibinfo {author} {\bibfnamefont {Nicolas}\ \bibnamefont {Sanchis-Gual}},\ }\bibfield  {title} {\enquote {\bibinfo {title} {{The imitation game: Proca stars that can mimic the Schwarzschild shadow}},}\ }\href {\doibase 10.1088/1475-7516/2021/04/051} {\bibfield  {journal} {\bibinfo  {journal} {JCAP}\ }\textbf {\bibinfo {volume} {04}},\ \bibinfo {pages} {051} (\bibinfo {year} {2021})},\ \Eprint {http://arxiv.org/abs/2102.01703} {arXiv:2102.01703 [gr-qc]} \BibitemShut {NoStop}%
\bibitem [{\citenamefont {Rosa}\ and\ \citenamefont {Rubiera-Garcia}(2022)}]{Rosa:2022tfv}%
  \BibitemOpen
  \bibfield  {author} {\bibinfo {author} {\bibfnamefont {Jo\~ao~Lu\'\i{}s}\ \bibnamefont {Rosa}}\ and\ \bibinfo {author} {\bibfnamefont {Diego}\ \bibnamefont {Rubiera-Garcia}},\ }\bibfield  {title} {\enquote {\bibinfo {title} {{Shadows of boson and Proca stars with thin accretion disks}},}\ }\href {\doibase 10.1103/PhysRevD.106.084004} {\bibfield  {journal} {\bibinfo  {journal} {Phys. Rev. D}\ }\textbf {\bibinfo {volume} {106}},\ \bibinfo {pages} {084004} (\bibinfo {year} {2022})},\ \Eprint {http://arxiv.org/abs/2204.12949} {arXiv:2204.12949 [gr-qc]} \BibitemShut {NoStop}%
\bibitem [{\citenamefont {Rosa}\ \emph {et~al.}(2023)\citenamefont {Rosa}, \citenamefont {Macedo},\ and\ \citenamefont {Rubiera-Garcia}}]{Rosa:2023qcv}%
  \BibitemOpen
  \bibfield  {author} {\bibinfo {author} {\bibfnamefont {Jo\~ao~Lu\'\i{}s}\ \bibnamefont {Rosa}}, \bibinfo {author} {\bibfnamefont {Caio F.~B.}\ \bibnamefont {Macedo}}, \ and\ \bibinfo {author} {\bibfnamefont {Diego}\ \bibnamefont {Rubiera-Garcia}},\ }\bibfield  {title} {\enquote {\bibinfo {title} {{Imaging compact boson stars with hot spots and thin accretion disks}},}\ }\href {\doibase 10.1103/PhysRevD.108.044021} {\bibfield  {journal} {\bibinfo  {journal} {Phys. Rev. D}\ }\textbf {\bibinfo {volume} {108}},\ \bibinfo {pages} {044021} (\bibinfo {year} {2023})},\ \Eprint {http://arxiv.org/abs/2303.17296} {arXiv:2303.17296 [gr-qc]} \BibitemShut {NoStop}%
\bibitem [{\citenamefont {Sengo}\ \emph {et~al.}(2024)\citenamefont {Sengo}, \citenamefont {Cunha}, \citenamefont {Herdeiro},\ and\ \citenamefont {Radu}}]{Sengo:2024pwk}%
  \BibitemOpen
  \bibfield  {author} {\bibinfo {author} {\bibfnamefont {Ivo}\ \bibnamefont {Sengo}}, \bibinfo {author} {\bibfnamefont {Pedro V.~P.}\ \bibnamefont {Cunha}}, \bibinfo {author} {\bibfnamefont {Carlos A.~R.}\ \bibnamefont {Herdeiro}}, \ and\ \bibinfo {author} {\bibfnamefont {Eugen}\ \bibnamefont {Radu}},\ }\bibfield  {title} {\enquote {\bibinfo {title} {{The imitation game reloaded: effective shadows of dynamically robust spinning Proca stars}},}\ }\href@noop {} {\  (\bibinfo {year} {2024})},\ \Eprint {http://arxiv.org/abs/2402.14919} {arXiv:2402.14919 [gr-qc]} \BibitemShut {NoStop}%
\bibitem [{\citenamefont {Adam}\ \emph {et~al.}(2010)\citenamefont {Adam}, \citenamefont {Grandi}, \citenamefont {Klimas}, \citenamefont {Sanchez-Guillen},\ and\ \citenamefont {Wereszczynski}}]{Adam:2010rrj}%
  \BibitemOpen
  \bibfield  {author} {\bibinfo {author} {\bibfnamefont {C.}~\bibnamefont {Adam}}, \bibinfo {author} {\bibfnamefont {N.}~\bibnamefont {Grandi}}, \bibinfo {author} {\bibfnamefont {P.}~\bibnamefont {Klimas}}, \bibinfo {author} {\bibfnamefont {J.}~\bibnamefont {Sanchez-Guillen}}, \ and\ \bibinfo {author} {\bibfnamefont {A.}~\bibnamefont {Wereszczynski}},\ }\bibfield  {title} {\enquote {\bibinfo {title} {{Compact boson stars in K field theories}},}\ }\href {\doibase 10.1007/s10714-010-1006-4} {\bibfield  {journal} {\bibinfo  {journal} {Gen. Rel. Grav.}\ }\textbf {\bibinfo {volume} {42}},\ \bibinfo {pages} {2663--2701} (\bibinfo {year} {2010})},\ \Eprint {http://arxiv.org/abs/0908.0218} {arXiv:0908.0218 [hep-th]} \BibitemShut {NoStop}%
\bibitem [{\citenamefont {Adam}\ \emph {et~al.}(2024)\citenamefont {Adam}, \citenamefont {Mourelle}, \citenamefont {dos Santos Costa~Filho}, \citenamefont {Herdeiro},\ and\ \citenamefont {Wereszczynski}}]{Adam:2024zqr}%
  \BibitemOpen
  \bibfield  {author} {\bibinfo {author} {\bibfnamefont {Christoph}\ \bibnamefont {Adam}}, \bibinfo {author} {\bibfnamefont {Jorge~Castelo}\ \bibnamefont {Mourelle}}, \bibinfo {author} {\bibfnamefont {Etevaldo}\ \bibnamefont {dos Santos Costa~Filho}}, \bibinfo {author} {\bibfnamefont {Carlos A.~R.}\ \bibnamefont {Herdeiro}}, \ and\ \bibinfo {author} {\bibfnamefont {Andrzej}\ \bibnamefont {Wereszczynski}},\ }\bibfield  {title} {\enquote {\bibinfo {title} {{Universal Relations for Rotating Scalar and Vector Boson Stars}},}\ }\href@noop {} {\  (\bibinfo {year} {2024})},\ \Eprint {http://arxiv.org/abs/2406.07613} {arXiv:2406.07613 [gr-qc]} \BibitemShut {NoStop}%
\bibitem [{\citenamefont {Bustillo}\ \emph {et~al.}(2021)\citenamefont {Bustillo}, \citenamefont {Sanchis-Gual}, \citenamefont {Torres-Forn\'e}, \citenamefont {Font}, \citenamefont {Vajpeyi}, \citenamefont {Smith}, \citenamefont {Herdeiro}, \citenamefont {Radu},\ and\ \citenamefont {Leong}}]{Bustillo:2020syj}%
  \BibitemOpen
  \bibfield  {author} {\bibinfo {author} {\bibfnamefont {Juan~Calder\'on}\ \bibnamefont {Bustillo}}, \bibinfo {author} {\bibfnamefont {Nicolas}\ \bibnamefont {Sanchis-Gual}}, \bibinfo {author} {\bibfnamefont {Alejandro}\ \bibnamefont {Torres-Forn\'e}}, \bibinfo {author} {\bibfnamefont {Jos\'e~A.}\ \bibnamefont {Font}}, \bibinfo {author} {\bibfnamefont {Avi}\ \bibnamefont {Vajpeyi}}, \bibinfo {author} {\bibfnamefont {Rory}\ \bibnamefont {Smith}}, \bibinfo {author} {\bibfnamefont {Carlos}\ \bibnamefont {Herdeiro}}, \bibinfo {author} {\bibfnamefont {Eugen}\ \bibnamefont {Radu}}, \ and\ \bibinfo {author} {\bibfnamefont {Samson H.~W.}\ \bibnamefont {Leong}},\ }\bibfield  {title} {\enquote {\bibinfo {title} {{GW190521 as a Merger of Proca Stars: A Potential New Vector Boson of $8.7\times 10^{-13}$ eV}},}\ }\href {\doibase 10.1103/PhysRevLett.126.081101} {\bibfield  {journal} {\bibinfo  {journal} {Phys. Rev. Lett.}\ }\textbf {\bibinfo {volume} {126}},\ \bibinfo {pages} {081101} (\bibinfo {year} {2021})},\ \Eprint
  {http://arxiv.org/abs/2009.05376} {arXiv:2009.05376 [gr-qc]} \BibitemShut {NoStop}%
\bibitem [{\citenamefont {Luna}\ \emph {et~al.}(2024)\citenamefont {Luna}, \citenamefont {Llorens-Monteagudo}, \citenamefont {Lorenzo-Medina}, \citenamefont {Bustillo}, \citenamefont {Sanchis-Gual}, \citenamefont {Torres-Forn\'e}, \citenamefont {Font}, \citenamefont {Herdeiro},\ and\ \citenamefont {Radu}}]{Luna:2024kof}%
  \BibitemOpen
  \bibfield  {author} {\bibinfo {author} {\bibfnamefont {Raimon}\ \bibnamefont {Luna}}, \bibinfo {author} {\bibfnamefont {Miquel}\ \bibnamefont {Llorens-Monteagudo}}, \bibinfo {author} {\bibfnamefont {Ana}\ \bibnamefont {Lorenzo-Medina}}, \bibinfo {author} {\bibfnamefont {Juan~Calder\'on}\ \bibnamefont {Bustillo}}, \bibinfo {author} {\bibfnamefont {Nicolas}\ \bibnamefont {Sanchis-Gual}}, \bibinfo {author} {\bibfnamefont {Alejandro}\ \bibnamefont {Torres-Forn\'e}}, \bibinfo {author} {\bibfnamefont {Jos\'e~A.}\ \bibnamefont {Font}}, \bibinfo {author} {\bibfnamefont {Carlos A.~R.}\ \bibnamefont {Herdeiro}}, \ and\ \bibinfo {author} {\bibfnamefont {Eugen}\ \bibnamefont {Radu}},\ }\bibfield  {title} {\enquote {\bibinfo {title} {{Numerical relativity surrogate models for exotic compact objects: the case of head-on mergers of equal-mass Proca stars}},}\ }\href@noop {} {\  (\bibinfo {year} {2024})},\ \Eprint {http://arxiv.org/abs/2404.01395} {arXiv:2404.01395 [gr-qc]} \BibitemShut {NoStop}%
\bibitem [{\citenamefont {Siemonsen}\ and\ \citenamefont {East}(2023)}]{Siemonsen:2023hko}%
  \BibitemOpen
  \bibfield  {author} {\bibinfo {author} {\bibfnamefont {Nils}\ \bibnamefont {Siemonsen}}\ and\ \bibinfo {author} {\bibfnamefont {William~E.}\ \bibnamefont {East}},\ }\bibfield  {title} {\enquote {\bibinfo {title} {{Binary boson stars: Merger dynamics and formation of rotating remnant stars}},}\ }\href {\doibase 10.1103/PhysRevD.107.124018} {\bibfield  {journal} {\bibinfo  {journal} {Phys. Rev. D}\ }\textbf {\bibinfo {volume} {107}},\ \bibinfo {pages} {124018} (\bibinfo {year} {2023})},\ \Eprint {http://arxiv.org/abs/2302.06627} {arXiv:2302.06627 [gr-qc]} \BibitemShut {NoStop}%
\bibitem [{\citenamefont {Schunck}(1998)}]{Schunck:1998nq}%
  \BibitemOpen
  \bibfield  {author} {\bibinfo {author} {\bibfnamefont {Franz~E.}\ \bibnamefont {Schunck}},\ }\bibfield  {title} {\enquote {\bibinfo {title} {{A Scalar field matter model for dark halos of galaxies and gravitational redshift}},}\ }\href@noop {} {\  (\bibinfo {year} {1998})},\ \Eprint {http://arxiv.org/abs/astro-ph/9802258} {arXiv:astro-ph/9802258} \BibitemShut {NoStop}%
\bibitem [{\citenamefont {Urena-Lopez}\ and\ \citenamefont {Bernal}(2010)}]{Urena-Lopez:2010zva}%
  \BibitemOpen
  \bibfield  {author} {\bibinfo {author} {\bibfnamefont {L.~Arturo}\ \bibnamefont {Urena-Lopez}}\ and\ \bibinfo {author} {\bibfnamefont {Argelia}\ \bibnamefont {Bernal}},\ }\bibfield  {title} {\enquote {\bibinfo {title} {{Bosonic gas as a Galactic Dark Matter Halo}},}\ }\href {\doibase 10.1103/PhysRevD.82.123535} {\bibfield  {journal} {\bibinfo  {journal} {Phys. Rev. D}\ }\textbf {\bibinfo {volume} {82}},\ \bibinfo {pages} {123535} (\bibinfo {year} {2010})},\ \Eprint {http://arxiv.org/abs/1008.1231} {arXiv:1008.1231 [gr-qc]} \BibitemShut {NoStop}%
\bibitem [{\citenamefont {Broadhurst}\ \emph {et~al.}(2020)\citenamefont {Broadhurst}, \citenamefont {de~Martino}, \citenamefont {Luu}, \citenamefont {Smoot},\ and\ \citenamefont {Tye}}]{Broadhurst:2019fsl}%
  \BibitemOpen
  \bibfield  {author} {\bibinfo {author} {\bibfnamefont {Tom}\ \bibnamefont {Broadhurst}}, \bibinfo {author} {\bibfnamefont {Ivan}\ \bibnamefont {de~Martino}}, \bibinfo {author} {\bibfnamefont {Hoang~Nhan}\ \bibnamefont {Luu}}, \bibinfo {author} {\bibfnamefont {George~F.}\ \bibnamefont {Smoot}}, \ and\ \bibinfo {author} {\bibfnamefont {S.~H.~Henry}\ \bibnamefont {Tye}},\ }\bibfield  {title} {\enquote {\bibinfo {title} {{Ghostly Galaxies as Solitons of Bose-Einstein Dark Matter}},}\ }\href {\doibase 10.1103/PhysRevD.101.083012} {\bibfield  {journal} {\bibinfo  {journal} {Phys. Rev. D}\ }\textbf {\bibinfo {volume} {101}},\ \bibinfo {pages} {083012} (\bibinfo {year} {2020})},\ \Eprint {http://arxiv.org/abs/1902.10488} {arXiv:1902.10488 [astro-ph.CO]} \BibitemShut {NoStop}%
\bibitem [{\citenamefont {Chen}\ \emph {et~al.}(2021)\citenamefont {Chen}, \citenamefont {Du}, \citenamefont {Lentz}, \citenamefont {Marsh},\ and\ \citenamefont {Niemeyer}}]{Chen:2020cef}%
  \BibitemOpen
  \bibfield  {author} {\bibinfo {author} {\bibfnamefont {Jiajun}\ \bibnamefont {Chen}}, \bibinfo {author} {\bibfnamefont {Xiaolong}\ \bibnamefont {Du}}, \bibinfo {author} {\bibfnamefont {Erik~W.}\ \bibnamefont {Lentz}}, \bibinfo {author} {\bibfnamefont {David J.~E.}\ \bibnamefont {Marsh}}, \ and\ \bibinfo {author} {\bibfnamefont {Jens~C.}\ \bibnamefont {Niemeyer}},\ }\bibfield  {title} {\enquote {\bibinfo {title} {{New insights into the formation and growth of boson stars in dark matter halos}},}\ }\href {\doibase 10.1103/PhysRevD.104.083022} {\bibfield  {journal} {\bibinfo  {journal} {Phys. Rev. D}\ }\textbf {\bibinfo {volume} {104}},\ \bibinfo {pages} {083022} (\bibinfo {year} {2021})},\ \Eprint {http://arxiv.org/abs/2011.01333} {arXiv:2011.01333 [astro-ph.CO]} \BibitemShut {NoStop}%
\bibitem [{\citenamefont {Annulli}\ \emph {et~al.}(2020)\citenamefont {Annulli}, \citenamefont {Cardoso},\ and\ \citenamefont {Vicente}}]{Annulli:2020ilw}%
  \BibitemOpen
  \bibfield  {author} {\bibinfo {author} {\bibfnamefont {Lorenzo}\ \bibnamefont {Annulli}}, \bibinfo {author} {\bibfnamefont {Vitor}\ \bibnamefont {Cardoso}}, \ and\ \bibinfo {author} {\bibfnamefont {Rodrigo}\ \bibnamefont {Vicente}},\ }\bibfield  {title} {\enquote {\bibinfo {title} {{Stirred and shaken: Dynamical behavior of boson stars and dark matter cores}},}\ }\href {\doibase 10.1016/j.physletb.2020.135944} {\bibfield  {journal} {\bibinfo  {journal} {Phys. Lett. B}\ }\textbf {\bibinfo {volume} {811}},\ \bibinfo {pages} {135944} (\bibinfo {year} {2020})},\ \Eprint {http://arxiv.org/abs/2007.03700} {arXiv:2007.03700 [astro-ph.HE]} \BibitemShut {NoStop}%
\bibitem [{\citenamefont {Amruth}\ \emph {et~al.}(2023)\citenamefont {Amruth} \emph {et~al.}}]{Amruth:2023xqj}%
  \BibitemOpen
  \bibfield  {author} {\bibinfo {author} {\bibfnamefont {A.}~\bibnamefont {Amruth}} \emph {et~al.},\ }\bibfield  {title} {\enquote {\bibinfo {title} {{Einstein rings modulated by wavelike dark matter from anomalies in gravitationally lensed images}},}\ }\href {\doibase 10.1038/s41550-023-01943-9} {\bibfield  {journal} {\bibinfo  {journal} {Nature Astron.}\ }\textbf {\bibinfo {volume} {7}},\ \bibinfo {pages} {736--747} (\bibinfo {year} {2023})},\ \Eprint {http://arxiv.org/abs/2304.09895} {arXiv:2304.09895 [astro-ph.CO]} \BibitemShut {NoStop}%
\bibitem [{\citenamefont {Pozo}\ \emph {et~al.}(2024)\citenamefont {Pozo}, \citenamefont {Broadhurst}, \citenamefont {Smoot}, \citenamefont {Chiueh}, \citenamefont {Luu}, \citenamefont {Vogelsberger},\ and\ \citenamefont {Mocz}}]{Pozo:2023zmx}%
  \BibitemOpen
  \bibfield  {author} {\bibinfo {author} {\bibfnamefont {Alvaro}\ \bibnamefont {Pozo}}, \bibinfo {author} {\bibfnamefont {Tom}\ \bibnamefont {Broadhurst}}, \bibinfo {author} {\bibfnamefont {George~F.}\ \bibnamefont {Smoot}}, \bibinfo {author} {\bibfnamefont {Tzihong}\ \bibnamefont {Chiueh}}, \bibinfo {author} {\bibfnamefont {Hoang~Nhan}\ \bibnamefont {Luu}}, \bibinfo {author} {\bibfnamefont {Mark}\ \bibnamefont {Vogelsberger}}, \ and\ \bibinfo {author} {\bibfnamefont {Philip}\ \bibnamefont {Mocz}},\ }\bibfield  {title} {\enquote {\bibinfo {title} {{Dwarf galaxies united by dark bosons}},}\ }\href {\doibase 10.1103/PhysRevD.109.083532} {\bibfield  {journal} {\bibinfo  {journal} {Phys. Rev. D}\ }\textbf {\bibinfo {volume} {109}},\ \bibinfo {pages} {083532} (\bibinfo {year} {2024})},\ \Eprint {http://arxiv.org/abs/2302.00181} {arXiv:2302.00181 [astro-ph.CO]} \BibitemShut {NoStop}%
\bibitem [{\citenamefont {Mourelle}\ and\ \citenamefont {Adam}(2024)}]{Mourelle:2024dlt}%
  \BibitemOpen
  \bibfield  {author} {\bibinfo {author} {\bibfnamefont {Jorge~Castelo}\ \bibnamefont {Mourelle}}\ and\ \bibinfo {author} {\bibfnamefont {Christoph}\ \bibnamefont {Adam}},\ }\bibfield  {title} {\enquote {\bibinfo {title} {{Galactic Halos and rotating bosonic dark matter}},}\ }\href@noop {} {\  (\bibinfo {year} {2024})},\ \Eprint {http://arxiv.org/abs/2407.07839} {arXiv:2407.07839 [astro-ph.GA]} \BibitemShut {NoStop}%
\bibitem [{\citenamefont {Cotner}\ and\ \citenamefont {Kusenko}(2017)}]{Cotner:2017tir}%
  \BibitemOpen
  \bibfield  {author} {\bibinfo {author} {\bibfnamefont {Eric}\ \bibnamefont {Cotner}}\ and\ \bibinfo {author} {\bibfnamefont {Alexander}\ \bibnamefont {Kusenko}},\ }\bibfield  {title} {\enquote {\bibinfo {title} {{Primordial black holes from scalar field evolution in the early universe}},}\ }\href {\doibase 10.1103/PhysRevD.96.103002} {\bibfield  {journal} {\bibinfo  {journal} {Phys. Rev. D}\ }\textbf {\bibinfo {volume} {96}},\ \bibinfo {pages} {103002} (\bibinfo {year} {2017})},\ \Eprint {http://arxiv.org/abs/1706.09003} {arXiv:1706.09003 [astro-ph.CO]} \BibitemShut {NoStop}%
\bibitem [{\citenamefont {Kusenko}\ \emph {et~al.}(2020)\citenamefont {Kusenko}, \citenamefont {Takhistov}, \citenamefont {Yamada},\ and\ \citenamefont {Yamazaki}}]{Kusenko:2019kcu}%
  \BibitemOpen
  \bibfield  {author} {\bibinfo {author} {\bibfnamefont {Alexander}\ \bibnamefont {Kusenko}}, \bibinfo {author} {\bibfnamefont {Volodymyr}\ \bibnamefont {Takhistov}}, \bibinfo {author} {\bibfnamefont {Masaki}\ \bibnamefont {Yamada}}, \ and\ \bibinfo {author} {\bibfnamefont {Masahito}\ \bibnamefont {Yamazaki}},\ }\bibfield  {title} {\enquote {\bibinfo {title} {{Fundamental Forces and Scalar Field Dynamics in the Early Universe}},}\ }\href {\doibase 10.1016/j.physletb.2020.135369} {\bibfield  {journal} {\bibinfo  {journal} {Phys. Lett. B}\ }\textbf {\bibinfo {volume} {804}},\ \bibinfo {pages} {135369} (\bibinfo {year} {2020})},\ \Eprint {http://arxiv.org/abs/1908.10930} {arXiv:1908.10930 [hep-th]} \BibitemShut {NoStop}%
\bibitem [{\citenamefont {Fodor}(2019)}]{Fodor:2019ftc}%
  \BibitemOpen
  \bibfield  {author} {\bibinfo {author} {\bibfnamefont {Gyula}\ \bibnamefont {Fodor}},\ }\emph {\bibinfo {title} {{A review on radiation of oscillons and oscillatons}}},\ \href@noop {} {Ph.D. thesis},\ \bibinfo  {school} {Wigner RCP, Budapest} (\bibinfo {year} {2019}),\ \Eprint {http://arxiv.org/abs/1911.03340} {arXiv:1911.03340 [hep-th]} \BibitemShut {NoStop}%
\bibitem [{\citenamefont {Schunck}\ and\ \citenamefont {Mielke}(2003)}]{Schunck:2003kk}%
  \BibitemOpen
  \bibfield  {author} {\bibinfo {author} {\bibfnamefont {Franz~E.}\ \bibnamefont {Schunck}}\ and\ \bibinfo {author} {\bibfnamefont {Eckehard~W.}\ \bibnamefont {Mielke}},\ }\bibfield  {title} {\enquote {\bibinfo {title} {{General relativistic boson stars}},}\ }\href {\doibase 10.1088/0264-9381/20/20/201} {\bibfield  {journal} {\bibinfo  {journal} {Class. Quant. Grav.}\ }\textbf {\bibinfo {volume} {20}},\ \bibinfo {pages} {R301--R356} (\bibinfo {year} {2003})},\ \Eprint {http://arxiv.org/abs/0801.0307} {arXiv:0801.0307 [astro-ph]} \BibitemShut {NoStop}%
\bibitem [{\citenamefont {Grandclement}\ \emph {et~al.}(2014)\citenamefont {Grandclement}, \citenamefont {Som\'e},\ and\ \citenamefont {Gourgoulhon}}]{Grandclement:2014msa}%
  \BibitemOpen
  \bibfield  {author} {\bibinfo {author} {\bibfnamefont {Philippe}\ \bibnamefont {Grandclement}}, \bibinfo {author} {\bibfnamefont {Claire}\ \bibnamefont {Som\'e}}, \ and\ \bibinfo {author} {\bibfnamefont {Eric}\ \bibnamefont {Gourgoulhon}},\ }\bibfield  {title} {\enquote {\bibinfo {title} {{Models of rotating boson stars and geodesics around them: new type of orbits}},}\ }\href {\doibase 10.1103/PhysRevD.90.024068} {\bibfield  {journal} {\bibinfo  {journal} {Phys. Rev. D}\ }\textbf {\bibinfo {volume} {90}},\ \bibinfo {pages} {024068} (\bibinfo {year} {2014})},\ \Eprint {http://arxiv.org/abs/1405.4837} {arXiv:1405.4837 [gr-qc]} \BibitemShut {NoStop}%
\bibitem [{\citenamefont {Choi}\ \emph {et~al.}(2019)\citenamefont {Choi}, \citenamefont {He},\ and\ \citenamefont {Schiappacasse}}]{Choi:2019mva}%
  \BibitemOpen
  \bibfield  {author} {\bibinfo {author} {\bibfnamefont {Gongjun}\ \bibnamefont {Choi}}, \bibinfo {author} {\bibfnamefont {Hong-Jian}\ \bibnamefont {He}}, \ and\ \bibinfo {author} {\bibfnamefont {Enrico~D.}\ \bibnamefont {Schiappacasse}},\ }\bibfield  {title} {\enquote {\bibinfo {title} {{Probing Dynamics of Boson Stars by Fast Radio Bursts and Gravitational Wave Detection}},}\ }\href {\doibase 10.1088/1475-7516/2019/10/043} {\bibfield  {journal} {\bibinfo  {journal} {JCAP}\ }\textbf {\bibinfo {volume} {10}},\ \bibinfo {pages} {043} (\bibinfo {year} {2019})},\ \Eprint {http://arxiv.org/abs/1906.02094} {arXiv:1906.02094 [astro-ph.CO]} \BibitemShut {NoStop}%
\bibitem [{\citenamefont {Guerra}\ \emph {et~al.}(2019)\citenamefont {Guerra}, \citenamefont {Macedo},\ and\ \citenamefont {Pani}}]{Guerra:2019srj}%
  \BibitemOpen
  \bibfield  {author} {\bibinfo {author} {\bibfnamefont {Davide}\ \bibnamefont {Guerra}}, \bibinfo {author} {\bibfnamefont {Caio F.~B.}\ \bibnamefont {Macedo}}, \ and\ \bibinfo {author} {\bibfnamefont {Paolo}\ \bibnamefont {Pani}},\ }\bibfield  {title} {\enquote {\bibinfo {title} {{Axion boson stars}},}\ }\href {\doibase 10.1088/1475-7516/2019/09/061} {\bibfield  {journal} {\bibinfo  {journal} {JCAP}\ }\textbf {\bibinfo {volume} {09}},\ \bibinfo {pages} {061} (\bibinfo {year} {2019})},\ \bibinfo {note} {[Erratum: JCAP 06, E01 (2020)]},\ \Eprint {http://arxiv.org/abs/1909.05515} {arXiv:1909.05515 [gr-qc]} \BibitemShut {NoStop}%
\bibitem [{\citenamefont {Delgado}\ \emph {et~al.}(2020)\citenamefont {Delgado}, \citenamefont {Herdeiro},\ and\ \citenamefont {Radu}}]{Delgado:2020udb}%
  \BibitemOpen
  \bibfield  {author} {\bibinfo {author} {\bibfnamefont {Jorge F.~M.}\ \bibnamefont {Delgado}}, \bibinfo {author} {\bibfnamefont {Carlos A.~R.}\ \bibnamefont {Herdeiro}}, \ and\ \bibinfo {author} {\bibfnamefont {Eugen}\ \bibnamefont {Radu}},\ }\bibfield  {title} {\enquote {\bibinfo {title} {{Rotating Axion Boson Stars}},}\ }\href {\doibase 10.1088/1475-7516/2020/06/037} {\bibfield  {journal} {\bibinfo  {journal} {JCAP}\ }\textbf {\bibinfo {volume} {06}},\ \bibinfo {pages} {037} (\bibinfo {year} {2020})},\ \Eprint {http://arxiv.org/abs/2005.05982} {arXiv:2005.05982 [gr-qc]} \BibitemShut {NoStop}%
\bibitem [{\citenamefont {Adam}\ \emph {et~al.}(2022)\citenamefont {Adam}, \citenamefont {Castelo}, \citenamefont {Garc\'\i{}a Mart\'\i{}n-Caro}, \citenamefont {Huidobro}, \citenamefont {V\'azquez},\ and\ \citenamefont {Wereszczynski}}]{Adam:2022nlq}%
  \BibitemOpen
  \bibfield  {author} {\bibinfo {author} {\bibfnamefont {Christoph}\ \bibnamefont {Adam}}, \bibinfo {author} {\bibfnamefont {Jorge}\ \bibnamefont {Castelo}}, \bibinfo {author} {\bibfnamefont {Alberto}\ \bibnamefont {Garc\'\i{}a Mart\'\i{}n-Caro}}, \bibinfo {author} {\bibfnamefont {Miguel}\ \bibnamefont {Huidobro}}, \bibinfo {author} {\bibfnamefont {Ricardo}\ \bibnamefont {V\'azquez}}, \ and\ \bibinfo {author} {\bibfnamefont {Andrzej}\ \bibnamefont {Wereszczynski}},\ }\bibfield  {title} {\enquote {\bibinfo {title} {{Universal relations for rotating boson stars}},}\ }\href {\doibase 10.1103/PhysRevD.106.123022} {\bibfield  {journal} {\bibinfo  {journal} {Phys. Rev. D}\ }\textbf {\bibinfo {volume} {106}},\ \bibinfo {pages} {123022} (\bibinfo {year} {2022})},\ \Eprint {http://arxiv.org/abs/2203.16558} {arXiv:2203.16558 [gr-qc]} \BibitemShut {NoStop}%
\bibitem [{\citenamefont {Adam}\ \emph {et~al.}(2023)\citenamefont {Adam}, \citenamefont {Castelo}, \citenamefont {Garc\'\i{}a Mart\'\i{}n-Caro}, \citenamefont {Huidobro},\ and\ \citenamefont {Wereszczynski}}]{Adam:2023qxj}%
  \BibitemOpen
  \bibfield  {author} {\bibinfo {author} {\bibfnamefont {Christoph}\ \bibnamefont {Adam}}, \bibinfo {author} {\bibfnamefont {Jorge}\ \bibnamefont {Castelo}}, \bibinfo {author} {\bibfnamefont {Alberto}\ \bibnamefont {Garc\'\i{}a Mart\'\i{}n-Caro}}, \bibinfo {author} {\bibfnamefont {Miguel}\ \bibnamefont {Huidobro}}, \ and\ \bibinfo {author} {\bibfnamefont {Andrzej}\ \bibnamefont {Wereszczynski}},\ }\bibfield  {title} {\enquote {\bibinfo {title} {{Effective no-hair relations for spinning boson stars}},}\ }\href {\doibase 10.1103/PhysRevD.108.043015} {\bibfield  {journal} {\bibinfo  {journal} {Phys. Rev. D}\ }\textbf {\bibinfo {volume} {108}},\ \bibinfo {pages} {043015} (\bibinfo {year} {2023})},\ \Eprint {http://arxiv.org/abs/2305.06181} {arXiv:2305.06181 [gr-qc]} \BibitemShut {NoStop}%
\bibitem [{\citenamefont {Cano}\ \emph {et~al.}(2024)\citenamefont {Cano}, \citenamefont {Machet},\ and\ \citenamefont {Myin}}]{Cano:2023bpe}%
  \BibitemOpen
  \bibfield  {author} {\bibinfo {author} {\bibfnamefont {Pablo~A.}\ \bibnamefont {Cano}}, \bibinfo {author} {\bibfnamefont {Ludovico}\ \bibnamefont {Machet}}, \ and\ \bibinfo {author} {\bibfnamefont {Charlotte}\ \bibnamefont {Myin}},\ }\bibfield  {title} {\enquote {\bibinfo {title} {{Boson stars with nonlinear sigma models}},}\ }\href {\doibase 10.1103/PhysRevD.109.044043} {\bibfield  {journal} {\bibinfo  {journal} {Phys. Rev. D}\ }\textbf {\bibinfo {volume} {109}},\ \bibinfo {pages} {044043} (\bibinfo {year} {2024})},\ \Eprint {http://arxiv.org/abs/2311.03433} {arXiv:2311.03433 [gr-qc]} \BibitemShut {NoStop}%
\bibitem [{\citenamefont {Herdeiro}\ \emph {et~al.}(2019)\citenamefont {Herdeiro}, \citenamefont {Perapechka}, \citenamefont {Radu},\ and\ \citenamefont {Shnir}}]{Herdeiro:2018djx}%
  \BibitemOpen
  \bibfield  {author} {\bibinfo {author} {\bibfnamefont {C.}~\bibnamefont {Herdeiro}}, \bibinfo {author} {\bibfnamefont {I.}~\bibnamefont {Perapechka}}, \bibinfo {author} {\bibfnamefont {E.}~\bibnamefont {Radu}}, \ and\ \bibinfo {author} {\bibfnamefont {Ya.}\ \bibnamefont {Shnir}},\ }\bibfield  {title} {\enquote {\bibinfo {title} {{Gravitating solitons and black holes with synchronised hair in the four dimensional O(3) sigma-model}},}\ }\href {\doibase 10.1007/JHEP02(2019)111} {\bibfield  {journal} {\bibinfo  {journal} {JHEP}\ }\textbf {\bibinfo {volume} {02}},\ \bibinfo {pages} {111} (\bibinfo {year} {2019})},\ \Eprint {http://arxiv.org/abs/1811.11799} {arXiv:1811.11799 [gr-qc]} \BibitemShut {NoStop}%
\bibitem [{\citenamefont {Verbin}(2007)}]{Verbin:2007fa}%
  \BibitemOpen
  \bibfield  {author} {\bibinfo {author} {\bibfnamefont {Y.}~\bibnamefont {Verbin}},\ }\bibfield  {title} {\enquote {\bibinfo {title} {{Sigma model Q-balls and Q-stars}},}\ }\href {\doibase 10.1103/PhysRevD.76.085018} {\bibfield  {journal} {\bibinfo  {journal} {Phys. Rev. D}\ }\textbf {\bibinfo {volume} {76}},\ \bibinfo {pages} {085018} (\bibinfo {year} {2007})},\ \Eprint {http://arxiv.org/abs/0708.3283} {arXiv:0708.3283 [gr-qc]} \BibitemShut {NoStop}%
\bibitem [{\citenamefont {Leung}\ and\ \citenamefont {Vafa}(1998)}]{Leung:1997tw}%
  \BibitemOpen
  \bibfield  {author} {\bibinfo {author} {\bibfnamefont {Naichung~Conan}\ \bibnamefont {Leung}}\ and\ \bibinfo {author} {\bibfnamefont {Cumrun}\ \bibnamefont {Vafa}},\ }\bibfield  {title} {\enquote {\bibinfo {title} {{Branes and toric geometry}},}\ }\href {\doibase 10.4310/ATMP.1998.v2.n1.a4} {\bibfield  {journal} {\bibinfo  {journal} {Adv. Theor. Math. Phys.}\ }\textbf {\bibinfo {volume} {2}},\ \bibinfo {pages} {91--118} (\bibinfo {year} {1998})},\ \Eprint {http://arxiv.org/abs/hep-th/9711013} {arXiv:hep-th/9711013} \BibitemShut {NoStop}%
\bibitem [{\citenamefont {Sawado}\ and\ \citenamefont {Yanai}(2020)}]{Sawado:2020ncc}%
  \BibitemOpen
  \bibfield  {author} {\bibinfo {author} {\bibfnamefont {Nobuyuki}\ \bibnamefont {Sawado}}\ and\ \bibinfo {author} {\bibfnamefont {Shota}\ \bibnamefont {Yanai}},\ }\bibfield  {title} {\enquote {\bibinfo {title} {{Compact, charged boson stars and shells in the $\mathbb{C}P^N$ gravitating nonlinear sigma model}},}\ }\href {\doibase 10.1103/PhysRevD.102.045007} {\bibfield  {journal} {\bibinfo  {journal} {Phys. Rev. D}\ }\textbf {\bibinfo {volume} {102}},\ \bibinfo {pages} {045007} (\bibinfo {year} {2020})},\ \Eprint {http://arxiv.org/abs/2006.03424} {arXiv:2006.03424 [hep-th]} \BibitemShut {NoStop}%
\bibitem [{\citenamefont {Kirichenkov}\ \emph {et~al.}(2024)\citenamefont {Kirichenkov}, \citenamefont {Kunz}, \citenamefont {Sawado},\ and\ \citenamefont {Shnir}}]{Kirichenkov:2023omy}%
  \BibitemOpen
  \bibfield  {author} {\bibinfo {author} {\bibfnamefont {R.}~\bibnamefont {Kirichenkov}}, \bibinfo {author} {\bibfnamefont {J.}~\bibnamefont {Kunz}}, \bibinfo {author} {\bibfnamefont {Nobuyuki}\ \bibnamefont {Sawado}}, \ and\ \bibinfo {author} {\bibfnamefont {Ya.}\ \bibnamefont {Shnir}},\ }\bibfield  {title} {\enquote {\bibinfo {title} {{Skyrmions and pion stars in the gauged U(1) Einstein-Skyrme model}},}\ }\href {\doibase 10.1103/PhysRevD.109.045002} {\bibfield  {journal} {\bibinfo  {journal} {Phys. Rev. D}\ }\textbf {\bibinfo {volume} {109}},\ \bibinfo {pages} {045002} (\bibinfo {year} {2024})},\ \Eprint {http://arxiv.org/abs/2311.12432} {arXiv:2311.12432 [hep-th]} \BibitemShut {NoStop}%
\bibitem [{\citenamefont {Adam}\ \emph {et~al.}(2015)\citenamefont {Adam}, \citenamefont {Naya}, \citenamefont {Sanchez-Guillen}, \citenamefont {Vazquez},\ and\ \citenamefont {Wereszczynski}}]{Adam:2015lpa}%
  \BibitemOpen
  \bibfield  {author} {\bibinfo {author} {\bibfnamefont {C.}~\bibnamefont {Adam}}, \bibinfo {author} {\bibfnamefont {C.}~\bibnamefont {Naya}}, \bibinfo {author} {\bibfnamefont {J.}~\bibnamefont {Sanchez-Guillen}}, \bibinfo {author} {\bibfnamefont {R.}~\bibnamefont {Vazquez}}, \ and\ \bibinfo {author} {\bibfnamefont {A.}~\bibnamefont {Wereszczynski}},\ }\bibfield  {title} {\enquote {\bibinfo {title} {{Neutron stars in the Bogomol'nyi-Prasad-Sommerfield Skyrme model: Mean-field limit versus full field theory}},}\ }\href {\doibase 10.1103/PhysRevC.92.025802} {\bibfield  {journal} {\bibinfo  {journal} {Phys. Rev. C}\ }\textbf {\bibinfo {volume} {92}},\ \bibinfo {pages} {025802} (\bibinfo {year} {2015})},\ \Eprint {http://arxiv.org/abs/1503.03095} {arXiv:1503.03095 [nucl-th]} \BibitemShut {NoStop}%
\bibitem [{\citenamefont {Adam}\ \emph {et~al.}(2020)\citenamefont {Adam}, \citenamefont {Garc\'\i{}a Mart\'\i{}n-Caro}, \citenamefont {Huidobro}, \citenamefont {V\'azquez},\ and\ \citenamefont {Wereszczynski}}]{Adam:2020yfv}%
  \BibitemOpen
  \bibfield  {author} {\bibinfo {author} {\bibfnamefont {Christoph}\ \bibnamefont {Adam}}, \bibinfo {author} {\bibfnamefont {Alberto}\ \bibnamefont {Garc\'\i{}a Mart\'\i{}n-Caro}}, \bibinfo {author} {\bibfnamefont {Miguel}\ \bibnamefont {Huidobro}}, \bibinfo {author} {\bibfnamefont {Ricardo}\ \bibnamefont {V\'azquez}}, \ and\ \bibinfo {author} {\bibfnamefont {Andrzej}\ \bibnamefont {Wereszczynski}},\ }\bibfield  {title} {\enquote {\bibinfo {title} {{A new consistent neutron star equation of state from a generalized Skyrme model}},}\ }\href {\doibase 10.1016/j.physletb.2020.135928} {\bibfield  {journal} {\bibinfo  {journal} {Phys. Lett. B}\ }\textbf {\bibinfo {volume} {811}},\ \bibinfo {pages} {135928} (\bibinfo {year} {2020})},\ \Eprint {http://arxiv.org/abs/2006.07983} {arXiv:2006.07983 [hep-th]} \BibitemShut {NoStop}%
\bibitem [{\citenamefont {Adam}\ \emph {et~al.}(2021{\natexlab{a}})\citenamefont {Adam}, \citenamefont {Garc\'\i{}a Mart\'\i{}n-Caro}, \citenamefont {Huidobro}, \citenamefont {V\'azquez},\ and\ \citenamefont {Wereszczynski}}]{Adam:2020aza}%
  \BibitemOpen
  \bibfield  {author} {\bibinfo {author} {\bibfnamefont {Christoph}\ \bibnamefont {Adam}}, \bibinfo {author} {\bibfnamefont {Alberto}\ \bibnamefont {Garc\'\i{}a Mart\'\i{}n-Caro}}, \bibinfo {author} {\bibfnamefont {Miguel}\ \bibnamefont {Huidobro}}, \bibinfo {author} {\bibfnamefont {Ricardo}\ \bibnamefont {V\'azquez}}, \ and\ \bibinfo {author} {\bibfnamefont {Andrzej}\ \bibnamefont {Wereszczynski}},\ }\bibfield  {title} {\enquote {\bibinfo {title} {{Quasiuniversal relations for generalized Skyrme stars}},}\ }\href {\doibase 10.1103/PhysRevD.103.023022} {\bibfield  {journal} {\bibinfo  {journal} {Phys. Rev. D}\ }\textbf {\bibinfo {volume} {103}},\ \bibinfo {pages} {023022} (\bibinfo {year} {2021}{\natexlab{a}})},\ \Eprint {http://arxiv.org/abs/2011.08573} {arXiv:2011.08573 [hep-th]} \BibitemShut {NoStop}%
\bibitem [{\citenamefont {Ortin}(2015)}]{Ortin:2015hya}%
  \BibitemOpen
  \bibfield  {author} {\bibinfo {author} {\bibfnamefont {Tomas}\ \bibnamefont {Ortin}},\ }\href {\doibase 10.1017/CBO9781139019750} {\emph {\bibinfo {title} {{Gravity and Strings}}}},\ \bibinfo {edition} {2nd}\ ed.,\ Cambridge Monographs on Mathematical Physics\ (\bibinfo  {publisher} {Cambridge University Press},\ \bibinfo {year} {2015})\BibitemShut {NoStop}%
\bibitem [{\citenamefont {Roest}(2005)}]{Roest:2004aqa}%
  \BibitemOpen
  \bibfield  {author} {\bibinfo {author} {\bibfnamefont {Diederik}\ \bibnamefont {Roest}},\ }\bibfield  {title} {\enquote {\bibinfo {title} {{M-theory and gauged supergravities}},}\ }\href {\doibase 10.1002/prop.200410192} {\bibfield  {journal} {\bibinfo  {journal} {Fortsch. Phys.}\ }\textbf {\bibinfo {volume} {53}},\ \bibinfo {pages} {119--230} (\bibinfo {year} {2005})},\ \Eprint {http://arxiv.org/abs/hep-th/0408175} {arXiv:hep-th/0408175} \BibitemShut {NoStop}%
\bibitem [{\citenamefont {Krippendorf}\ \emph {et~al.}(2018)\citenamefont {Krippendorf}, \citenamefont {Muia},\ and\ \citenamefont {Quevedo}}]{Krippendorf:2018tei}%
  \BibitemOpen
  \bibfield  {author} {\bibinfo {author} {\bibfnamefont {Sven}\ \bibnamefont {Krippendorf}}, \bibinfo {author} {\bibfnamefont {Francesco}\ \bibnamefont {Muia}}, \ and\ \bibinfo {author} {\bibfnamefont {Fernando}\ \bibnamefont {Quevedo}},\ }\bibfield  {title} {\enquote {\bibinfo {title} {{Moduli Stars}},}\ }\href {\doibase 10.1007/JHEP08(2018)070} {\bibfield  {journal} {\bibinfo  {journal} {JHEP}\ }\textbf {\bibinfo {volume} {08}},\ \bibinfo {pages} {070} (\bibinfo {year} {2018})},\ \Eprint {http://arxiv.org/abs/1806.04690} {arXiv:1806.04690 [hep-th]} \BibitemShut {NoStop}%
\bibitem [{\citenamefont {de~Rham}\ \emph {et~al.}(2016)\citenamefont {de~Rham}, \citenamefont {Tolley},\ and\ \citenamefont {Zhou}}]{deRham:2015ijs}%
  \BibitemOpen
  \bibfield  {author} {\bibinfo {author} {\bibfnamefont {Claudia}\ \bibnamefont {de~Rham}}, \bibinfo {author} {\bibfnamefont {Andrew~J.}\ \bibnamefont {Tolley}}, \ and\ \bibinfo {author} {\bibfnamefont {Shuang-Yong}\ \bibnamefont {Zhou}},\ }\bibfield  {title} {\enquote {\bibinfo {title} {{Non-compact nonlinear sigma models}},}\ }\href {\doibase 10.1016/j.physletb.2016.07.035} {\bibfield  {journal} {\bibinfo  {journal} {Phys. Lett. B}\ }\textbf {\bibinfo {volume} {760}},\ \bibinfo {pages} {579--583} (\bibinfo {year} {2016})},\ \Eprint {http://arxiv.org/abs/1512.06838} {arXiv:1512.06838 [hep-th]} \BibitemShut {NoStop}%
\bibitem [{\citenamefont {Collodel}\ \emph {et~al.}(2020)\citenamefont {Collodel}, \citenamefont {Doneva},\ and\ \citenamefont {Yazadjiev}}]{Collodel:2019uns}%
  \BibitemOpen
  \bibfield  {author} {\bibinfo {author} {\bibfnamefont {Lucas~G.}\ \bibnamefont {Collodel}}, \bibinfo {author} {\bibfnamefont {Daniela~D.}\ \bibnamefont {Doneva}}, \ and\ \bibinfo {author} {\bibfnamefont {Stoytcho~S.}\ \bibnamefont {Yazadjiev}},\ }\bibfield  {title} {\enquote {\bibinfo {title} {{Rotating tensor-multiscalar solitons}},}\ }\href {\doibase 10.1103/PhysRevD.101.044021} {\bibfield  {journal} {\bibinfo  {journal} {Phys. Rev. D}\ }\textbf {\bibinfo {volume} {101}},\ \bibinfo {pages} {044021} (\bibinfo {year} {2020})},\ \Eprint {http://arxiv.org/abs/1912.02498} {arXiv:1912.02498 [gr-qc]} \BibitemShut {NoStop}%
\bibitem [{\citenamefont {Schunck}\ and\ \citenamefont {Mielke}(1998)}]{schunck1998rotating}%
  \BibitemOpen
  \bibfield  {author} {\bibinfo {author} {\bibfnamefont {Franz~E}\ \bibnamefont {Schunck}}\ and\ \bibinfo {author} {\bibfnamefont {Eckehard~W}\ \bibnamefont {Mielke}},\ }\bibfield  {title} {\enquote {\bibinfo {title} {Rotating boson star as an effective mass torus in general relativity},}\ }\href@noop {} {\bibfield  {journal} {\bibinfo  {journal} {Physics Letters A}\ }\textbf {\bibinfo {volume} {249}},\ \bibinfo {pages} {389--394} (\bibinfo {year} {1998})}\BibitemShut {NoStop}%
\bibitem [{\citenamefont {Yoshida}\ and\ \citenamefont {Eriguchi}(1997)}]{Yoshida:1997qf}%
  \BibitemOpen
  \bibfield  {author} {\bibinfo {author} {\bibfnamefont {Shijun}\ \bibnamefont {Yoshida}}\ and\ \bibinfo {author} {\bibfnamefont {Yoshiharu}\ \bibnamefont {Eriguchi}},\ }\bibfield  {title} {\enquote {\bibinfo {title} {{Rotating boson stars in general relativity}},}\ }\href {\doibase 10.1103/PhysRevD.56.762} {\bibfield  {journal} {\bibinfo  {journal} {Phys. Rev. D}\ }\textbf {\bibinfo {volume} {56}},\ \bibinfo {pages} {762--771} (\bibinfo {year} {1997})}\BibitemShut {NoStop}%
\bibitem [{\citenamefont {Lidsey}\ \emph {et~al.}(2000)\citenamefont {Lidsey}, \citenamefont {Wands},\ and\ \citenamefont {Copeland}}]{Lidsey:1999mc}%
  \BibitemOpen
  \bibfield  {author} {\bibinfo {author} {\bibfnamefont {James~E.}\ \bibnamefont {Lidsey}}, \bibinfo {author} {\bibfnamefont {David}\ \bibnamefont {Wands}}, \ and\ \bibinfo {author} {\bibfnamefont {Edmund~J.}\ \bibnamefont {Copeland}},\ }\bibfield  {title} {\enquote {\bibinfo {title} {{Superstring cosmology}},}\ }\href {\doibase 10.1016/S0370-1573(00)00064-8} {\bibfield  {journal} {\bibinfo  {journal} {Phys. Rept.}\ }\textbf {\bibinfo {volume} {337}},\ \bibinfo {pages} {343--492} (\bibinfo {year} {2000})},\ \Eprint {http://arxiv.org/abs/hep-th/9909061} {arXiv:hep-th/9909061} \BibitemShut {NoStop}%
\bibitem [{\citenamefont {Vaglio}\ \emph {et~al.}(2022)\citenamefont {Vaglio}, \citenamefont {Pacilio}, \citenamefont {Maselli},\ and\ \citenamefont {Pani}}]{Vaglio:2022flq}%
  \BibitemOpen
  \bibfield  {author} {\bibinfo {author} {\bibfnamefont {Massimo}\ \bibnamefont {Vaglio}}, \bibinfo {author} {\bibfnamefont {Costantino}\ \bibnamefont {Pacilio}}, \bibinfo {author} {\bibfnamefont {Andrea}\ \bibnamefont {Maselli}}, \ and\ \bibinfo {author} {\bibfnamefont {Paolo}\ \bibnamefont {Pani}},\ }\bibfield  {title} {\enquote {\bibinfo {title} {{The multipolar structure of rotating boson stars}},}\ }\href@noop {} {\  (\bibinfo {year} {2022})},\ \Eprint {http://arxiv.org/abs/2203.07442} {arXiv:2203.07442 [gr-qc]} \BibitemShut {NoStop}%
\bibitem [{\citenamefont {Ryan}(1997)}]{Ryan:1996nk}%
  \BibitemOpen
  \bibfield  {author} {\bibinfo {author} {\bibfnamefont {Fintan~D.}\ \bibnamefont {Ryan}},\ }\bibfield  {title} {\enquote {\bibinfo {title} {{Spinning boson stars with large selfinteraction}},}\ }\href {\doibase 10.1103/PhysRevD.55.6081} {\bibfield  {journal} {\bibinfo  {journal} {Phys. Rev. D}\ }\textbf {\bibinfo {volume} {55}},\ \bibinfo {pages} {6081--6091} (\bibinfo {year} {1997})}\BibitemShut {NoStop}%
\bibitem [{\citenamefont {Herdeiro}\ and\ \citenamefont {Radu}(2015)}]{Herdeiro:2015gia}%
  \BibitemOpen
  \bibfield  {author} {\bibinfo {author} {\bibfnamefont {Carlos}\ \bibnamefont {Herdeiro}}\ and\ \bibinfo {author} {\bibfnamefont {Eugen}\ \bibnamefont {Radu}},\ }\bibfield  {title} {\enquote {\bibinfo {title} {{Construction and physical properties of Kerr black holes with scalar hair}},}\ }\href {\doibase 10.1088/0264-9381/32/14/144001} {\bibfield  {journal} {\bibinfo  {journal} {Class. Quant. Grav.}\ }\textbf {\bibinfo {volume} {32}},\ \bibinfo {pages} {144001} (\bibinfo {year} {2015})},\ \Eprint {http://arxiv.org/abs/1501.04319} {arXiv:1501.04319 [gr-qc]} \BibitemShut {NoStop}%
\bibitem [{\citenamefont {Thorne}(1980)}]{RevModPhys.52.299}%
  \BibitemOpen
  \bibfield  {author} {\bibinfo {author} {\bibfnamefont {Kip~S.}\ \bibnamefont {Thorne}},\ }\bibfield  {title} {\enquote {\bibinfo {title} {Multipole expansions of gravitational radiation},}\ }\href {\doibase 10.1103/RevModPhys.52.299} {\bibfield  {journal} {\bibinfo  {journal} {Rev. Mod. Phys.}\ }\textbf {\bibinfo {volume} {52}},\ \bibinfo {pages} {299--339} (\bibinfo {year} {1980})}\BibitemShut {NoStop}%
\bibitem [{\citenamefont {Kidder}(1995)}]{PhysRevD.52.821}%
  \BibitemOpen
  \bibfield  {author} {\bibinfo {author} {\bibfnamefont {Lawrence~E.}\ \bibnamefont {Kidder}},\ }\bibfield  {title} {\enquote {\bibinfo {title} {Coalescing binary systems of compact objects to (post${)}^{5/2}$-newtonian order. v. spin effects},}\ }\href {\doibase 10.1103/PhysRevD.52.821} {\bibfield  {journal} {\bibinfo  {journal} {Phys. Rev. D}\ }\textbf {\bibinfo {volume} {52}},\ \bibinfo {pages} {821--847} (\bibinfo {year} {1995})}\BibitemShut {NoStop}%
\bibitem [{\citenamefont {Geroch}(1970)}]{Geroch:1970cd}%
  \BibitemOpen
  \bibfield  {author} {\bibinfo {author} {\bibfnamefont {Robert~P.}\ \bibnamefont {Geroch}},\ }\bibfield  {title} {\enquote {\bibinfo {title} {{Multipole moments. II. Curved space}},}\ }\href {\doibase 10.1063/1.1665427} {\bibfield  {journal} {\bibinfo  {journal} {J. Math. Phys.}\ }\textbf {\bibinfo {volume} {11}},\ \bibinfo {pages} {2580--2588} (\bibinfo {year} {1970})}\BibitemShut {NoStop}%
\bibitem [{\citenamefont {Hansen}(1974)}]{Hansen:1974zz}%
  \BibitemOpen
  \bibfield  {author} {\bibinfo {author} {\bibfnamefont {R.~O.}\ \bibnamefont {Hansen}},\ }\bibfield  {title} {\enquote {\bibinfo {title} {{Multipole moments of stationary space-times}},}\ }\href {\doibase 10.1063/1.1666501} {\bibfield  {journal} {\bibinfo  {journal} {J. Math. Phys.}\ }\textbf {\bibinfo {volume} {15}},\ \bibinfo {pages} {46--52} (\bibinfo {year} {1974})}\BibitemShut {NoStop}%
\bibitem [{\citenamefont {Fodor}\ \emph {et~al.}(1989)\citenamefont {Fodor}, \citenamefont {Hoenselaers},\ and\ \citenamefont {Perj{\'e}s}}]{fodor1989multipole}%
  \BibitemOpen
  \bibfield  {author} {\bibinfo {author} {\bibfnamefont {G}~\bibnamefont {Fodor}}, \bibinfo {author} {\bibfnamefont {C}~\bibnamefont {Hoenselaers}}, \ and\ \bibinfo {author} {\bibfnamefont {Z}~\bibnamefont {Perj{\'e}s}},\ }\bibfield  {title} {\enquote {\bibinfo {title} {Multipole moments of axisymmetric systems in relativity},}\ }\href@noop {} {\bibfield  {journal} {\bibinfo  {journal} {Journal of Mathematical Physics}\ }\textbf {\bibinfo {volume} {30}},\ \bibinfo {pages} {2252--2257} (\bibinfo {year} {1989})}\BibitemShut {NoStop}%
\bibitem [{\citenamefont {Pappas}\ \emph {et~al.}(2019)\citenamefont {Pappas}, \citenamefont {Doneva}, \citenamefont {Sotiriou}, \citenamefont {Yazadjiev},\ and\ \citenamefont {Kokkotas}}]{Pappas:2018csu}%
  \BibitemOpen
  \bibfield  {author} {\bibinfo {author} {\bibfnamefont {George}\ \bibnamefont {Pappas}}, \bibinfo {author} {\bibfnamefont {Daniela~D.}\ \bibnamefont {Doneva}}, \bibinfo {author} {\bibfnamefont {Thomas~P.}\ \bibnamefont {Sotiriou}}, \bibinfo {author} {\bibfnamefont {Stoytcho~S.}\ \bibnamefont {Yazadjiev}}, \ and\ \bibinfo {author} {\bibfnamefont {Kostas~D.}\ \bibnamefont {Kokkotas}},\ }\bibfield  {title} {\enquote {\bibinfo {title} {{Multipole moments and universal relations for scalarized neutron stars}},}\ }\href {\doibase 10.1103/PhysRevD.99.104014} {\bibfield  {journal} {\bibinfo  {journal} {Phys. Rev. D}\ }\textbf {\bibinfo {volume} {99}},\ \bibinfo {pages} {104014} (\bibinfo {year} {2019})},\ \Eprint {http://arxiv.org/abs/1812.01117} {arXiv:1812.01117 [gr-qc]} \BibitemShut {NoStop}%
\bibitem [{\citenamefont {{Butterworth}}\ and\ \citenamefont {{Ipser}}(1976)}]{1976ApJ...204..200B}%
  \BibitemOpen
  \bibfield  {author} {\bibinfo {author} {\bibfnamefont {E.~M.}\ \bibnamefont {{Butterworth}}}\ and\ \bibinfo {author} {\bibfnamefont {J.~R.}\ \bibnamefont {{Ipser}}},\ }\bibfield  {title} {\enquote {\bibinfo {title} {{On the structure and stability of rapidly rotating fluid bodies in general relativity. I. The numerical method for computing structure and its application to uniformly rotating homogeneous bodies.}}}\ }\href {\doibase 10.1086/154163} {\bibfield  {journal} {\bibinfo  {journal} {\apj}\ }\textbf {\bibinfo {volume} {204}},\ \bibinfo {pages} {200--223} (\bibinfo {year} {1976})}\BibitemShut {NoStop}%
\bibitem [{\citenamefont {Morse}\ and\ \citenamefont {Feshbach}(1954)}]{morse1954methods}%
  \BibitemOpen
  \bibfield  {author} {\bibinfo {author} {\bibfnamefont {Philip~M}\ \bibnamefont {Morse}}\ and\ \bibinfo {author} {\bibfnamefont {Herman}\ \bibnamefont {Feshbach}},\ }\bibfield  {title} {\enquote {\bibinfo {title} {Methods of theoretical physics},}\ }\href@noop {} {\bibfield  {journal} {\bibinfo  {journal} {American Journal of Physics}\ }\textbf {\bibinfo {volume} {22}},\ \bibinfo {pages} {410--413} (\bibinfo {year} {1954})}\BibitemShut {NoStop}%
\bibitem [{\citenamefont {Doneva}\ and\ \citenamefont {Pappas}(2018)}]{Doneva:2017jop}%
  \BibitemOpen
  \bibfield  {author} {\bibinfo {author} {\bibfnamefont {Daniela~D.}\ \bibnamefont {Doneva}}\ and\ \bibinfo {author} {\bibfnamefont {George}\ \bibnamefont {Pappas}},\ }\bibfield  {title} {\enquote {\bibinfo {title} {{Universal Relations and Alternative Gravity Theories}},}\ }\href {\doibase 10.1007/978-3-319-97616-7_13} {\bibfield  {journal} {\bibinfo  {journal} {Astrophys. Space Sci. Libr.}\ }\textbf {\bibinfo {volume} {457}},\ \bibinfo {pages} {737--806} (\bibinfo {year} {2018})},\ \Eprint {http://arxiv.org/abs/1709.08046} {arXiv:1709.08046 [gr-qc]} \BibitemShut {NoStop}%
\bibitem [{\citenamefont {Pappas}\ and\ \citenamefont {Sotiriou}(2015)}]{Pappas:2014gca}%
  \BibitemOpen
  \bibfield  {author} {\bibinfo {author} {\bibfnamefont {George}\ \bibnamefont {Pappas}}\ and\ \bibinfo {author} {\bibfnamefont {Thomas~P.}\ \bibnamefont {Sotiriou}},\ }\bibfield  {title} {\enquote {\bibinfo {title} {{Multipole moments in scalar-tensor theory of gravity}},}\ }\href {\doibase 10.1103/PhysRevD.91.044011} {\bibfield  {journal} {\bibinfo  {journal} {Phys. Rev. D}\ }\textbf {\bibinfo {volume} {91}},\ \bibinfo {pages} {044011} (\bibinfo {year} {2015})},\ \Eprint {http://arxiv.org/abs/1412.3494} {arXiv:1412.3494 [gr-qc]} \BibitemShut {NoStop}%
\bibitem [{\citenamefont {{Yagi}}\ and\ \citenamefont {{Yunes}}(2013)}]{2013Sci...341..365Y}%
  \BibitemOpen
  \bibfield  {author} {\bibinfo {author} {\bibfnamefont {Kent}\ \bibnamefont {{Yagi}}}\ and\ \bibinfo {author} {\bibfnamefont {Nicol{\'a}s}\ \bibnamefont {{Yunes}}},\ }\bibfield  {title} {\enquote {\bibinfo {title} {{I-Love-Q: Unexpected Universal Relations for Neutron Stars and Quark Stars}},}\ }\href {\doibase 10.1126/science.1236462} {\bibfield  {journal} {\bibinfo  {journal} {Science}\ }\textbf {\bibinfo {volume} {341}},\ \bibinfo {pages} {365--368} (\bibinfo {year} {2013})},\ \Eprint {http://arxiv.org/abs/1302.4499} {arXiv:1302.4499 [gr-qc]} \BibitemShut {NoStop}%
\bibitem [{\citenamefont {Yagi}\ \emph {et~al.}(2014)\citenamefont {Yagi}, \citenamefont {Kyutoku}, \citenamefont {Pappas}, \citenamefont {Yunes},\ and\ \citenamefont {Apostolatos}}]{Yagi:2014bxa}%
  \BibitemOpen
  \bibfield  {author} {\bibinfo {author} {\bibfnamefont {Kent}\ \bibnamefont {Yagi}}, \bibinfo {author} {\bibfnamefont {Koutarou}\ \bibnamefont {Kyutoku}}, \bibinfo {author} {\bibfnamefont {George}\ \bibnamefont {Pappas}}, \bibinfo {author} {\bibfnamefont {Nicol\'as}\ \bibnamefont {Yunes}}, \ and\ \bibinfo {author} {\bibfnamefont {Theocharis~A.}\ \bibnamefont {Apostolatos}},\ }\bibfield  {title} {\enquote {\bibinfo {title} {{Effective No-Hair Relations for Neutron Stars and Quark Stars: Relativistic Results}},}\ }\href {\doibase 10.1103/PhysRevD.89.124013} {\bibfield  {journal} {\bibinfo  {journal} {Phys. Rev. D}\ }\textbf {\bibinfo {volume} {89}},\ \bibinfo {pages} {124013} (\bibinfo {year} {2014})},\ \Eprint {http://arxiv.org/abs/1403.6243} {arXiv:1403.6243 [gr-qc]} \BibitemShut {NoStop}%
\bibitem [{\citenamefont {Pappas}\ and\ \citenamefont {Apostolatos}(2012{\natexlab{a}})}]{Pappas:2012ns}%
  \BibitemOpen
  \bibfield  {author} {\bibinfo {author} {\bibfnamefont {George}\ \bibnamefont {Pappas}}\ and\ \bibinfo {author} {\bibfnamefont {Theocharis~A.}\ \bibnamefont {Apostolatos}},\ }\bibfield  {title} {\enquote {\bibinfo {title} {{Revising the multipole moments of numerical spacetimes, and its consequences}},}\ }\href {\doibase 10.1103/PhysRevLett.108.231104} {\bibfield  {journal} {\bibinfo  {journal} {Phys. Rev. Lett.}\ }\textbf {\bibinfo {volume} {108}},\ \bibinfo {pages} {231104} (\bibinfo {year} {2012}{\natexlab{a}})},\ \Eprint {http://arxiv.org/abs/1201.6067} {arXiv:1201.6067 [gr-qc]} \BibitemShut {NoStop}%
\bibitem [{\citenamefont {Pappas}\ and\ \citenamefont {Apostolatos}(2012{\natexlab{b}})}]{Pappas:2012qg}%
  \BibitemOpen
  \bibfield  {author} {\bibinfo {author} {\bibfnamefont {George}\ \bibnamefont {Pappas}}\ and\ \bibinfo {author} {\bibfnamefont {Theocharis~A}\ \bibnamefont {Apostolatos}},\ }\bibfield  {title} {\enquote {\bibinfo {title} {{Multipole Moments of numerical spacetimes}},}\ }\href@noop {} {\  (\bibinfo {year} {2012}{\natexlab{b}})},\ \Eprint {http://arxiv.org/abs/1211.6299} {arXiv:1211.6299 [gr-qc]} \BibitemShut {NoStop}%
\bibitem [{\citenamefont {Sukhov}(2023)}]{Sukhov:2023rln}%
  \BibitemOpen
  \bibfield  {author} {\bibinfo {author} {\bibfnamefont {Nikolay}\ \bibnamefont {Sukhov}},\ }\emph {\bibinfo {title} {{Explorations of Boson Stars}}},\ \href@noop {} {Ph.D. thesis},\ \bibinfo  {school} {Princeton U.} (\bibinfo {year} {2023})\BibitemShut {NoStop}%
\bibitem [{\citenamefont {Di~Giovanni}\ \emph {et~al.}(2020)\citenamefont {Di~Giovanni}, \citenamefont {Sanchis-Gual}, \citenamefont {Cerd\'a-Dur\'an}, \citenamefont {Zilh\~ao}, \citenamefont {Herdeiro}, \citenamefont {Font},\ and\ \citenamefont {Radu}}]{DiGiovanni:2020ror}%
  \BibitemOpen
  \bibfield  {author} {\bibinfo {author} {\bibfnamefont {Fabrizio}\ \bibnamefont {Di~Giovanni}}, \bibinfo {author} {\bibfnamefont {Nicolas}\ \bibnamefont {Sanchis-Gual}}, \bibinfo {author} {\bibfnamefont {Pablo}\ \bibnamefont {Cerd\'a-Dur\'an}}, \bibinfo {author} {\bibfnamefont {Miguel}\ \bibnamefont {Zilh\~ao}}, \bibinfo {author} {\bibfnamefont {Carlos}\ \bibnamefont {Herdeiro}}, \bibinfo {author} {\bibfnamefont {Jos\'e~A.}\ \bibnamefont {Font}}, \ and\ \bibinfo {author} {\bibfnamefont {Eugen}\ \bibnamefont {Radu}},\ }\bibfield  {title} {\enquote {\bibinfo {title} {{Dynamical bar-mode instability in spinning bosonic stars}},}\ }\href {\doibase 10.1103/PhysRevD.102.124009} {\bibfield  {journal} {\bibinfo  {journal} {Phys. Rev. D}\ }\textbf {\bibinfo {volume} {102}},\ \bibinfo {pages} {124009} (\bibinfo {year} {2020})},\ \Eprint {http://arxiv.org/abs/2010.05845} {arXiv:2010.05845 [gr-qc]} \BibitemShut {NoStop}%
\bibitem [{\citenamefont {Silveira}\ and\ \citenamefont {de~Sousa}(1995)}]{Silveira:1995dh}%
  \BibitemOpen
  \bibfield  {author} {\bibinfo {author} {\bibfnamefont {Vanda}\ \bibnamefont {Silveira}}\ and\ \bibinfo {author} {\bibfnamefont {Claudio M.~G.}\ \bibnamefont {de~Sousa}},\ }\bibfield  {title} {\enquote {\bibinfo {title} {{Boson star rotation: A Newtonian approximation}},}\ }\href {\doibase 10.1103/PhysRevD.52.5724} {\bibfield  {journal} {\bibinfo  {journal} {Phys. Rev. D}\ }\textbf {\bibinfo {volume} {52}},\ \bibinfo {pages} {5724--5728} (\bibinfo {year} {1995})},\ \Eprint {http://arxiv.org/abs/astro-ph/9508034} {arXiv:astro-ph/9508034} \BibitemShut {NoStop}%
\bibitem [{\citenamefont {Ferrell}\ and\ \citenamefont {Gleiser}(1989)}]{Ferrell:1989kz}%
  \BibitemOpen
  \bibfield  {author} {\bibinfo {author} {\bibfnamefont {Robert}\ \bibnamefont {Ferrell}}\ and\ \bibinfo {author} {\bibfnamefont {Marcelo}\ \bibnamefont {Gleiser}},\ }\bibfield  {title} {\enquote {\bibinfo {title} {{Gravitational Atoms. 1. Gravitational Radiation From Excited Boson Stars}},}\ }\href {\doibase 10.1103/PhysRevD.40.2524} {\bibfield  {journal} {\bibinfo  {journal} {Phys. Rev. D}\ }\textbf {\bibinfo {volume} {40}},\ \bibinfo {pages} {2524} (\bibinfo {year} {1989})}\BibitemShut {NoStop}%
\bibitem [{\citenamefont {Vaglio}(2023)}]{Vaglio:2023zpm}%
  \BibitemOpen
  \bibfield  {author} {\bibinfo {author} {\bibfnamefont {Massimo}\ \bibnamefont {Vaglio}},\ }\emph {\bibinfo {title} {{Modelling and phenomenology of boson stars as gravitational sources for future ground- and space-based interferometers}}},\ \href@noop {} {Ph.D. thesis},\ \bibinfo  {school} {Rome U.} (\bibinfo {year} {2023})\BibitemShut {NoStop}%
\bibitem [{\citenamefont {Khlopov}\ \emph {et~al.}(1985)\citenamefont {Khlopov}, \citenamefont {Malomed}, \citenamefont {Zeldovich},\ and\ \citenamefont {Zeldovich}}]{Khlopov:1985fch}%
  \BibitemOpen
  \bibfield  {author} {\bibinfo {author} {\bibfnamefont {M.~Yu.}\ \bibnamefont {Khlopov}}, \bibinfo {author} {\bibfnamefont {B.~A.}\ \bibnamefont {Malomed}}, \bibinfo {author} {\bibfnamefont {Ia.~B.}\ \bibnamefont {Zeldovich}}, \ and\ \bibinfo {author} {\bibfnamefont {Ya.~B.}\ \bibnamefont {Zeldovich}},\ }\bibfield  {title} {\enquote {\bibinfo {title} {{Gravitational instability of scalar fields and formation of primordial black holes}},}\ }\href {\doibase 10.1093/mnras/215.4.575} {\bibfield  {journal} {\bibinfo  {journal} {Mon. Not. Roy. Astron. Soc.}\ }\textbf {\bibinfo {volume} {215}},\ \bibinfo {pages} {575--589} (\bibinfo {year} {1985})}\BibitemShut {NoStop}%
\bibitem [{\citenamefont {Herdeiro}\ \emph {et~al.}(2015)\citenamefont {Herdeiro}, \citenamefont {Radu},\ and\ \citenamefont {R\'unarsson}}]{Herdeiro:2015tia}%
  \BibitemOpen
  \bibfield  {author} {\bibinfo {author} {\bibfnamefont {Carlos A.~R.}\ \bibnamefont {Herdeiro}}, \bibinfo {author} {\bibfnamefont {Eugen}\ \bibnamefont {Radu}}, \ and\ \bibinfo {author} {\bibfnamefont {Helgi}\ \bibnamefont {R\'unarsson}},\ }\bibfield  {title} {\enquote {\bibinfo {title} {{Kerr black holes with self-interacting scalar hair: hairier but not heavier}},}\ }\href {\doibase 10.1103/PhysRevD.92.084059} {\bibfield  {journal} {\bibinfo  {journal} {Phys. Rev. D}\ }\textbf {\bibinfo {volume} {92}},\ \bibinfo {pages} {084059} (\bibinfo {year} {2015})},\ \Eprint {http://arxiv.org/abs/1509.02923} {arXiv:1509.02923 [gr-qc]} \BibitemShut {NoStop}%
\bibitem [{\citenamefont {Kleihaus}\ \emph {et~al.}(2008)\citenamefont {Kleihaus}, \citenamefont {Kunz}, \citenamefont {List},\ and\ \citenamefont {Schaffer}}]{Kleihaus:2007vk}%
  \BibitemOpen
  \bibfield  {author} {\bibinfo {author} {\bibfnamefont {Burkhard}\ \bibnamefont {Kleihaus}}, \bibinfo {author} {\bibfnamefont {Jutta}\ \bibnamefont {Kunz}}, \bibinfo {author} {\bibfnamefont {Meike}\ \bibnamefont {List}}, \ and\ \bibinfo {author} {\bibfnamefont {Isabell}\ \bibnamefont {Schaffer}},\ }\bibfield  {title} {\enquote {\bibinfo {title} {{Rotating Boson Stars and Q-Balls. II. Negative Parity and Ergoregions}},}\ }\href {\doibase 10.1103/PhysRevD.77.064025} {\bibfield  {journal} {\bibinfo  {journal} {Phys. Rev. D}\ }\textbf {\bibinfo {volume} {77}},\ \bibinfo {pages} {064025} (\bibinfo {year} {2008})},\ \Eprint {http://arxiv.org/abs/0712.3742} {arXiv:0712.3742 [gr-qc]} \BibitemShut {NoStop}%
\bibitem [{\citenamefont {Tsokaros}\ \emph {et~al.}(2019)\citenamefont {Tsokaros}, \citenamefont {Ruiz}, \citenamefont {Sun}, \citenamefont {Shapiro},\ and\ \citenamefont {Ury\={u}}}]{Tsokaros:2019mlz}%
  \BibitemOpen
  \bibfield  {author} {\bibinfo {author} {\bibfnamefont {Antonios}\ \bibnamefont {Tsokaros}}, \bibinfo {author} {\bibfnamefont {Milton}\ \bibnamefont {Ruiz}}, \bibinfo {author} {\bibfnamefont {Lunan}\ \bibnamefont {Sun}}, \bibinfo {author} {\bibfnamefont {Stuart~L.}\ \bibnamefont {Shapiro}}, \ and\ \bibinfo {author} {\bibfnamefont {K\={o}ji}\ \bibnamefont {Ury\={u}}},\ }\bibfield  {title} {\enquote {\bibinfo {title} {{Dynamically stable ergostars exist: General relativistic models and simulations}},}\ }\href {\doibase 10.1103/PhysRevLett.123.231103} {\bibfield  {journal} {\bibinfo  {journal} {Phys. Rev. Lett.}\ }\textbf {\bibinfo {volume} {123}},\ \bibinfo {pages} {231103} (\bibinfo {year} {2019})},\ \Eprint {http://arxiv.org/abs/1907.03765} {arXiv:1907.03765 [gr-qc]} \BibitemShut {NoStop}%
\bibitem [{\citenamefont {Cardoso}\ \emph {et~al.}(2008)\citenamefont {Cardoso}, \citenamefont {Pani}, \citenamefont {Cadoni},\ and\ \citenamefont {Cavaglia}}]{Cardoso:2007az}%
  \BibitemOpen
  \bibfield  {author} {\bibinfo {author} {\bibfnamefont {Vitor}\ \bibnamefont {Cardoso}}, \bibinfo {author} {\bibfnamefont {Paolo}\ \bibnamefont {Pani}}, \bibinfo {author} {\bibfnamefont {Mariano}\ \bibnamefont {Cadoni}}, \ and\ \bibinfo {author} {\bibfnamefont {Marco}\ \bibnamefont {Cavaglia}},\ }\bibfield  {title} {\enquote {\bibinfo {title} {{Ergoregion instability of ultracompact astrophysical objects}},}\ }\href {\doibase 10.1103/PhysRevD.77.124044} {\bibfield  {journal} {\bibinfo  {journal} {Phys. Rev. D}\ }\textbf {\bibinfo {volume} {77}},\ \bibinfo {pages} {124044} (\bibinfo {year} {2008})},\ \Eprint {http://arxiv.org/abs/0709.0532} {arXiv:0709.0532 [gr-qc]} \BibitemShut {NoStop}%
\bibitem [{\citenamefont {Maggio}\ \emph {et~al.}(2019)\citenamefont {Maggio}, \citenamefont {Cardoso}, \citenamefont {Dolan},\ and\ \citenamefont {Pani}}]{Maggio:2018ivz}%
  \BibitemOpen
  \bibfield  {author} {\bibinfo {author} {\bibfnamefont {Elisa}\ \bibnamefont {Maggio}}, \bibinfo {author} {\bibfnamefont {Vitor}\ \bibnamefont {Cardoso}}, \bibinfo {author} {\bibfnamefont {Sam~R.}\ \bibnamefont {Dolan}}, \ and\ \bibinfo {author} {\bibfnamefont {Paolo}\ \bibnamefont {Pani}},\ }\bibfield  {title} {\enquote {\bibinfo {title} {{Ergoregion instability of exotic compact objects: electromagnetic and gravitational perturbations and the role of absorption}},}\ }\href {\doibase 10.1103/PhysRevD.99.064007} {\bibfield  {journal} {\bibinfo  {journal} {Phys. Rev. D}\ }\textbf {\bibinfo {volume} {99}},\ \bibinfo {pages} {064007} (\bibinfo {year} {2019})},\ \Eprint {http://arxiv.org/abs/1807.08840} {arXiv:1807.08840 [gr-qc]} \BibitemShut {NoStop}%
\bibitem [{\citenamefont {Aranguren}\ \emph {et~al.}(2024)\citenamefont {Aranguren}, \citenamefont {Font}, \citenamefont {Sanchis-Gual},\ and\ \citenamefont {Vera}}]{Aranguren:2024hds}%
  \BibitemOpen
  \bibfield  {author} {\bibinfo {author} {\bibfnamefont {Eneko}\ \bibnamefont {Aranguren}}, \bibinfo {author} {\bibfnamefont {Jos\'e~A.}\ \bibnamefont {Font}}, \bibinfo {author} {\bibfnamefont {Nicolas}\ \bibnamefont {Sanchis-Gual}}, \ and\ \bibinfo {author} {\bibfnamefont {Ra\"ul}\ \bibnamefont {Vera}},\ }\bibfield  {title} {\enquote {\bibinfo {title} {{I-Love-Q, and \ensuremath{\delta}M too: The role of the mass in universal relations of compact stars}},}\ }\href {\doibase 10.1103/PhysRevD.110.084027} {\bibfield  {journal} {\bibinfo  {journal} {Phys. Rev. D}\ }\textbf {\bibinfo {volume} {110}},\ \bibinfo {pages} {084027} (\bibinfo {year} {2024})},\ \Eprint {http://arxiv.org/abs/2407.20151} {arXiv:2407.20151 [gr-qc]} \BibitemShut {NoStop}%
\bibitem [{\citenamefont {Aranguren}\ \emph {et~al.}(2023)\citenamefont {Aranguren}, \citenamefont {Font}, \citenamefont {Sanchis-Gual},\ and\ \citenamefont {Vera}}]{Aranguren:2023ujo}%
  \BibitemOpen
  \bibfield  {author} {\bibinfo {author} {\bibfnamefont {Eneko}\ \bibnamefont {Aranguren}}, \bibinfo {author} {\bibfnamefont {Jos\'e~A.}\ \bibnamefont {Font}}, \bibinfo {author} {\bibfnamefont {Nicolas}\ \bibnamefont {Sanchis-Gual}}, \ and\ \bibinfo {author} {\bibfnamefont {Ra\"ul}\ \bibnamefont {Vera}},\ }\bibfield  {title} {\enquote {\bibinfo {title} {{Revisiting the I-Love-Q relations for superfluid neutron stars}},}\ }\href {\doibase 10.1103/PhysRevD.108.104065} {\bibfield  {journal} {\bibinfo  {journal} {Phys. Rev. D}\ }\textbf {\bibinfo {volume} {108}},\ \bibinfo {pages} {104065} (\bibinfo {year} {2023})},\ \Eprint {http://arxiv.org/abs/2309.03816} {arXiv:2309.03816 [gr-qc]} \BibitemShut {NoStop}%
\bibitem [{\citenamefont {Yagi}\ and\ \citenamefont {Yunes}(2013)}]{Yagi:2013awa}%
  \BibitemOpen
  \bibfield  {author} {\bibinfo {author} {\bibfnamefont {Kent}\ \bibnamefont {Yagi}}\ and\ \bibinfo {author} {\bibfnamefont {Nicolas}\ \bibnamefont {Yunes}},\ }\bibfield  {title} {\enquote {\bibinfo {title} {{I-Love-Q Relations in Neutron Stars and their Applications to Astrophysics, Gravitational Waves and Fundamental Physics}},}\ }\href {\doibase 10.1103/PhysRevD.88.023009} {\bibfield  {journal} {\bibinfo  {journal} {Phys. Rev. D}\ }\textbf {\bibinfo {volume} {88}},\ \bibinfo {pages} {023009} (\bibinfo {year} {2013})},\ \Eprint {http://arxiv.org/abs/1303.1528} {arXiv:1303.1528 [gr-qc]} \BibitemShut {NoStop}%
\bibitem [{\citenamefont {Adam}\ \emph {et~al.}(2021{\natexlab{b}})\citenamefont {Adam}, \citenamefont {Mart{\'\i}n-Caro}, \citenamefont {Huidobro}, \citenamefont {V{\'a}zquez},\ and\ \citenamefont {Wereszczynski}}]{adam2021quasiuniversal}%
  \BibitemOpen
  \bibfield  {author} {\bibinfo {author} {\bibfnamefont {Christoph}\ \bibnamefont {Adam}}, \bibinfo {author} {\bibfnamefont {Alberto~Garc{\'\i}a}\ \bibnamefont {Mart{\'\i}n-Caro}}, \bibinfo {author} {\bibfnamefont {Miguel}\ \bibnamefont {Huidobro}}, \bibinfo {author} {\bibfnamefont {Ricardo}\ \bibnamefont {V{\'a}zquez}}, \ and\ \bibinfo {author} {\bibfnamefont {Andrzej}\ \bibnamefont {Wereszczynski}},\ }\bibfield  {title} {\enquote {\bibinfo {title} {Quasiuniversal relations for generalized skyrme stars},}\ }\href@noop {} {\bibfield  {journal} {\bibinfo  {journal} {Physical Review D}\ }\textbf {\bibinfo {volume} {103}},\ \bibinfo {pages} {023022} (\bibinfo {year} {2021}{\natexlab{b}})}\BibitemShut {NoStop}%
\bibitem [{\citenamefont {Lee}\ and\ \citenamefont {Pang}(1989)}]{LEE1989477}%
  \BibitemOpen
  \bibfield  {author} {\bibinfo {author} {\bibfnamefont {T.D.}\ \bibnamefont {Lee}}\ and\ \bibinfo {author} {\bibfnamefont {Yang}\ \bibnamefont {Pang}},\ }\bibfield  {title} {\enquote {\bibinfo {title} {Stability of mini-boson stars},}\ }\href {\doibase https://doi.org/10.1016/0550-3213(89)90365-9} {\bibfield  {journal} {\bibinfo  {journal} {Nuclear Physics B}\ }\textbf {\bibinfo {volume} {315}},\ \bibinfo {pages} {477--516} (\bibinfo {year} {1989})}\BibitemShut {NoStop}%
\bibitem [{\citenamefont {Gleiser}\ and\ \citenamefont {Watkins}(1989)}]{GLEISER1989733}%
  \BibitemOpen
  \bibfield  {author} {\bibinfo {author} {\bibfnamefont {Marcelo}\ \bibnamefont {Gleiser}}\ and\ \bibinfo {author} {\bibfnamefont {Richard}\ \bibnamefont {Watkins}},\ }\bibfield  {title} {\enquote {\bibinfo {title} {{Gravitational Stability of Scalar Matter}},}\ }\href {\doibase 10.1016/0550-3213(89)90627-5} {\bibfield  {journal} {\bibinfo  {journal} {Nucl. Phys. B}\ }\textbf {\bibinfo {volume} {319}},\ \bibinfo {pages} {733--746} (\bibinfo {year} {1989})}\BibitemShut {NoStop}%
\bibitem [{\citenamefont {Kusmartsev}\ \emph {et~al.}(1991{\natexlab{a}})\citenamefont {Kusmartsev}, \citenamefont {Mielke},\ and\ \citenamefont {Schunck}}]{Kusmartsev:1990cr}%
  \BibitemOpen
  \bibfield  {author} {\bibinfo {author} {\bibfnamefont {Fjodor~V.}\ \bibnamefont {Kusmartsev}}, \bibinfo {author} {\bibfnamefont {Eckehard~W.}\ \bibnamefont {Mielke}}, \ and\ \bibinfo {author} {\bibfnamefont {Franz~E.}\ \bibnamefont {Schunck}},\ }\bibfield  {title} {\enquote {\bibinfo {title} {{Gravitational stability of boson stars}},}\ }\href {\doibase 10.1103/PhysRevD.43.3895} {\bibfield  {journal} {\bibinfo  {journal} {Phys. Rev. D}\ }\textbf {\bibinfo {volume} {43}},\ \bibinfo {pages} {3895--3901} (\bibinfo {year} {1991}{\natexlab{a}})},\ \Eprint {http://arxiv.org/abs/0810.0696} {arXiv:0810.0696 [astro-ph]} \BibitemShut {NoStop}%
\bibitem [{\citenamefont {Kusmartsev}\ \emph {et~al.}(1991{\natexlab{b}})\citenamefont {Kusmartsev}, \citenamefont {Mielke},\ and\ \citenamefont {Schunck}}]{PhysRevD.43.3895}%
  \BibitemOpen
  \bibfield  {author} {\bibinfo {author} {\bibfnamefont {Fjodor~V.}\ \bibnamefont {Kusmartsev}}, \bibinfo {author} {\bibfnamefont {Eckehard~W.}\ \bibnamefont {Mielke}}, \ and\ \bibinfo {author} {\bibfnamefont {Franz~E.}\ \bibnamefont {Schunck}},\ }\bibfield  {title} {\enquote {\bibinfo {title} {Gravitational stability of boson stars},}\ }\href {\doibase 10.1103/PhysRevD.43.3895} {\bibfield  {journal} {\bibinfo  {journal} {Phys. Rev. D}\ }\textbf {\bibinfo {volume} {43}},\ \bibinfo {pages} {3895--3901} (\bibinfo {year} {1991}{\natexlab{b}})}\BibitemShut {NoStop}%
\bibitem [{\citenamefont {Tamaki}\ and\ \citenamefont {Sakai}(2011)}]{PhysRevD.83.044027}%
  \BibitemOpen
  \bibfield  {author} {\bibinfo {author} {\bibfnamefont {Takashi}\ \bibnamefont {Tamaki}}\ and\ \bibinfo {author} {\bibfnamefont {Nobuyuki}\ \bibnamefont {Sakai}},\ }\bibfield  {title} {\enquote {\bibinfo {title} {How does gravity save or kill $q$-balls?}}\ }\href {\doibase 10.1103/PhysRevD.83.044027} {\bibfield  {journal} {\bibinfo  {journal} {Phys. Rev. D}\ }\textbf {\bibinfo {volume} {83}},\ \bibinfo {pages} {044027} (\bibinfo {year} {2011})}\BibitemShut {NoStop}%
\bibitem [{\citenamefont {Kleihaus}\ \emph {et~al.}(2012)\citenamefont {Kleihaus}, \citenamefont {Kunz},\ and\ \citenamefont {Schneider}}]{Kleihaus:2011sx}%
  \BibitemOpen
  \bibfield  {author} {\bibinfo {author} {\bibfnamefont {Burkhard}\ \bibnamefont {Kleihaus}}, \bibinfo {author} {\bibfnamefont {Jutta}\ \bibnamefont {Kunz}}, \ and\ \bibinfo {author} {\bibfnamefont {Stefanie}\ \bibnamefont {Schneider}},\ }\bibfield  {title} {\enquote {\bibinfo {title} {{Stable Phases of Boson Stars}},}\ }\href {\doibase 10.1103/PhysRevD.85.024045} {\bibfield  {journal} {\bibinfo  {journal} {Phys. Rev. D}\ }\textbf {\bibinfo {volume} {85}},\ \bibinfo {pages} {024045} (\bibinfo {year} {2012})},\ \Eprint {http://arxiv.org/abs/1109.5858} {arXiv:1109.5858 [gr-qc]} \BibitemShut {NoStop}%
\bibitem [{\citenamefont {Liebling}\ and\ \citenamefont {Palenzuela}(2012)}]{Liebling:2012fv}%
  \BibitemOpen
  \bibfield  {author} {\bibinfo {author} {\bibfnamefont {Steven~L.}\ \bibnamefont {Liebling}}\ and\ \bibinfo {author} {\bibfnamefont {Carlos}\ \bibnamefont {Palenzuela}},\ }\bibfield  {title} {\enquote {\bibinfo {title} {{Dynamical Boson Stars}},}\ }\href {\doibase 10.12942/lrr-2012-6} {\bibfield  {journal} {\bibinfo  {journal} {Living Rev. Rel.}\ }\textbf {\bibinfo {volume} {15}},\ \bibinfo {pages} {6} (\bibinfo {year} {2012})},\ \Eprint {http://arxiv.org/abs/1202.5809} {arXiv:1202.5809 [gr-qc]} \BibitemShut {NoStop}%
\bibitem [{\citenamefont {Valdez-Alvarado}\ \emph {et~al.}(2013)\citenamefont {Valdez-Alvarado}, \citenamefont {Palenzuela}, \citenamefont {Alic},\ and\ \citenamefont {Ure\~na L\'opez}}]{PhysRevD.87.084040}%
  \BibitemOpen
  \bibfield  {author} {\bibinfo {author} {\bibfnamefont {Susana}\ \bibnamefont {Valdez-Alvarado}}, \bibinfo {author} {\bibfnamefont {Carlos}\ \bibnamefont {Palenzuela}}, \bibinfo {author} {\bibfnamefont {Daniela}\ \bibnamefont {Alic}}, \ and\ \bibinfo {author} {\bibfnamefont {L.~Arturo}\ \bibnamefont {Ure\~na L\'opez}},\ }\bibfield  {title} {\enquote {\bibinfo {title} {Dynamical evolution of fermion-boson stars},}\ }\href {\doibase 10.1103/PhysRevD.87.084040} {\bibfield  {journal} {\bibinfo  {journal} {Phys. Rev. D}\ }\textbf {\bibinfo {volume} {87}},\ \bibinfo {pages} {084040} (\bibinfo {year} {2013})}\BibitemShut {NoStop}%
\bibitem [{\citenamefont {Siemonsen}\ and\ \citenamefont {East}(2021)}]{Siemonsen:2020hcg}%
  \BibitemOpen
  \bibfield  {author} {\bibinfo {author} {\bibfnamefont {Nils}\ \bibnamefont {Siemonsen}}\ and\ \bibinfo {author} {\bibfnamefont {William~E.}\ \bibnamefont {East}},\ }\bibfield  {title} {\enquote {\bibinfo {title} {{Stability of rotating scalar boson stars with nonlinear interactions}},}\ }\href {\doibase 10.1103/PhysRevD.103.044022} {\bibfield  {journal} {\bibinfo  {journal} {Phys. Rev. D}\ }\textbf {\bibinfo {volume} {103}},\ \bibinfo {pages} {044022} (\bibinfo {year} {2021})},\ \Eprint {http://arxiv.org/abs/2011.08247} {arXiv:2011.08247 [gr-qc]} \BibitemShut {NoStop}%
\bibitem [{\citenamefont {Collodel}\ \emph {et~al.}(2017)\citenamefont {Collodel}, \citenamefont {Kleihaus},\ and\ \citenamefont {Kunz}}]{Collodel:2017biu}%
  \BibitemOpen
  \bibfield  {author} {\bibinfo {author} {\bibfnamefont {Lucas~G.}\ \bibnamefont {Collodel}}, \bibinfo {author} {\bibfnamefont {Burkhard}\ \bibnamefont {Kleihaus}}, \ and\ \bibinfo {author} {\bibfnamefont {Jutta}\ \bibnamefont {Kunz}},\ }\bibfield  {title} {\enquote {\bibinfo {title} {{Excited Boson Stars}},}\ }\href {\doibase 10.1103/PhysRevD.96.084066} {\bibfield  {journal} {\bibinfo  {journal} {Phys. Rev. D}\ }\textbf {\bibinfo {volume} {96}},\ \bibinfo {pages} {084066} (\bibinfo {year} {2017})},\ \Eprint {http://arxiv.org/abs/1708.02057} {arXiv:1708.02057 [gr-qc]} \BibitemShut {NoStop}%
\bibitem [{\citenamefont {Sanchis-Gual}\ \emph {et~al.}(2022)\citenamefont {Sanchis-Gual}, \citenamefont {Herdeiro},\ and\ \citenamefont {Radu}}]{Sanchis-Gual:2021phr}%
  \BibitemOpen
  \bibfield  {author} {\bibinfo {author} {\bibfnamefont {Nicolas}\ \bibnamefont {Sanchis-Gual}}, \bibinfo {author} {\bibfnamefont {Carlos}\ \bibnamefont {Herdeiro}}, \ and\ \bibinfo {author} {\bibfnamefont {Eugen}\ \bibnamefont {Radu}},\ }\bibfield  {title} {\enquote {\bibinfo {title} {{Self-interactions can stabilize excited boson stars}},}\ }\href {\doibase 10.1088/1361-6382/ac4b9b} {\bibfield  {journal} {\bibinfo  {journal} {Class. Quant. Grav.}\ }\textbf {\bibinfo {volume} {39}},\ \bibinfo {pages} {064001} (\bibinfo {year} {2022})},\ \Eprint {http://arxiv.org/abs/2110.03000} {arXiv:2110.03000 [gr-qc]} \BibitemShut {NoStop}%
\bibitem [{\citenamefont {Castelo~Mourelle}(2025)}]{CasteloMourelle:2025ujn}%
  \BibitemOpen
  \bibfield  {author} {\bibinfo {author} {\bibfnamefont {Jorge}\ \bibnamefont {Castelo~Mourelle}},\ }\emph {\bibinfo {title} {{Universal relations for bosonic stars}}},\ \href@noop {} {Ph.D. thesis},\ \bibinfo  {school} {U. Santiago de Compostela (main)} (\bibinfo {year} {2025})\BibitemShut {NoStop}%
\bibitem [{\citenamefont {Kleihaus}\ \emph {et~al.}(2005)\citenamefont {Kleihaus}, \citenamefont {Kunz},\ and\ \citenamefont {List}}]{Kleihaus:2005me}%
  \BibitemOpen
  \bibfield  {author} {\bibinfo {author} {\bibfnamefont {Burkhard}\ \bibnamefont {Kleihaus}}, \bibinfo {author} {\bibfnamefont {Jutta}\ \bibnamefont {Kunz}}, \ and\ \bibinfo {author} {\bibfnamefont {Meike}\ \bibnamefont {List}},\ }\bibfield  {title} {\enquote {\bibinfo {title} {{Rotating boson stars and Q-balls}},}\ }\href {\doibase 10.1103/PhysRevD.72.064002} {\bibfield  {journal} {\bibinfo  {journal} {Phys. Rev. D}\ }\textbf {\bibinfo {volume} {72}},\ \bibinfo {pages} {064002} (\bibinfo {year} {2005})},\ \Eprint {http://arxiv.org/abs/gr-qc/0505143} {arXiv:gr-qc/0505143} \BibitemShut {NoStop}%
\bibitem [{\citenamefont {Herdeiro}\ \emph {et~al.}(2024)\citenamefont {Herdeiro}, \citenamefont {Radu},\ and\ \citenamefont {dos Santos Costa~Filho}}]{Herdeiro:2024pmv}%
  \BibitemOpen
  \bibfield  {author} {\bibinfo {author} {\bibfnamefont {Carlos}\ \bibnamefont {Herdeiro}}, \bibinfo {author} {\bibfnamefont {Eugen}\ \bibnamefont {Radu}}, \ and\ \bibinfo {author} {\bibfnamefont {Etevaldo}\ \bibnamefont {dos Santos Costa~Filho}},\ }\bibfield  {title} {\enquote {\bibinfo {title} {{Spinning Proca-Higgs balls, stars and hairy black holes}},}\ }\href@noop {} {\  (\bibinfo {year} {2024})},\ \Eprint {http://arxiv.org/abs/2406.03552} {arXiv:2406.03552 [gr-qc]} \BibitemShut {NoStop}%
\bibitem [{\citenamefont {Sch\"{o}nauer}\ and\ \citenamefont {Schnepf}(1987)}]{fidisol}%
  \BibitemOpen
  \bibfield  {author} {\bibinfo {author} {\bibfnamefont {Willi}\ \bibnamefont {Sch\"{o}nauer}}\ and\ \bibinfo {author} {\bibfnamefont {Eric}\ \bibnamefont {Schnepf}},\ }\bibfield  {title} {\enquote {\bibinfo {title} {Software considerations for the “black box” solver fidisol for partial differential equations},}\ }\href {\doibase 10.1145/35078.35080} {\bibfield  {journal} {\bibinfo  {journal} {ACM Trans. Math. Softw.}\ }\textbf {\bibinfo {volume} {13}},\ \bibinfo {pages} {333–349} (\bibinfo {year} {1987})}\BibitemShut {NoStop}%
\bibitem [{\citenamefont {Sch{\"o}nauer}\ and\ \citenamefont {Wei$\beta$}(1989)}]{schonauer1989efficient}%
  \BibitemOpen
  \bibfield  {author} {\bibinfo {author} {\bibfnamefont {W}~\bibnamefont {Sch{\"o}nauer}}\ and\ \bibinfo {author} {\bibfnamefont {R}~\bibnamefont {Wei$\beta$}},\ }\bibfield  {title} {\enquote {\bibinfo {title} {Efficient vectorizable pde solvers},}\ }\href@noop {} {\bibfield  {journal} {\bibinfo  {journal} {Journal of computational and applied mathematics}\ }\textbf {\bibinfo {volume} {27}},\ \bibinfo {pages} {279--297} (\bibinfo {year} {1989})}\BibitemShut {NoStop}%
\bibitem [{\citenamefont {Sch{\"o}nauer}\ and\ \citenamefont {Adolph}(2001)}]{schonauer2001we}%
  \BibitemOpen
  \bibfield  {author} {\bibinfo {author} {\bibfnamefont {Willi}\ \bibnamefont {Sch{\"o}nauer}}\ and\ \bibinfo {author} {\bibfnamefont {Torsten}\ \bibnamefont {Adolph}},\ }\bibfield  {title} {\enquote {\bibinfo {title} {How we solve pdes},}\ }\href@noop {} {\bibfield  {journal} {\bibinfo  {journal} {Journal of computational and applied mathematics}\ }\textbf {\bibinfo {volume} {131}},\ \bibinfo {pages} {473--492} (\bibinfo {year} {2001})}\BibitemShut {NoStop}%
\bibitem [{\citenamefont {Delgado}(2022)}]{Delgado:2022pwo}%
  \BibitemOpen
  \bibfield  {author} {\bibinfo {author} {\bibfnamefont {Jorge F.~M.}\ \bibnamefont {Delgado}},\ }\emph {\bibinfo {title} {{Spinning Black Holes with Scalar Hair and Horizonless Compact Objects within and beyond General Relativity}}},\ \href@noop {} {Ph.D. thesis},\ \bibinfo  {school} {Aveiro U.} (\bibinfo {year} {2022}),\ \Eprint {http://arxiv.org/abs/2204.02419} {arXiv:2204.02419 [gr-qc]} \BibitemShut {NoStop}%
\end{thebibliography}%


\begin{appendix}
\section{Including unstable solutions}
\label{appendix}

\begin{figure*}[h!]
\centering
\hspace*{-0.7cm}\includegraphics[width=0.5\textwidth]{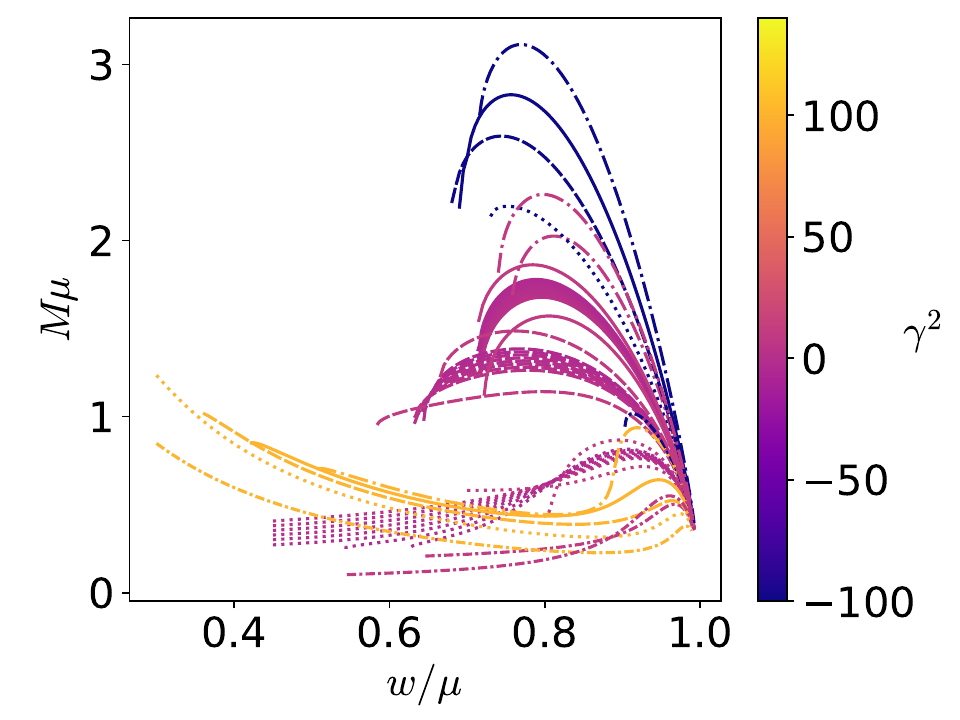}
\caption{Mass-frequency curves for $\gamma^2=-100,-10,-4,-3,-2,-1,0,1,2,3,10,100$ and $\Lambda=-100,-40,0,40,100$. We use the following linestyle codes: densely dashed for $\Lambda=0$, solid for $\Lambda=40$, dashdot for $\Lambda=100$, dotted for $\Lambda=-40$ and densely dashdoted for $\Lambda=-100$.}
\label{masastot}
\end{figure*}

As mentioned in \cref{results2}, as a consistency check we studied the universality for an even broader set of solutions, incorporating additional solution curves for $\gamma^2$ that include the unstable branches to the analysis. In \Cref{masastot}, we present the complete set of analyzed solutions.

\begin{figure}[h!]
\includegraphics[clip,width=1.0\columnwidth]{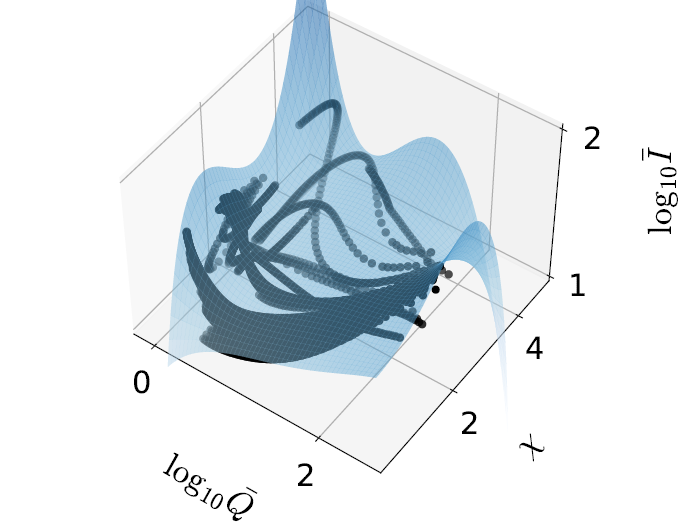}\\%
\includegraphics[clip,width=1.0\columnwidth]{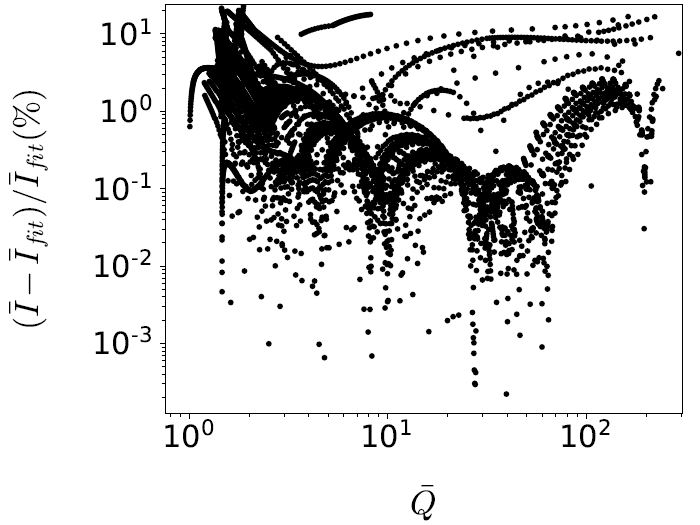}%
\caption{$\eta-\chi-\zeta$ surface for $n=1$ spinning BSs , fitting the data points shown in \cref{masastot}. The lower plot shows the relative difference between data and fitted value.  The maximum error ranges the $22\%$.}
\label{IXQ1_extreme}
\end{figure}

In \Cref{IXQ1_extreme}, and similarly to what was observed in \Cref{results2}, we analyze the behavior of the solutions in the $3D$ space defined by $\eta$. 
The data effectively form a surface that can be fitted with maximum errors of up to $22\%$, compared to $12\%$ in the case where purely unstable solutions were excluded. Despite this deterioration, it is noteworthy that even when including purely unstable and extreme values, we are able to obtain an adjustable surface with reasonably acceptable maximum errors. 

Furthermore, an examination of the error plot reveals that only a small fraction of the data points reach such high error values, with most remaining below the previously obtained $12\%$ threshold. A closer look shows that these larger errors are consistently associated with solutions located in the unstable regions of the mass-frequency curves. Therefore, our results suggest that stability plays a relevant role in the universality of these relations, as unstable solutions tend to be outside of the universality surface.

\subsection{About stability}\label{stability_app}
The stability of BS has been studied from multiple perspectives, depending on the type of configuration considered. For static BS, several approaches have been employed:
(a) Temporal analysis of linear perturbations or mode stability \cite{LEE1989477,GLEISER1989733};
(b) Catastrophe theory and thermodynamic stability methods \cite{Kusmartsev:1990cr,PhysRevD.43.3895,PhysRevD.83.044027,Kleihaus:2011sx};
(c) Nonlinear stability analysis through full numerical evolution of the Einstein-Klein-Gordon (EKG) system \cite{Liebling:2012fv,PhysRevD.87.084040}.

As discussed in \cite{Siemonsen:2020hcg}, these studies generally agree in the stability of spherically symmetric static BS transitions at the extrema of the mass curve. Denoting by $\phi_0$ the central field amplitude of a static BS, the stability changes occur at points where $dM/d\phi_0=0$ in the $M$–$\phi_0$ diagram. Since $\phi_0$ determines the field frequency $w$ in these static cases, equivalent conclusions can be drawn using the $M-w$ diagram.

The self-interaction potential of the scalar field plays a crucial role in stability, helping identify stable branches—similar to how mass–radius relations are used for neutron stars. The structure of the potential determines how many stable branches the model admits.

Turning point arguments used for static BS can also be extended to rotating BS \cite{Kleihaus:2011sx,Collodel:2017biu}, making the $M-w$ diagrams equally useful. According to the turning point criterion, stability transitions occur where $dM/dw$ changes sign.
In some models, this leads to a second or even third branch of solutions—typically more compact configurations of particular interest due to their unique properties.

However, it is essential to emphasize that turning point criteria do not always correspond to thermodynamic or dynamical stability. Recent work has shown that some configurations that appear stable under turning point analysis are, in fact, dynamically unstable. In such cases, nonaxisymmetric instabilities (NAI) can play a central role in the system’s evolution.

Overall, the stability of rotating BS remains a more complex and less understood problem than their static counterparts, warranting further investigation \cite{Siemonsen:2020hcg,Sanchis-Gual:2021phr}.

\subsection{Numerical errors}\label{errors_numerical}
The estimation of numerical errors is a subtle and multifaceted aspect of this study, as it involves several sources of uncertainty across different stages of the computation. First, each boson star solution is obtained via numerical integration and thus inherits residual errors intrinsic to the solver. To ensure high accuracy, we restrict our dataset to configurations exhibiting residual errors in the range of $10^{-9}$ to $10^{-12}$. This level of precision is achieved by increasing the number of solver iterations and carefully selecting initial guesses. It is important to emphasize that these residuals represent a general measure of accuracy and vary across different metric functions, scalar fields, and field frequencies $w$. A more detailed discussion can be found in \cite{CasteloMourelle:2025ujn}.
A particularly delicate source of numerical uncertainty lies in the extraction of multipole moments. The accuracy of the multipolar decomposition depends strongly on the spatial profile of each solution, which in turn is influenced by the field frequency $w$. Stars that are spatially extended contribute to the source at large radii, reducing the quality of the asymptotic fitting. Conversely, ultra-compact configurations lead to stronger spacetime deformations that also complicate the fitting process.
In our approach, multipole moments are not evaluated directly at spatial infinity, but rather inferred through radial fitting over a finite number of data points, based on the known radial behavior of each multipole. The number and distribution of these points depend on the specific configuration, as each solution may require different sampling to yield reliable fits. To quantify the accuracy of this procedure, we define the average statistical error (ASE) for each multipole, calculated as the root mean square deviation between the numerically projected values and the fitted function:

\begin{equation}
ASE = \sqrt{ \frac{1}{N} \sum_{i=1}^{N} \left(c^i_s(r) - c^i_f(r)\right)^2 },
\end{equation}

where $N$ is the number of radial points, $c^i_s(r)$ denotes the simulated source projection, and $c^i_f(r)$ is the corresponding fitted value at radius $r$. This estimator is standard in numerical analysis and offers a robust measure of the fitting accuracy for each multipole and configuration.
It is also essential to consider that the derived errors are for the coefficients that define the multiples, not for the multiples themselves, but the relation is straightforward.
In ~\Cref{ASE_table}, we provide the ASE values for a representative set of solutions with $\gamma^2 = 0$ and $\Lambda = 0$, which serve as a reference for other models studied in this work.

\begin{table}[h!]
\centering
\resizebox{\columnwidth}{!}{%
\begin{tabular}{|l|l|l|l|l|l|l|l|}
\hline
\diagbox{$w$}{ASE$\sim$}& $\nu_0$ & $\nu_2$ &  $\nu_4$  &$\omega_1$ & $\omega_3$  & $B_0$  & $B_2$  \\ \hline
 $0.95$& $1e-7$  &$1e-7$ &$5e-11$ & $2e-8$ & $7e-11$ & $3e-12$ &  $7e-12$\\ \hline
$0.89$ & $1e-7$ & $2e-7$ & $3e-11$& $4e-8$ & $4e-11$  & $2e-12$ & $1e-9$ \\ \hline
 $0.79$& $3e-7$  & $2e-7$ & $2e-11$& $5e-8$  & $2e-11$ & $1e-12$  & $1e-9$ \\ \hline
\end{tabular}%
}
\caption{ASE for the multipolar coefficients and three different stars. We see how the magnitudes can vary depending on the field frequency.}
\label{ASE_table}
\end{table}


\section{Further multipolar relations}
\label{app:multipoles}

\subsection{$S_3-\chi-Q$ and $M_4-\chi-Q$}

For the same data set shown at \Cref{masasc}, we will now present in detail the universal relation between $\chi,Q$ and the spin octupolar moment $S_3$ or the hexadecapolar moment $M_4$. 

\begin{figure}[]
\includegraphics[clip,width=1.0\columnwidth]{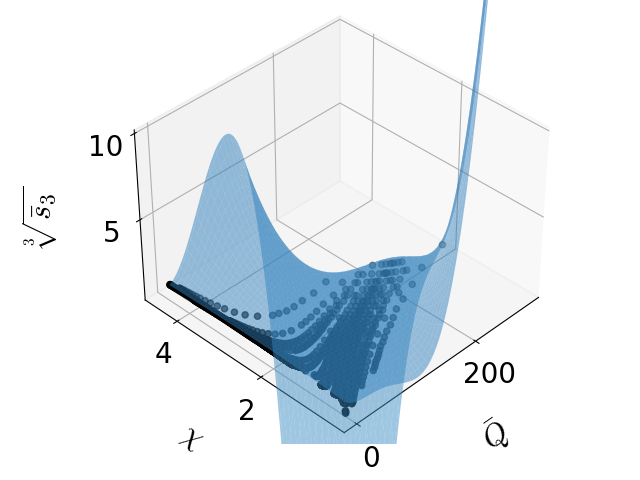}\\%
\includegraphics[clip,width=1.0\columnwidth]{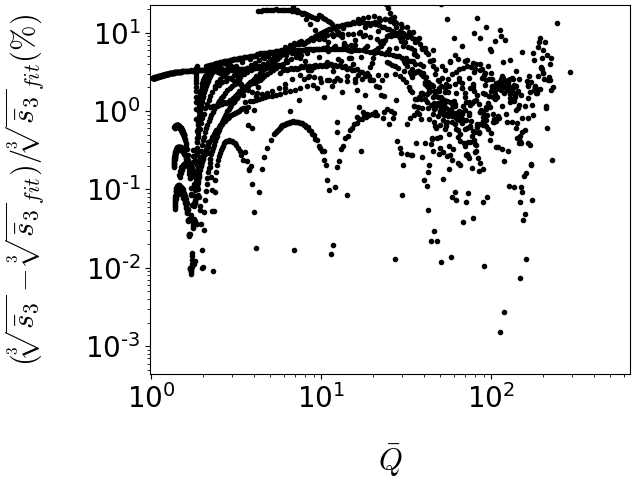}%
\caption{$\sqrt[3]{\bar{s}_3}-\chi-\bar{Q}$ surface for $n=1$ spinning BSs , fitting the data points. The lower plot shows the relative difference between data and fitted value.}
\label{s3xq}
\end{figure}

We begin with the $S_3-X-Q$ relation. Plotting the data in this space \Cref{s3xq}, we see how all the stars define a surface that can be fitted by using the following function,

\begin{equation}
   \sqrt[3]{\bar{s_3}}=A_0+A_s^j\chi^j\left(\Bar{Q}-B\right)^s, 
\end{equation}
where $s={1,2,3}$, $j={0,1,2,3,4,5}$. The fitting coefficients are shown in \Cref{s3QXBSn1fit_Kin}. With a maximum error under the $15\%$, we can also ensure that the universality for this higher-order multipole holds.

An interesting additional feature can be extracted from the upper plot of \Cref{s3xq}. $\bar{Q}$ and $\bar{s_3}$ tend to be a constant value near the Kerr limit in some regions. This phenomenon suggests that, for some models, the space-time resembles the Kerr BH one. This was also studied for usual bosonic star models in \cite{Adam:2024zqr}.
\begin{figure}[]
\includegraphics[clip,width=1.0\columnwidth]{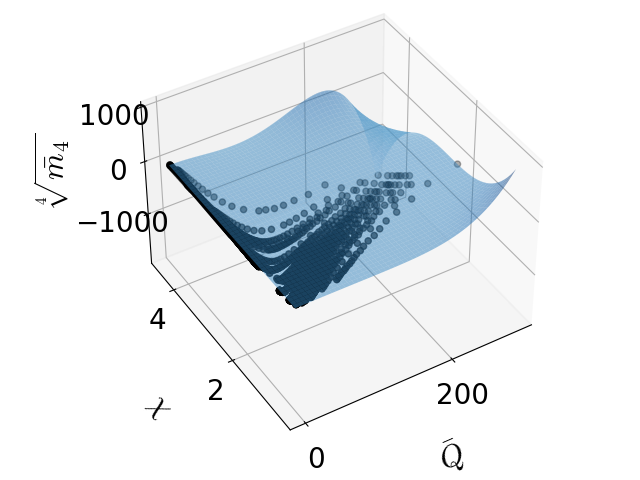}\\%
\includegraphics[clip,width=1.0\columnwidth]{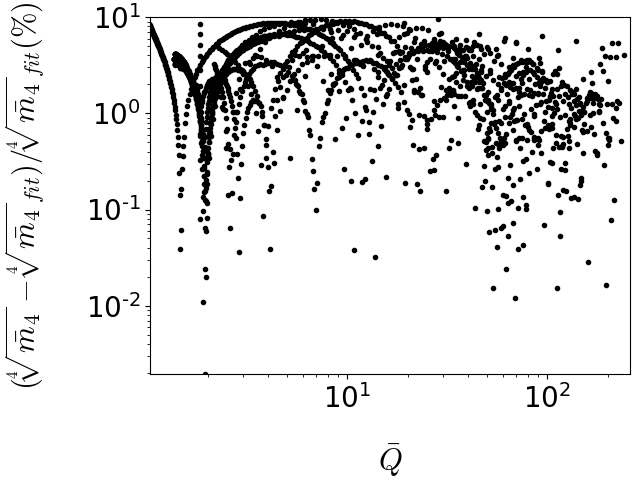}%
\caption{$\sqrt[4]{\bar{m}_4}-\chi-\bar{Q}$ surface for $n=1$ spinning BSs , fitting the data points. The lower plot shows the relative difference between data and fitted value.}
\label{m4xq}
\end{figure}

We perform the same analysis for the mass hexadecapolar moment. As before we take a function of the multipole and plot the stars in the space of parameters defined by $\sqrt[4]{\bar{m}_4}-\chi-\bar{Q}$ in \cref{m4xq}. The solutions define a surface which can be fitted in the following form
\begin{equation}
   \sqrt[4]{\bar{m_4}}=A_0+A_s^j\chi^j\left(\Bar{Q}-B\right)^s, 
\end{equation}
with $s={1,2,3}$, $j={0,1,2,3,4,5}$. The fitting coefficients are shown again in \cref{m4QXBSn1fit_Kin}. With a maximum error under $10\%$, this is a less precise relation. However, it is also the highest-order relation, and these multipole calculations carry out a higher amount of numerical errors. As explained in \cite{Adam:2023qxj}, fitting higher-order multipoles becomes increasingly challenging. This difficulty arises from numerical contamination caused by dominant lower-order multipoles. Extracting higher-order multipoles introduces systematic errors, since the numerical strategy relies on subtracting the lower-order contributions from the source before the projection with the corresponding angular polynomial basis. Nevertheless, without this subtraction technique, the resulting contamination would obscure the true values of the higher-order multipoles \cite{Adam:2023qxj,CasteloMourelle:2025ujn}.Thus, the fact that the error is still under $15\%$ makes this relation acceptable.


\subsection{Compactness}
\begin{figure}[b!]
\includegraphics[clip,width=1.0\columnwidth]{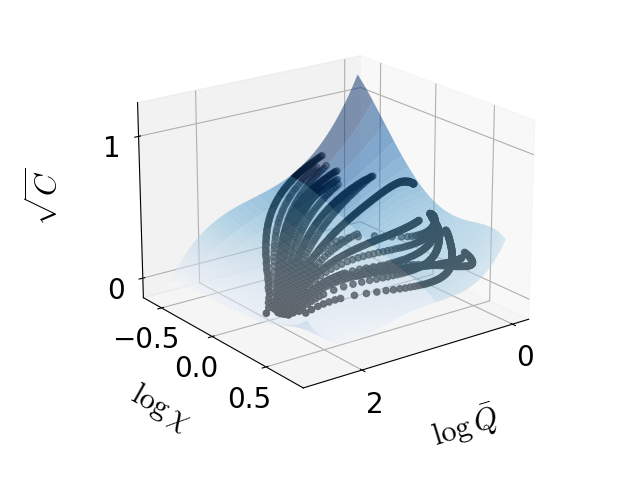}\\%
\includegraphics[clip,width=1.0\columnwidth]{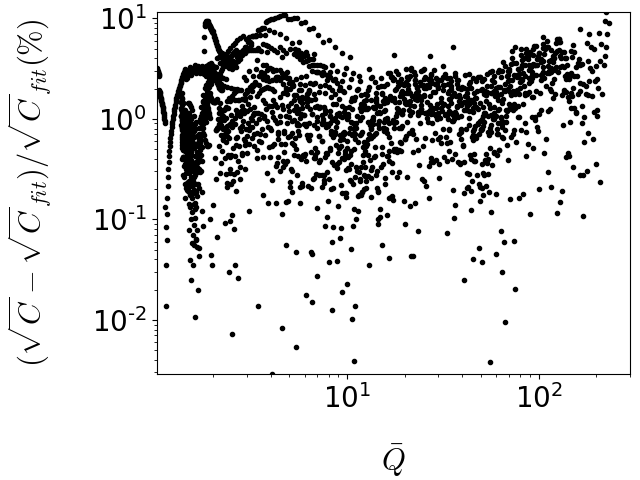}%
\caption{$\sqrt{C}-\chi-\bar{Q}$ surface for $n=1$ spinning BSs, fitting the data points. The lower plot shows the relative difference between the data and the fitted value.}
\label{cxq}
\end{figure}

The last universal relation we showed links the compactness with the quadrupolar and dimensionless spin moments. Again, for the data set shown at \Cref{masasc} In \Cref{cxq} we plot the square root of the compactness against the decimal logarithms of the quadrupolar and spin moments. The data for all the stars in this space define a surface well described by the following function,
\begin{equation}  \sqrt{C}=A_0+A_s^j\log_{10}\chi^j\left(\log_{10}\Bar{Q}-B\right)^s, 
\end{equation}
with $s={1,2,3}$, $j={0,1,2,3,4}$. The values of the fitting coefficients are in  \Cref{CQXBSn1fit_Kin}.

Looking at the lower plot in \cref{cxq}, it is clear that the maximum error is about $\sim 11\%$. This result is quite good, taking into account that the definition of compactness carries some errors from the calculation of $R_{99}$. It is clear that the universality in this space can be important from the empirical perspective since the mass and radii (even with the mentioned difficulty) are measurable quantities. This relation could be used as a tool helping to describe the quadrupolar and inertia moments for some observed GW signals.

\vspace{2cm}

\section{Numerical implementation}

\subsection{Units and boundary conditions }\label{numerical}

In this appendix, we shall review the methodology to perform the numerical integration of the relevant Einstein equations that have also been used in
\cite{Adam:2022nlq,Adam:2023qxj} for the case of the standard flat target space Boson Stars. Further research studying BS or distinct compact objects, uses the same numerical approach \cite{Kleihaus:2005me,Kleihaus:2007vk,Herdeiro:2015gia,Herdeiro:2024pmv}. Radial distances and angular frequencies are rescaled by the mass $\mu$ of the boson field, $
r\rightarrow r\mu, \hspace{0.2cm}w\rightarrow w/\mu$ so that the explicit $\mu$ dependence dissapears from the field equations (although it changes the definitions of coupling constants for the different potentials). 

Upon imposing the stationary, axially symmetric ansatz for the spacetime metric and the scalar field, the Einstein-$\sigma$ model system reduces to a set of five coupled, nonlinear partial differential equations (namely, $4$ equations for the metric functions plus the Euler Lagrange equation for the matter field). We also take into account the constraints $E^r_{\theta}=0, E^r_r-E^{\theta}_{\theta}=0$, where $E^{\mu}_{\nu}=R^{\mu}_{\nu}-\frac{1}{2}Rg^{\mu}_{\nu}-2 T^{\mu}_{\nu}$. 

In order to find a solution to the system, we must impose boundary conditions on the field profile and the metric functions. Asymptotic flatness leads to,
\begin{equation}
    \lim_{r\rightarrow\infty} \alpha =\lim_{r\rightarrow\infty} \beta =\lim_{r\rightarrow\infty} \nu =\lim_{r\rightarrow\infty} W =\lim_{r\rightarrow\infty} \phi = 0.
\end{equation}
Reflection on the rotation axis and  axial symmetry imply that at $\theta=0$ and $\theta=\pi$,
\begin{eqnarray}
    \partial_{\theta}\alpha=\partial_{\theta}\beta=\partial_{\theta}\nu=\partial_{\theta}W=\partial_{\theta}\phi=0. 
\end{eqnarray}
Since the solutions must be symmetric with respect to a reflection along the equatorial plane, this condition is also obeyed on the equatorial plane, $\theta=\pi/2$.
Eventually, regularity at the origin requires $\partial_r \alpha=\partial_r \beta=\partial_r \nu =W= \phi=0$ when $r \to 0$, and regularity on the symmetry axis further imposes $\left.  \alpha=\beta \right|_{\theta=0, \pi}$ \cite{Herdeiro:2015gia}.

Furthermore, the radial coordinate is compactified through the following redefinition $x\equiv r/(c+r)$, which takes $r\in [0,\infty)$ to a finite segment $x \in [0,1]$. Then, the differential equations are discretized on a $(401\times 40)$ grid for $(x,\theta)$, where $0\leq x\leq 1$ and $0\leq \theta\leq \pi/2$. The considerable size of the grid allows us to fix $c=1$.

The system of equations plus boundary conditions on the finite grid specified above is then solved using the FIDISOL/CADSOL package \cite{fidisol,schonauer1989efficient,schonauer2001we}, a Newton-Raphson-based code with an arbitrary grid and
consistency order. It also provides an error estimate for each unknown function. 
More details about the solver can be found in \cite{Delgado:2022pwo,Adam:2022nlq}.

\subsection{The Einstein-Klein Gordon system}

The solver requires the equations to be written in the following specific form:
\begin{equation}  \sin^2{\theta}\Big (r^2F_{i,rr} + F_{i,\theta\theta}\Big) + \mathcal{F}_i(r,\theta;F_j;\partial {F}_j) =  0,\label{generic_form}
\end{equation}
where $F_i(r,\theta)$ denotes the different metric potentials and the scalar field ($i=(\nu,\alpha,\beta,W,\phi)$), and $\mathcal{F}_i$ denotes the remaining (source) terms, containing only the fields and their first derivatives with respect to $r$ and $\theta$.

We manage to obtain the EKG system in the particular form above presented by performing particular linear combinations of the Einstein equations $ {E^{\nu}_{\mu}\equiv G^{\mu}_{\nu}-2\kappa^2T_{\mu}^{\nu}=0}$ together with the Klein-Gordon equation.  We also multiply the equations by appropriate factors in order to avoid terms that may cause numerical divergences such as $1/r$ or $1/\sin(\theta)$. The required combinations of the Einstein equations are: 
\begin{equation}
\begin{split}
&-e^{2\alpha}\frac{r^2}{2}\sin^2(\theta)\left(-E_{t}^{t}+E_{r}^{r}+E_{\theta}^{\theta}-E_{\psi}^{\psi}\right)=0\\
&e^{2\alpha}\frac{r^2}{2}\sin^2(\theta)\left(E_{t}^{t}+E_{r}^{r}+E_{\theta}^{\theta}-E_{\psi}^{\psi}+\frac{2}{r}WE_{\psi}^{t}\right)=0\\
&e^{2\alpha}\frac{r^2}{2}\sin^2(\theta)\left(-E_{t}^{t}+E_{r}^{r}+E_{\theta}^{\theta}-E_{\psi}^{\psi}-\frac{2}{r}WE_{\psi}^{t}\right)=0\\
&\qquad\quad\quad2re^{2\nu +2\alpha-2\beta}E_{\psi}^{t}=0,
\end{split}
\label{EKG-system}
\end{equation}
and the equation of motion for the scalar field is just \Cref{kg} with the prefactor $\frac{e^{2\alpha}r^2\sin^2(\theta)}{\phi}$.

These combinations allow to rewrite the EKG system in terms of 5 independent equations in the form \eqref{generic_form}, where the corresponding source terms are given by:
\vspace{0.4cm}
\begin{widetext}
\begin{align}
\mathcal{F}_\alpha=&\frac{e^{-2( \nu +  \beta)}}{4 ( \gamma^4 \phi^2-4)^2} \left[ -128 \kappa^2 e^{2 (\nu + \beta)} n^2 \phi^2 \right.  + e^{2 \beta} \sin^2\theta W^2 \left( 128 \kappa^2 e^{2 \beta} n^2 \phi^2 - e^{2 \beta} \sin^2\theta ( \gamma^4 \phi^2-4)^2 \right)\notag \\[2mm]
&\quad + 2 e^{2 \beta} r \sin^2\theta W \left( -128 \kappa^2 e^{2 \beta} n w \phi^2 + e^{2 \beta} \sin^2\theta ( \gamma^4 \phi^2-4)^2 W_{,r} \right) - e^{4 \beta} \sin^4\theta ( \gamma^4 \phi^2-4)^2 \left( W_{,\theta}^2 + r^2 W_{,r}^2 \right) \notag\\[2mm]
&\quad + 4 e^{2 \beta} \sin\theta \bigg( 32 \kappa^2 e^{2 \beta} r^2 w^2 \sin\theta \phi^2 + e^{2 \nu} \Big( -( \gamma^4 \phi^2-4)^2 \nu_{,\theta} (\cos\theta + \sin\theta \beta_{,\theta}) \notag\\
&\quad + \sin\theta  ( \gamma^4 \phi^2-4)^2 \big( r \beta_{,r} - \nu_{,r} (1 + r \beta_{,r}) \big) + 32 \sin\theta \kappa^2 \left( \phi_{,\theta}^2 + r^2 \phi_{,r}^2 \right) \Big)\bigg) \bigg];
\end{align}
\vspace{0.5cm}
\begin{align}
\mathcal{F}_\beta&= e^{-2 \nu + 2 \beta} \frac{1}{2}\Big[\sin^4\theta W^2 + \frac{128 \kappa^2  n^2 \phi^2}{( \gamma^4 \phi^2-4)^2} - 2  r \sin^4\theta W W_{,r} +  \sin^4\theta \left( W_{,\theta}^2 + r^2 W_{,r}^2 \right)\Big]+  \sin\theta\cos\theta \left( \nu_{,\theta} + 2 \beta_{,\theta} \right) \notag \\[2mm]
&\quad +  \sin^2\theta \Big[\beta_{,\theta} (\nu_{,\theta} + \beta_{,\theta})   + r \Big( \nu_{,r} + 3 \beta_{,r} + r \big( \kappa^2 e^{2 \alpha} \phi^2 (2 \mu^2 + \Lambda \phi^2) + \beta_{,r} (\nu_{,r} + \beta_{,r}) \big) \Big)  \Big];
\end{align}

\vspace{0.5cm}

\begin{align}
\mathcal{F}_\nu&=\frac{e^{-2 \nu} \sin\theta }{2 ( \gamma^4 \phi^2-4)^2} \left[ -16 e^{2 \beta} \sin^3\theta W^2 - 128 \kappa^2 e^{2 \beta} r^2 w^2 \sin\theta \phi^2 \right. + 256 \kappa^2 e^{2 \beta} n r w \sin\theta W \phi^2 - 128 \kappa^2 e^{2 \alpha} n^2 \sin\theta W^2 \phi^2 \notag\\[2mm]
&\quad + 8 e^{2 \beta} \gamma^2 \sin^3\theta W^2 \phi^2 - e^{2 \beta} \gamma^4 \sin^3\theta W^2 \phi^4 - 16 e^{2 \beta} \sin^3\theta W_{,\theta}^2 + 8 e^{2 \beta} \gamma^2 \sin^3\theta \phi^2 W_{,\theta}^2 \notag\\[2mm]
&\quad - e^{2 \beta} \gamma^4 \sin^3\theta \phi^4 W_{,\theta}^2 + 32 e^{2 \beta} r \sin^3\theta W W_{,r} - 16 e^{2 \beta} \gamma^2 r \sin^3\theta W \phi^2 W_{,r} + 2 e^{2 \beta} \gamma^4 r \sin^3\theta W \phi^4 W_{,r} \notag\\[2mm]
&\quad - 16 e^{2 \beta} r^2 \sin^3\theta W_{,r}^2 + 8 e^{2 \beta} \gamma r^4 \sin^3\theta \phi^2 W_{,r}^2 - e^{2 \beta} \gamma^4 r^2 \sin^3\theta \phi^4 W_{,r}^2 + 2 e^{2 \nu} ( \gamma^4 \phi^2-4)^2 \left( \cos\theta \nu_{,\theta} + \sin\theta \right.\notag \\[2mm]
&\quad \left( \nu_{,\theta} (\nu_{,\theta} + \beta_{,\theta})  + r (2 \nu_{,r} + r (\kappa^2 e^{2 \beta} \phi^2 (2 \mu^2 + \Lambda \phi^2)\right.+ \nu_{,r} (\nu_{,r} + \beta_{,r}) )) ) ) ];
\end{align}

\vspace{0.5cm}

\begin{align}
\mathcal{F}_W=\frac{128 \kappa^2 e^{2( \alpha-\beta)}   \phi^2}{( \gamma^4 \phi^2-4)^2}(  nr w - n^2 W ) + \sin\theta &\Big[ 3 \cos\theta W_{,\theta} + \sin\theta \Big(   (r \nu_{,r}-2 - 3 r \beta_{,r}) (W - r W_{,r})-(\nu_{,\theta} - 3 \beta_{,\theta}) W_{,\theta}\Big) \Big]=0;
\end{align}

\vspace{0.5cm}

\begin{align}
\mathcal{F}_\phi&=\frac{e^{-2 (\nu + \beta)}}{16 ( \gamma^4 \phi^2-4)}  \left[ 16 e^{2 (\nu + \beta)} n^2 \phi (4 + \gamma^4 \phi^2) \right. - e^{2 \beta} \sin\theta \left( 16 e^{2 \beta} \sin\theta (r w - n W)^2 \phi (4 + \gamma^4 \phi^2) \right. \notag\\
&\quad + e^{2 \nu} \bigg( e^{2 \beta} r^2 \sin\theta \phi (\mu^2 + \Lambda \phi^2) ( \gamma^4 \phi^2-4)^3  + 16 \left[ -( \gamma^4 \phi^2-4) (\cos\theta + \sin\theta (\nu_{,\theta} + \beta_{,\theta})) \phi_{,\theta} \right. \notag\\[2mm]
&\quad + 2 \gamma^2 \sin\theta \phi \phi_{,\theta}^2 + \sin\theta \left(   r \phi_{,r}   \left( -( \gamma^4 \phi^2-4) \left( 2 + r (\alpha_{,r} + \beta_{,r}) \right) 
 + 2 \gamma^2 r \phi \phi_{,r}  \right) \right) \bigg)  \bigg]=0;
\end{align}
\section{Fitting coefficients}
\label{app:Coefficients}
\begin{table}[h!]
	\centering
		\begin{tabular}{|c|c|c|c|c|}
			\hline
		 Coeffs & $A_0=  1.47795615$ & $B=    2.14941469$ & $\sim$& $\sim$\\ \hline
   $A_1^0=0.34739609$	& $A_1^1=-0.16420366$  & $A_1^2=   -0.23371246$ & $A_1^3= 0.2392852$ & $A_1^4= -0.04898879$ \\ \hline
	$A_2^0=  -0.16645427$ & $A_2^1= -0.11136905$ &  $A_2^2= -0.0908392$&  $A_2^3=  0.20073476 $&  $A_2^4= -0.04924137 $\\  \hline
	$A_3^0=  -0.08061319$ & $A_3^1=  0.06648312 $& $A_3^2= -0.12334708$& $A_3^3=0.02485992$& $A_3^4=-0.01105319$\\ \hline
		\end{tabular}
		\caption{Coefficients that fit the universal $\eta-\chi-\xi$ surface for $n=1$ BSs, corresponding to the set of data and fitting surface shown in \cref{IXQ1}. }
\label{IQXBSn1fit_Kin}
\end{table}

\begin{table}[h!]
\centering\resizebox{14cm}{!} {
\begin{tabular}{|llll|l|}
\hline
\multicolumn{4}{|l|}{Coeffs}                                         &$A_0=1.64827010$  \\ \hline
\multicolumn{1}{|l|}{$A_{1}^{0}=3.11103883e-1$}    & \multicolumn{1}{l|}{$A_{1}^{1}=-5.24448612e-1$}     & $A_{1}^{2}=2.46947136e-1$      & $A_{1}^{3}=-3.22482380e-2$    &$B_{1}=-5.79441647e1$  \\ \hline
\multicolumn{1}{|l|}{$A_{2}^{0}=1.34076725e-2$}   & \multicolumn{1}{l|}{$A_{2}^{1}=-2.14875646e-2$}      & $A_{2}^{2}= 9.97722836e-3$     & $A_{2}^{3}=-1.27530181e-3$      &$B_{2}=-1.63096747e1$ \\ \hline
\multicolumn{1}{|l|}{$A_{3}^{0}=-4.58643882e-5$}   & \multicolumn{1}{l|}{$A_{3}^{1}=7.44049998e-5$}      & $A_{3}^{2}=-3.48491831e-5$     & $A_{3}^{3}=4.52643968e-6$   &$B_{3}= -7.86809203e1$  \\ \hline
\multicolumn{1}{|l|}{$A_{4}^{0}=1.67865227e-7$}   & \multicolumn{1}{l|}{$A_{4}^{1}=-2.88101454e-7$}      & $A_{4}^{2}=1.40403489e-7$     &  $A_{4}^{3}=-1.92129453e-8$  &$B_{4}=-1.49230288e1$  \\ \hline
\multicolumn{1}{|l|}{$A_{5}^{0}=-1.72132140e-10$}    & \multicolumn{1}{l|}{$A_{5}^{1}=3.61075428e-10$}     & $A_{5}^{2}=-1.90879132e-10$     & $A_{5}^{3}=2.69820708e-11$   &$B_{5}=-8.97730034e-1$  \\ \hline
\end{tabular}
}
\caption{Coefficients that fit the universal $\sqrt[3]{\bar{s_3}}-\chi-Q$ surface for $n=1$ BSs, corresponding to the set of data and fitting surface shown in \cref{s3xq}. }
\label{s3QXBSn1fit_Kin}
\end{table}

\begin{table}[h!]
\centering\resizebox{14cm}{!} {
\begin{tabular}{|lllll|l|}
\hline
\multicolumn{5}{|l|}{Coeffs}                                                                      &$ A_0= 1.18488064$  \\ \hline
\multicolumn{1}{|l|}{$A_{1}^{0}=-2.80071883e-1$} & \multicolumn{1}{l|}{$A_{1}^{1}=1.03200297$} & \multicolumn{1}{l|}{$A_{1}^{2}=-9.06482891e-1$} & $A_{1}^{3}= 3.11089138e-1$ & $A_{1}^{4}=-3.65779871e-2$ & $B_{1}=-3.93371042e-1$ \\ \hline
\multicolumn{1}{|l|}{$A_{2}^{0}=3.46841426e-3$} & \multicolumn{1}{l|}{$A_{2}^{1}=-6.82356132e-3$} & \multicolumn{1}{l|}{$A_{2}^{2}=1.81198713e-03$} & $A_{2}^{3}= 9.62710011e-4$ & $A_{2}^{4}=-3.51019689e-4$ & $B_{2}= 6.66984613$  \\ \hline
\multicolumn{1}{|l|}{$A_{3}^{0}=-1.65389431e-4$} & \multicolumn{1}{l|}{$A_{3}^{1}=6.74954356e-4$} & \multicolumn{1}{l|}{$A_{3}^{2}=-6.17611647e-4$} & $A_{3}^{3}= 2.10380800e-4$ & $A_{3}^{4}=-2.40059423e-5$ & $B_{3}=2.36574372$   \\ \hline
\multicolumn{1}{|l|}{$A_{4}^{0}=4.77843486e-7$} & \multicolumn{1}{l|}{$A_{4}^{1}=-2.37903676e-6$} & \multicolumn{1}{l|}{$A_{4}^{2}=2.35777811e-6$} & $A_{4}^{3}=-8.61590487e-7$ & $A_{4}^{4}=1.06738419e-7$ & $B_{4}=-2.91812198e1$ \\ \hline
\multicolumn{1}{|l|}{$A_{5}^{0}=4.10987473e-10$} & \multicolumn{1}{l|}{$A_{5}^{1}=2.99280392e-9$} & \multicolumn{1}{l|}{$A_{5}^{2}=-4.12556631e-9$} & $A_{5}^{3}=1.75799728e-9$ & $A_{5}^{4}=-2.44406304e-10$ & $B_{5}=9.11658542e-1$  \\ \hline
\end{tabular}
}
\caption{Coefficients that fit the universal $\sqrt[4]{\bar{m_4}}-\chi-Q$ surface for $n=1$ BSs, corresponding to the set of data and fitting surface shown in \cref{m4xq}.
}
\label{m4QXBSn1fit_Kin}
\end{table}

 \begin{table}[h!]
	\centering
		\begin{tabular}{|c|c|c|c|}
			\hline
		 Coeffs & $A_0= 0.0995$ & $B=  2.3671$ &$\sim$\\ \hline
   $A_1^0= -0.2865$	& $A_1^1=  0.1931$  & $A_1^2=  0.0015$  & $A_1^3=  1.6727$ \\ \hline
	$A_2^0=    -0.1449$& $A_2^1=-0.1312$ &  $A_2^2=0.8725 $&  $A_2^3= 3.7680 $\\  \hline
	$A_3^0=  -0.2036$ & $A_3^1=    -0.2946 $& $A_3^2= 0.5836$ & $A_3^3=  2.0845$\\ \hline
        $A_4^0=  -0.0445$ & $A_4^1=  -0.0778$& $A_4^2=  0.1903$ & $A_4^3= 0.4011$\\ \hline
\end{tabular}
\caption{Coefficients that fit the universal $\sqrt{C}-\chi-Q$ surface for $n=1$ BSs, corresponding to the set of data and fitting surface shown in \cref{cxq}.}
\label{CQXBSn1fit_Kin}
\end{table}
\begin{table}[h!]
	\centering
		\begin{tabular}{|c|c|c|c|c|}
			\hline
		 Coeffs & $A_0= 1.58343163$ & $B=  2.40214426$ & $\sim$& $\sim$\\ \hline
   $A_1^0= 3.74934079e-1$	& $A_1^1= -2.31572443e-1$  & $A_1^2=    5.07754830e-2$ & $A_1^3=    -4.43298655e-3$ & $A_1^4=  1.20824431e-4$ \\ \hline
	$A_2^0=   1.98238162e-1$ & $A_2^1=-2.57798417e-1$ &  $A_2^2=  6.90393237e-2 $&  $A_2^3=  -6.33530647e-3 $&  $A_2^4= 1.75569086e-4 $\\  \hline
	$A_3^0=    -9.45481987e-2$ & $A_3^1=  2.58285085e-2 $& $A_3^2=  8.41948922e-2$& $A_3^3= -2.35882788e-3$& $A_3^4=  1.78556670$\\ \hline
		\end{tabular}
		\caption{Coefficients that fit the universal $\eta-\chi-\xi$ surface for $n=1$ BSs  in the extreme region, corresponding to the set of data and fitting surface shown in \cref{high_beta_IXQ}.} 
\label{IQXBSn1fit_Kin_long}
\end{table}

\end{widetext}

\end{appendix}



\end{document}